\documentclass[12pt, a4paper]{article}

\usepackage[total={39pc, 55.5pc}, top=30mm, centering]{geometry}

\usepackage{setspace}
\usepackage{amsmath, amsthm, amssymb, amsfonts, mathrsfs, amsbsy, bm}

\usepackage{graphicx}
\usepackage{xcolor} 

\usepackage{algorithm}
\usepackage{algpseudocode}

\usepackage[nosort]{cite}

\usepackage[hidelinks]{hyperref}

\usepackage{placeins} 

\newcommand{\bbC}{\mathbb{C}}
\newcommand{\bbR}{\mathbb{R}}
\newcommand{\bbZ}{\mathbb{Z}}

\newcommand{\calD}{\mathcal{D}}
\newcommand{\calF}{\mathcal{F}}
\newcommand{\calJ}{\mathcal{J}}
\newcommand{\calK}{\mathcal{K}}
\newcommand{\calL}{\mathcal{L}}
\newcommand{\calM}{\mathcal{M}}
\newcommand{\calO}{\mathcal{O}}
\newcommand{\calP}{\mathcal{P}}
\newcommand{\calR}{\mathcal{R}}
\newcommand{\calS}{\mathcal{S}}

\newcommand{\ReS}{\mathrm{Re}\,S}
\newcommand{\ImS}{\mathrm{Im}\,S}

\newcommand{\circdot}[1]{\overset{\circ}{#1}}

\newcommand{\vev}[1]{\langle #1 \rangle}
\newcommand{\bigvev}[1]{\bigl\langle #1 \bigr\rangle}
\newcommand{\Bigvev}[1]{\Bigl\langle #1 \Bigr\rangle}

\newcommand{\tr}{\mathrm{tr}\,}
\newcommand{\re}{\mathrm{Re}\,}
\newcommand{\im}{\mathrm{Im}\,}

\newcommand{\LieG}{\mathfrak{g}}
\newcommand{\LieGC}{\mathfrak{g}^\bbC}

\newcommand{\dimG}{\dim G}
\newcommand{\with}{{~~\mbox{with}~~}}

\newtheorem{thm}{Theorem}

\makeatletter
\@addtoreset{equation}{section}

\def\@seccntformat#1{\csname the#1\endcsname.~~}
\makeatother

\begin{document}

\begin{titlepage} 

\renewcommand{\thefootnote}{\fnsymbol{footnote}}
\begin{flushright}
  KUNS-3050
\end{flushright}
\vspace*{1.0cm}

\begin{center}
{\Large \bf
Worldvolume Hybrid Monte Carlo algorithm for group manifolds
}
\vspace{1.0cm}

{Masafumi Fukuma}%
\footnote{
  E-mail address: fukuma@gauge.scphys.kyoto-u.ac.jp
}

\vskip 0.5cm
  {\it Department of Physics, Kyoto University,
  Kyoto 606-8502, Japan}

\end{center}

\begin{abstract}
The Worldvolume Hybrid Monte Carlo (WV-HMC) method [arXiv:2012.08468] is a reliable and versatile algorithm for addressing the numerical sign problem. It resolves the ergodicity issues commonly encountered in Lefschetz thimble-based approaches without requiring explicit evaluation of the Jacobian during configuration generation. In this paper, as a general framework for applying WV-HMC to lattice gauge theories, we extend the algorithm to systems defined on connected compact group manifolds. The key is to introduce a symplectic structure on the tangent bundle of the worldvolume and formulate molecular dynamics on it. The validity of the proposed algorithm is demonstrated using one-site $SU(2)$ and $SU(3)$ models for purely imaginary couplings.
\end{abstract}

\end{titlepage}

\pagestyle{empty}
\pagestyle{plain}

\tableofcontents
\setcounter{footnote}{0}

\section{Introduction}
\label{sec:introduction}

The numerical sign problem has long been a major obstacle 
in first-principles computations of various physical systems. 
While a variety of techniques have been developed to address it, 
they are typically tailored to specific systems of interest. 
Over the past fifteen years, 
there has been a growing effort to pursue more versatile solutions. 
Among these, methods based on Lefschetz thimbles 
have attracted considerable attention 
owing to their mathematical rigor rooted in Picard-Lefschetz theory 
\cite{Witten:2010cx,Cristoforetti:2012su,Cristoforetti:2013wha,
Fujii:2013sra,Fujii:2015bua,Fujii:2015vha,
Alexandru:2015xva,Alexandru:2015sua,Alexandru:2017lqr}. 
Nevertheless, it soon became clear that 
the original Lefschetz thimble method generally 
suffers from an ergodicity problem 
\cite{Fujii:2015bua,Fujii:2015vha,Alexandru:2015xva}. 
This issue arises when the integration surface is deformed 
significantly into the complexified configuration space 
in order to sufficiently tame the oscillatory behavior of the path integral. 

A general solution to the tension 
between mitigating the sign problem 
and avoiding the ergodicity problem  
was first provided by the \emph{tempered Lefschetz thimble} 
(TLT) method \cite{Fukuma:2017fjq} 
(see also Ref.~\cite{Alexandru:2017oyw}). 
This established a general algorithmic framework 
for simultaneously addressing both the sign and ergodicity problems. 
The TLT method was subsequently applied to the Hubbard model 
away from half filling \cite{Fukuma:2019wbv}. 
In this approach, the parallel tempering algorithm is incorporated 
into the Lefschetz thimble framework 
by utilizing the deformation parameter as the tempering variable. 
However, a significant limitation is its high computational cost. 
Indeed, 
in order to take into account the difference of volume elements 
between adjacent replicas, 
one needs to compute the Jacobian of the deformation 
at each configuration exchange, 
which requires $O(N^3)$ operations, 
where $N$ is the number of degrees of freedom. 
Furthermore, the number of replicas must be increased 
to maintain an adequate acceptance rate during the exchange process.  
These Jacobian- and replica-related computational limitations 
of the TLT method are avoided 
in the \emph{Worldvolume Hybrid Monte Carlo} (WV-HMC) method 
\cite{Fukuma:2020fez} 
(see also Refs.~\cite{Fukuma:2021aoo,Fukuma:2023eru}), 
in which HMC updates are performed 
on a continuous union of deformed surfaces (the \emph{worldvolume}). 
WV-HMC does not require explicit evaluation of the Jacobian
during configuration generation.
This is because the path integral is reformulated 
as a phase-space integral over the tangent bundle of the worldvolume, 
endowed with a symplectic structure; 
thus, one does not need to take into account changes in the phase-space volume 
as long as the associated molecular dynamics is designed to preserve it.

The application of Lefschetz thimble-based methods to Yang-Mills theories 
was first discussed in the seminal work 
by Cristoforetti \emph{et al.}\ \cite{Cristoforetti:2012su}. 
However, as noted above, 
the original Lefschetz thimble method introduces ergodicity issues. 
The main objective of this paper is 
to extend the WV-HMC algorithm to group manifolds, 
thereby providing a natural framework 
for applications to lattice gauge theories. 
We first develop the framework of complex analysis 
and contour deformation on complexified group manifolds, 
and subsequently introduce the path integral over the worldvolume. 
This is further reformulated as an integral over the tangent bundle 
of the worldvolume, endowed with a symplectic structure. 
The validity of the proposed algorithm is verified 
using one-site $SU(2)$ and $SU(3)$ models with purely imaginary couplings. 
The application of the present formalism to Yang-Mills theories
requires no fundamental modification to the algorithmic structure
and will be presented separately. 

In parallel with the extension of WV-HMC to group manifolds, 
we also discuss the corresponding extension 
of the generalized thimble Hybrid Monte Carlo (GT-HMC) method 
\cite{Alexandru:2019,Fukuma:2019uot}. 
GT-HMC is an algorithm
that performs HMC updates on a fixed deformed integration surface,
using its tangent bundle as the phase space.
This method can suffer from an ergodicity problem for large deformations,
but it is nevertheless useful for evaluating observables on a fixed surface
without evaluating the Jacobian during configuration generation.

This paper is organized as follows. 
In Sect.~\ref{sec:picard-lefschetz_wv-hmc}, 
we briefly review Picard-Lefschetz theory
and the basic ideas underlying 
GT-HMC \cite{Alexandru:2019,Fukuma:2019uot}
and WV-HMC \cite{Fukuma:2020fez} in flat space. 
In Sect.~\ref{sec:GbbC}, 
we define the complexified group $G^\bbC$ of a connected compact Lie group $G$ 
and develop the framework for complex analysis on $G^\bbC$, 
including Cauchy's theorem. 
Section~\ref{sec:outline} introduces the anti-holomorphic gradient flow 
used to deform the integration surface within $G^\bbC$. 
Here, we also provide an outline of the GT-HMC and WV-HMC methods 
for group manifolds. 
Before detailing these algorithms, 
we develop a general theory of molecular dynamics on $G^\bbC$ 
(Sect.~\ref{sec:unconstrained}) 
and on constrained submanifolds of $G^\bbC$ (Sect.~\ref{sec:constrained}). 
The explicit algorithms for GT-HMC and WV-HMC 
are then presented in Sects.~\ref{sec:gt} and~\ref{sec:wv}, respectively. 
In Sect.~\ref{sec:1site}, 
we validate the proposed algorithm 
using one-site $SU(2)$ and $SU(3)$ models with purely imaginary couplings. 
Section~\ref{sec:conclusion} is devoted to conclusions and future perspectives. 
Appendix~\ref{sec:hamiltonian_dynamics} provides 
a general discussion of Hamiltonian dynamics on symplectic manifolds, 
including a brief introduction to the HMC algorithm \cite{Duane:1987de}. 
Appendix~\ref{sec:G} reviews the HMC algorithm for compact groups 
to clarify the correspondence 
between the concepts introduced in the main text 
and those in conventional HMC sampling on a compact group. 
Appendix~\ref{sec:Gaussian} discusses 
a Gaussian sampling of Lie algebra elements for $G$ and $G^\bbC$. 
The remaining appendices contain proofs of various formulas 
used in the main text.

\section{Picard-Lefschetz theory and the WV-HMC algorithm}
\label{sec:picard-lefschetz_wv-hmc}

In this section, 
we briefly review Picard-Lefschetz theory
and the basic ideas underlying the generalized thimble HMC (GT-HMC) 
and Worldvolume HMC (WV-HMC) methods in flat space. 
See Refs.~\cite{Alexandru:2019,Fukuma:2019uot,Fukuma:2023eru} (GT-HMC) 
and~\cite{Fukuma:2020fez,Fukuma:2021aoo,Fukuma:2023eru} (WV-HMC) 
for details. 
In this section only, 
we use the notation 
\begin{align}
  \vev{u,v} \equiv \re u^\dagger v = \re \overline{u^i} v^i
\end{align}
for complex vectors $u = (u^i)$ and $v = (v^i)$.

\subsection{Picard-Lefschetz theory}
\label{sec:picard-lefschetz}

Picard-Lefschetz theory provides 
a systematic framework for the method of steepest descent 
in multidimensional complex integrals, 
a subject with a long history 
\cite{Howls:1997,Pham:1967}. 
The theory has been applied in a variety of contexts, 
including analytic continuation of Chern-Simons theory \cite{Witten:2010cx}, 
real-time quantum dynamics, tunneling, and path integrals 
\cite{Tanizaki:2014xba,Ai:2019fri,Feldbrugge:2025alg}, 
connections with resurgence \cite{Dunne:2016nmc}, 
Lorentzian quantum cosmology \cite{Feldbrugge:2017kzv,Asante:2021phx}, 
and wave-optics problems in astrophysics \cite{Feldbrugge:2019fjs}. 
Its application to the numerical sign problem is another example, 
and will be discussed in the following sections. 

Let us consider an integral of the form 
\begin{align}
  Z_\calO = \int_{\bbR^N} dx\,e^{-S(x)}\,\calO(x).
\end{align}
Here, $x=(x^i)\in \bbR^N$ is a dynamical variable of $N$ degrees of freedom, 
$dx \equiv dx^1\wedge\cdots\wedge dx^N = \prod_i dx^i$ is the flat measure,%
\footnote{
  We omit the symbol $\wedge$ (wedge)
  when no confusion is expected. 
} 
and $e^{-S(x)}$ is the dominant factor in the integrand 
and can be complex-valued. 
We assume that $e^{-S(x)}$ decreases sufficiently rapidly as $|x|\to \infty$ 
so that the integral converges for the observable $\calO(x)$ under consideration. 
In this paper, 
$S(x)$ is the action of a system, 
$e^{-S(x)} / \int dx\,e^{-S(x)}$ the Boltzmann weight, 
and $\calO(x)$ an observable.

As in the saddle-point method, 
we complexify the configuration space $\bbR^N = \{ x = (x^i) \}$ 
to $\bbC^N = \{ z = (z^i) \}$, 
and define \emph{critical points} $\zeta$ 
as points at which the gradient of the action vanishes, 
$\partial S(\zeta) = (\partial_i S(\zeta)) = 0$. 
The Lefschetz thimble associated to a critical point $\zeta$ 
is a higher-dimensional counterpart to the steepest descent path 
and is defined as follows. 

\begin{figure}[t]
  \centering
  \includegraphics[width=90mm]{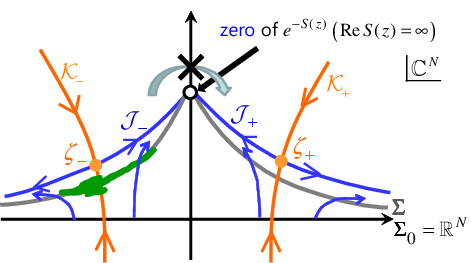}
  \caption{
    Deformation of the integration surface and the ergodicity problem 
    (figure adapted from Ref.~\cite{Fukuma:2023eru}). 
    The original integration surface $\Sigma_0 = \bbR^N$ flows 
    into $\Sigma_t = \Sigma$ within $\bbC^N$. 
    $\zeta_\pm$ are critical points, 
    and $\calJ_\pm$ ($\calK_\pm$) 
    are their associated thimbles (anti-thimbles),
    on each of which $\ImS(z)$ is constant 
    ($\ImS(z) = \ImS(\zeta_\pm)$). 
    The white circle schematically represents the zero set of $e^{-S(z)}$, 
    which acts as an infinitely high barrier 
    for a Monte Carlo walker.
    Note that the arrows represent flow directions 
    (not the orientations of surfaces). 
  }
  \label{fig:ergodicity_problem}
\end{figure}

With respect to the standard Hermitian metric on \(\mathbb C^N\), 
$ds^2 = \overline{dz^i} dz^i$, 
the anti-holomorphic gradient flow 
is defined by the differential equation
with respect to the \emph{flow time} $t$,%
\footnote{ 
  The equation can be generalized as 
  $\dot{z}^i = g^{i\bar{j}}\, \overline{\partial_j S(z)}$,
  where $g^{i\bar{j}}$ denotes (the inverse of) a Hermitian metric 
  satisfying 
  $g^{i\bar{j}} v_i \overline{v_j} \geq 0$ 
  ($\forall v = (v_i) \in \bbC^N$) 
  with equality only when $v = 0$. 
  See, e.g., Ref.~\cite{Fukuma:2020fez} 
  for a possible use of this generalization. 
} 
\begin{align}
  \dot{z}^i = \overline{\partial_i S(z)} \with z^i|_{t=0} = x^i.
\label{ah-flow}
\end{align}
This defines the orbit $z = z(t,x) \in \bbC^N$ 
starting from $x \in \bbR^N$, 
and deforms the original integration surface $\Sigma_0 = \bbR^N$
into $\Sigma_t \equiv \{ z(t,x)\,|\,x \in \bbR^N \}$ 
within $\bbC^N$ 
(see Fig.~\ref{fig:ergodicity_problem}). 
The flow \eqref{ah-flow} satisfies the monotonicity relation 
\begin{align}
  [S(z)]^{\centerdot} = \dot{z} \cdot \partial S(z) = |\partial S(z)|^2 \geq 0,
\end{align}
or equivalently, 
\begin{align}
  [\ReS(z)]^{\centerdot} = |\partial S(z)|^2 \geq 0,
  \qquad
  [\ImS(z)]^{\centerdot} = 0.
\end{align}
This implies that the real part of the action, $\ReS(z)$, 
always increases along the flow except at critical points, 
while the imaginary part $\ImS(z)$ is kept constant along the flow. 
The flow is in the steepest-descent direction 
of $|e^{-S(z)}| = e^{-\ReS(z)}$ 
because, upon writing $z^i = X^i + iY^i$, 
the flow equation \eqref{ah-flow} can be rewritten as 
$\dot{X}^i = \partial_{X^i} \ReS$ 
and $\dot{Y}^i = \partial_{Y^i} \ReS$. 
The Lefschetz thimble $\calJ_\sigma$ 
(anti-thimble $\calK_\sigma$) associated to a critical point $\zeta_\sigma$ 
is defined as the union of orbits flowing out of (flowing into) $\zeta_\sigma$. 
For a nondegenerate critical point,
both $\calJ_\sigma$ and $\calK_\sigma$ are 
real $N$-dimensional submanifolds of $\bbC^N \simeq \bbR^{2N}$,
corresponding respectively to the unstable and stable manifolds of the flow.
Due to the invariance of $\ImS(z)$ along the flow, 
the oscillatory factor $e^{-i\,\ImS(z)}$ is constant over each Lefschetz thimble, 
$e^{-i\,\ImS(z)} = e^{-i\,\ImS(\zeta_\sigma)}$ $(z \in \calJ_\sigma)$. 

In this paper, 
we assume that $e^{-S(z)}$ and $e^{-S(z)}\,\calO(z)$ 
are holomorphic on $\bbC^N$, 
as is often the case in physical applications.%
\footnote{ 
  When $e^{-S(z)}$ has zeros, 
  $S(z)$ itself need not be globally holomorphic. 
  In this case, $\partial S=-\partial\log e^{-S}$
  is understood away from the zero set, 
  which may form part of the boundary of the deformed integration surface.
} 
Then, Cauchy's theorem ensures that 
the integrals $Z_\calO$ and $Z \equiv Z_1$ do not change under the flow:
\begin{align}
  Z_\calO = \int_\Sigma dz\,e^{-S(z)}\,\calO(z)
  = \int_\Sigma dz\, e^{-\ReS(z)}\, e^{-i\,\ImS(z)}\,\calO(z),
\end{align} 
where $dz \equiv dz^1\wedge\cdots\wedge dz^N$ 
is the holomorphic $N$-form. 
In particular, convergence is preserved under the flow 
because $\ReS(z)$ is nondecreasing, 
so that the vanishing of $e^{-S(z)}$ 
at the ends of the original integration surface 
is maintained under the deformation. 
In the limit $t \to \infty$,
the deformed integration surface $\Sigma_t$
approaches a sum of Lefschetz thimbles,
schematically expressed as
\begin{align}
  \Sigma_t \to \sum_\sigma n_\sigma\,\calJ_\sigma,
\end{align}
where the integer coefficients $n_\sigma$ are given
by the intersection numbers between the original integration surface $\bbR^N$
and the anti-thimbles $\calK_\sigma$,%
\footnote{
  We call $\zeta_\sigma$ and $\calJ_\sigma$ \emph{relevant}
  when $n_\sigma \neq 0$.
}
\begin{align}
  n_\sigma = \langle \bbR^N,\,\calK_\sigma \rangle.
\end{align}
Since \(\ImS\) is constant on each Lefschetz thimble, 
the rapidly fluctuating phase originating from \(e^{-i\,\ImS}\) 
is absent within each thimble, 
and thus the integral can be written as
\begin{align}
  Z_\calO = \sum_\sigma\, n_\sigma
  \int_{\calJ_\sigma} dz\,e^{-S(z)}\,\calO(z)
  = \sum_\sigma\, n_\sigma\,e^{-i\,\ImS(\zeta_\sigma)}
  \int_{\calJ_\sigma} dz\,e^{-\ReS(z)}\,\calO(z).
\end{align}
At finite flow time, $\ImS$ is not generally constant on $\Sigma_t$, 
but its fluctuations can be substantially reduced 
as $\Sigma_t$ is deformed toward the relevant thimbles.

\subsection{Lefschetz thimble method and its generalizations}
\label{sec:lt_gt}

Our aim is to numerically estimate 
the expectation value of an observable $\calO(x)$ 
defined by the following integrals over $\mathbb{R}^N = \{ x = (x^a) \}$,%
\footnote{ 
  The indices $a,b,\ldots$ and $i,j,\ldots$
  both run over the same set $\{1,\,2,\,\ldots,\,N\}$,
  the former being often used
  for indices associated with $x\in\bbR^N$.
} 
\begin{align}
  \vev{\calO}
  \equiv \frac{\int_{\bbR^N} dx \, e^{-S(x)} \, \calO(x)}
  {\int_{\bbR^N} dx \, e^{-S(x)}} 
  = \frac{Z_\calO}{Z}.
\label{vev}
\end{align}
For a complex action, 
the Boltzmann weight cannot be regarded as a probability density.  
Naive reweighting with the positive weight $\propto e^{-\ReS(x)}$ gives
\begin{align}
  \vev{ \mathcal{O} }
  = \frac{
    \int_{\bbR^N} dx \, e^{-\ReS(x)} \,  e^{-i\,\ImS(x)}\,\calO(x) 
    \,/\,\int_{\bbR^N} dx \, e^{-\ReS(x)} 
  }{ 
    \int_{\bbR^N} dx \, e^{-\ReS(x)} \, e^{-i\,\ImS(x)}
    \,/\,\int_{\bbR^N} dx \, e^{-\ReS(x)} 
  }
  \equiv \frac{ \vev{e^{-i\,\ImS(x)}\,\calO(x) }_{\mathrm{rewt}} }
         { \vev{e^{-i\,\ImS(x)} }_{\mathrm{rewt}} }.
\label{observable_naive_reweighting}
\end{align}
For a severe sign problem,
both reweighted averages in the numerator and the denominator
are typically of order $e^{-O(N)}$, 
although their ratio can remain $O(1)$ 
for an $O(1)$ observable $\calO$. 
Since their Monte Carlo statistical errors scale as
$O(1/\sqrt{N_{\textrm{conf}}})$,
resolving these exponentially small signals requires
an exponentially large sample size, 
$N_{\textrm{conf}} \sim e^{O(N)}$.

The original Lefschetz thimble method 
is intended to evaluate the integral 
directly over the sum of the relevant thimbles.
However, there are two fundamental issues in this approach: 
(i) the difficulty of identifying all relevant critical points 
$\zeta_\sigma$, thimbles $\calJ_\sigma$, 
and intersection numbers $n_\sigma$, 
and 
(ii) loss of ergodicity due to the infinitely high barriers 
separating different thimbles.%
\footnote{ 
  This multimodal problem is actually important. 
  For example, 
  in the one-dimensional lattice Thirring model, 
  one needs to take into account the contributions from multiple thimbles 
  to reproduce the exact results for observables 
  in the crossover region \cite{Fujii:2015bua}. 
} 
These barriers arise where $e^{-S(z)}$ vanishes 
[$\ReS(z)\to+\infty$], 
either at finite $z$ or asymptotically at infinity. 
Some flow trajectories starting from $\Sigma_0 = \bbR^N$ 
can reach such zeros or diverge to infinity at finite flow times. 
As a result, transitions between different regions of $\Sigma_t$ 
can be completely blocked even at finite $t$.%
\footnote{ 
  For a given flow time $t$, let $B_t\subset\bbR^N$ 
  be the set of initial configurations 
  whose flow trajectories reach a zero of $e^{-S(z)}$ 
  or diverge to infinity at or before time $t$. 
  We then parametrize $\Sigma_t$ 
  by $x\in\bbR^N\setminus B_t$ 
  through the flow map $x\mapsto z(t,x)$. 
  In what follows, this restriction on $x$ is understood 
  whenever the flow map is used to parametrize $\Sigma_t$. 
} 

The \emph{generalized thimble method} \cite{Alexandru:2015sua} instead 
performs the integration over a finite-flow manifold
rather than directly over the thimbles. 
The flow time is chosen large enough to suppress
the fluctuations of the action phase $e^{-i\,\ImS}$, 
but not so large that the multimodal structure leads to loss of ergodicity. 
However, finding a flow time that satisfies 
both requirements is difficult in practice 
\cite{Fukuma:2019wbv}. 

For the present choice of finite-flow manifold, 
namely the flowed surface $\Sigma_t$ defined above, 
the original configurations $x\in\Sigma_0=\bbR^N$ 
naturally parametrize $\Sigma_t$ 
by the flow map $x\mapsto z(t,x)$ at fixed $t$.%
\footnote{ 
  The relative homology class 
  $[\bbR^N] = \sum_\sigma n_\sigma [\calJ_\sigma]$ 
  can be represented by different integration manifolds. 
  In the original proposal \cite{Alexandru:2015sua}, 
  the finite-flow integration manifolds were constructed 
  by flowing the tangent space at a critical point 
  (parallel to $\bbR^N$ in the model studied there), 
  and the algorithm was called the \emph{contraction algorithm}. 
  In the TLT formulation \cite{Fukuma:2017fjq}, 
  the original integration surface $\Sigma_0=\bbR^N$ 
  was used directly as the initial manifold. 
} 
Writing $z(t,x)=z_t(x)$, we have 
\begin{align}
  \vev{\calO} 
  &= \frac{\int_{\Sigma_t} dz\, e^{-S(z)}\,\calO(z)}
     {\int_{\Sigma_t} dz\, e^{-S(z)}}
\nonumber
\\
  &= \frac{\int_{\bbR^N} dx\, \det E_t(x)\, e^{-S(z_t(x))}\,\calO(z_t(x))}
     {\int_{\bbR^N} dx\, \det E_t(x)\, e^{-S(z_t(x))}}
\nonumber
\\
  &\equiv \frac{\vev{e^{i\,\theta_t(x)}\,\calO(z_t(x))}_{\Sigma_t}}
          {\vev{e^{i\,\theta_t(x)}}_{\Sigma_t}}.
\label{gt_start}
\end{align}
Here, $E_t=((E_t)^i{}_a)$ is the Jacobian matrix
with $(E_t)^i{}_a\equiv\partial z_t^i/\partial x^a$,
and the reweighted average $\vev{\cdots}_{\Sigma_t}$ is defined as 
\begin{align}
  \vev{g(x)}_{\Sigma_t} \equiv 
  \frac{\int_{\bbR^N} dx\, e^{-S_t(x)}\,g(x)}
  {\int_{\bbR^N} dx\, e^{-S_t(x)}}
\end{align}
with 
\begin{align}
  e^{-S_t(x)} &\equiv e^{-\ReS(z_t(x))}\, |\det E_t(x)|,
\\
  e^{i\,\theta_t(x)} &\equiv e^{-i\,\ImS(z_t(x))}\, 
  \frac{\det E_t(x)}{|\det E_t(x)|}.
\end{align}
Note the appearance of $|\det E_t(x)|$ 
in the weight $\propto e^{-S_t(x)}$. 
Its direct evaluation for a dense $N\times N$ matrix
requires $O(N^3)$ operations. 

The tempered Lefschetz thimble (TLT) method 
\cite{Fukuma:2017fjq,Fukuma:2019wbv} 
(see also Ref.~\cite{Alexandru:2017oyw}) 
resolves the multimodal problem 
of the generalized thimble method 
by implementing the parallel tempering algorithm 
with respect to the flow time, 
although it still requires the evaluation of $|\det E|$. 
One introduces replicas at $t = t_r$ $(r = 0, 1, \ldots,R)$
with $t_0=0 < t_1 < \cdots < t_R=T$, 
where $T$ is chosen such that the sign problem
associated with fluctuations of the reweighting factor
$e^{i\theta_T(x)}$ is sufficiently alleviated. 
One then performs updates within each replica with weight
$\propto e^{-S_{t_r}(x_r)}$,
and exchanges configurations between adjacent replicas. 
The exchange probability for two adjacent replicas $r$ and $s$ 
is given by%
\footnote{ 
  If the flow map is not defined for either configuration 
  at its proposed new flow time, 
  the exchange is rejected. 
} 
\begin{align}
  \min\left(1,
  \frac{e^{-S_{t_r}(x_s)-S_{t_s}(x_r)}}
  {e^{-S_{t_r}(x_r)-S_{t_s}(x_s)}}\right).
\end{align}
Configurations can therefore move between modes 
through replicas at smaller flow times, 
while measurements can be made at sufficiently large flow times 
where the sign problem is mild.
This was the first algorithm to address 
the sign and ergodicity problems simultaneously. 
Its main drawback is the computational cost
associated with $|\det E_t(x)|$,
which enters the exchange probability 
to account for the difference of volume elements between adjacent replicas.
In addition, a sufficient number of replicas is required
to maintain adequate acceptance rates for replica exchange.

\subsection{Generalized thimble Hybrid Monte Carlo (GT-HMC)}
\label{sec:gt-hmc}

The explicit evaluation of $|\det E|$ during configuration generation 
in the generalized thimble method 
can be avoided by formulating the HMC algorithm \cite{Duane:1987de} 
on the deformed integration surface. 
We refer to this algorithm as 
\emph{generalized thimble Hybrid Monte Carlo} (GT-HMC) 
\cite{Alexandru:2019,Fukuma:2019uot} 
(see also Ref.~\cite{Fukuma:2023eru}).%
\footnote{ 
  The GT-HMC approach was first presented, 
  together with numerical results, in Ref.~\cite{Alexandru:2019}. 
  An explicit algorithmic description, 
  including the projection procedure, 
  was given in Ref.~\cite{Fukuma:2019uot} 
  in the context of a partial improvement over the TLT algorithm. 
} 
The key idea is to shift the sampling problem 
from the configuration space, 
whose induced volume element varies with the configuration, 
to the corresponding phase space, 
whose symplectic volume element is preserved 
during the numerical integration of Hamilton's equations 
(molecular dynamics) if a symplectic integrator is employed. 
Historically, HMC updates were first introduced 
to the Lefschetz thimble method 
in the pioneering paper~\cite{Fujii:2013sra} 
to sample a Lefschetz thimble directly. 
See Appendices~\ref{sec:hamiltonian_dynamics} 
and~\ref{sec:G} for reviews of the standard HMC algorithm 
in flat space and on compact groups, respectively. 
Both reviews are organized to parallel the presentation 
in the main text. 

In this subsection, 
we will drop the index $t$ for quantities on $\Sigma = \Sigma_t$ 
unless explicit dependence on $t$ needs to be indicated. 
We start from the expression 
\begin{align}
  \vev{\calO} 
  &=
  \frac{\int_\Sigma dz\,e^{-S(z)}\,\calO(z)}
  {\int_\Sigma dz\,e^{-S(z)}}. 
\label{vev_gt_start}
\end{align}
With the embedding $z=z(x)$,
we have $dz^i(x)=E^i_a\,dx^a$,
and thus the holomorphic $N$-form $dz\equiv dz^1\wedge\cdots\wedge dz^N$ 
is given by 
\begin{align}
  dz = \det E\,dx,
\end{align}
as in Eq.~\eqref{gt_start}. 
The Jacobian matrix $E = (E^i_a)$ can be computed 
by integrating the flow equations 
for the vectors $E_a \equiv (E^i_a) \in T_z \Sigma$,%
\footnote{ 
  Here and below, for a submanifold $\calS$ of a manifold $\calM$,
  $T_z\calS$ denotes the tangent space to $\calS$ at $z$, 
  and $N_z\calS$ denotes the orthogonal complement of $T_z\calS$
  in $T_z\calM$ with respect to a prescribed real inner product.
  Tangent vectors $v = (v^i) \in T_z \Sigma_t$ 
  satisfy the flow equations  
  $\dot{v}^i = \overline{\partial_i \partial_j S(z)\, v^j}$, 
  which are obtained 
  by considering the flow of an adjacent configuration $z + \epsilon v$, 
  $(z^i + \epsilon v^i)^\centerdot = \overline{\partial_i S(z + \epsilon v)}$,
  and extracting terms linear in $\epsilon$ 
  \cite{Witten:2010cx,Cristoforetti:2012su}.
} 
\begin{align}
  \dot{E}_a^i = \overline{\partial_i \partial_j S(z)\,E_a^j}
  ~~\text{with}~~
  E^i_a|_{t=0} = \delta^i_a.
\end{align}
The induced metric on $\Sigma$ is then given by 
\begin{align}
  ds_\Sigma^2 = \overline{dz^i(x)} dz^i(x) = \gamma_{ab}(x)\,dx^a dx^b
\end{align}
with 
\begin{align}
  \gamma_{ab} \equiv \vev{E_a, E_b} = \re \overline{E_a^i} E_b^i.
\end{align}
Since the ``reality condition''
$\im \overline{E_a^i}E_b^i=0$
holds in flat space \cite{Fujii:2013sra}, 
the metric can be written as
$\gamma_{ab}=\overline{E_a^i}E_b^i$. 
We thus have the following identity 
for $\gamma\equiv\det(\gamma_{ab})$: 
\begin{align}
  \sqrt{\gamma} = |\det E|\quad \textrm{(flat space)},
\end{align}
from which the induced volume element on $\Sigma$ is given by 
\begin{align}
  |dz|_\Sigma \equiv \sqrt{\gamma}\, dx = |\det E|\,dx. 
\end{align}
Thus, the expression \eqref{vev_gt_start} can be rewritten 
as a ratio of reweighted averages on $\Sigma$,
\begin{align}
  \vev{\calO} 
  = \frac{\int_\Sigma |dz|_\Sigma\,e^{-\ReS(z)}\,\calF(z)\,\calO(z)}
    {\int_\Sigma |dz|_\Sigma\,e^{-\ReS(z)}\,\calF(z)}
  = \frac{\vev{\calF(z)\,\calO(z)}_\Sigma}
    {\vev{\calF(z)}_\Sigma},
\label{gt_rewt}
\end{align}
where the reweighted average $\vev{\cdots}_\Sigma$ is given by 
\begin{align}
  \vev{g(z)}_\Sigma
  \equiv
  \frac{\int_\Sigma |dz|_\Sigma\,e^{-\ReS(z)}\,g(z)}
  {\int_\Sigma |dz|_\Sigma\,e^{-\ReS(z)}},
\label{gt_rewt_avg}
\end{align}
and the reweighting factor $\calF(z)$ takes the form%
\footnote{ 
  The reweighting factor $\calF(z)$ includes the Jacobian phase, 
  so that the residual sign problem 
  (phase cancellation in the Jacobian phase) 
  can be checked 
  by monitoring its average. 
  A global sign problem 
  (phase cancellation among different thimbles) 
  can also be checked by monitoring the average reweighting factor 
  in algorithms such as TLT and WV-HMC, 
  which can sample the vicinities of different thimbles. 
} 
\begin{align}
  \calF(z) \equiv \frac{dz}{|dz|_\Sigma}\,e^{-i\,\ImS(z)} 
  = \frac{\det E}{|\det E|}\,e^{-i\,\ImS(z)}. 
\label{rewtfactor_gt}
\end{align}

To rewrite the integrals in Eq.~\eqref{gt_rewt_avg} 
as phase-space integrals, 
we introduce an auxiliary momentum $p = (p_a)$ 
with $dp = dp_1 \wedge \cdots \wedge dp_N \equiv \prod_a dp_a$. 
We further introduce the cotangent bundle 
$T^\ast \bbR^N = \{ (x,p) \}$ 
with the symplectic potential 
\begin{align}
  a_\Sigma \equiv p_a dx^a
\end{align}
and the symplectic form 
\begin{align}
  \omega_\Sigma \equiv da_\Sigma = dp_a \wedge dx^a,
\end{align}
the latter of which gives 
the symplectic volume form (up to irrelevant sign)
\begin{align}
  d\Omega_\Sigma \equiv \frac{\omega_\Sigma^N}{N!}
  = \pm dx \wedge dp. 
\end{align}
Then, using the identity%
\footnote{ 
  Overall Gaussian normalization factors, such as $(2\pi)^{-N/2}$ here,
  will be omitted below,
  since they cancel in normalized expectation values.
} 
\begin{align}
  \sqrt{\gamma}
  = \frac{1}{(2\pi)^{N/2}}
    \int dp\,e^{-(1/2)\,\gamma^{ab}p_a p_b} 
  \quad
  [(\gamma^{ab}) \equiv (\gamma_{ab})^{-1}],
\end{align}
the integral of an arbitrary function $f(z)$ over $\Sigma$ 
with the induced volume $|dz|_\Sigma$ 
can be replaced by an integral over $T^\ast \bbR^N$ as 
\begin{align}
  &\int_\Sigma |dz|_\Sigma\,f(z) 
  = \int_{\bbR^N} dx\,\sqrt{\gamma(x)}\,f(z(x))
\nonumber
\\
  &\propto
  \int_{T^\ast\bbR^N} dx \wedge dp\,
  e^{-(1/2)\,\gamma^{ab}(x)\,p_a p_b}\,f(z(x))
  = \pm \int_{T^\ast\bbR^N} d\Omega_\Sigma\,
  e^{-(1/2)\,\gamma^{ab}(x)\,p_a p_b}\,f(z(x)), 
\end{align}
and thus the average \eqref{gt_rewt_avg} 
can be rewritten as a ratio of phase-space integrals 
over $T^\ast \bbR^N$,
\begin{align}
  \vev{g(z)}_\Sigma
  = \frac{\int_{T^\ast \bbR^N} d\Omega_\Sigma\,
    e^{-(1/2)\,\gamma^{ab}(x)\,p_a p_b - \ReS(z(x))}\,g(z(x))}
    {\int_{T^\ast \bbR^N} d\Omega_\Sigma\,
    e^{-(1/2)\,\gamma^{ab}(x)\,p_a p_b - \ReS(z(x))}}.
\label{vev_Sigma_flat}
\end{align}

For practical Monte Carlo computations, 
it is more convenient to rewrite everything 
in terms of quantities in the ``target space'' $\bbC^N$ 
at the cost of projecting them onto $\Sigma$. 
In this paper,
we adopt the RATTLE algorithm
\cite{Andersen:1983,Leimkuhler:1994} for projection.
There, we introduce the lifted momentum (or more precisely ``velocity'') 
$\pi = (\pi^i) \in T_z \Sigma$ by 
\begin{align}
  \pi^i = E^i_a\, p^a
  \quad
  (p^a \equiv \gamma^{ab} p_b),
\end{align}
and consider the tangent bundle as the phase space, 
\begin{align}
  T\Sigma = \{ (z,\pi)\,|\,z \in \Sigma,\,\pi \in T_z \Sigma \}.
\end{align}
Then, one can show the identities 
\begin{align}
  a_\Sigma &= \vev{\pi, dz} = \re \overline{\pi^i} dz^i,
\\
  \omega_\Sigma &= d\vev{\pi, dz}= \re \overline{d\pi^i} \wedge dz^i,
\\
  \gamma^{ab}\,p_a p_b &= \vev{\pi,\pi},
\end{align}
from which the expression \eqref{vev_Sigma_flat} can be rewritten 
with integrals over $T\Sigma$ as
\begin{align}
  \vev{g(z)}_\Sigma
  = \frac{\int_{T\Sigma} d\Omega_\Sigma\,e^{-H(z,\pi)}\,g(z)}
  {\int_{T\Sigma} d\Omega_\Sigma\,e^{-H(z,\pi)}},
\end{align}
where the Hamiltonian $H(z,\pi)$ is given by 
\begin{align}
  H(z,\pi) = \frac{1}{2}\,\vev{\pi,\pi} + \ReS(z). 
\end{align}

We can now apply HMC using molecular dynamics on $T\Sigma$ 
\cite{Alexandru:2019,Fukuma:2019uot,Fukuma:2023eru}. 
The momentum $\pi \in T_z\Sigma$ 
is refreshed according to the Gaussian weight 
$e^{-\vev{\pi,\pi}/2}$, 
and molecular dynamics is implemented by the RATTLE algorithm 
\cite{Andersen:1983,Leimkuhler:1994}, 
which constrains $z$ and $\pi$ to satisfy 
$z\in\Sigma$ and $\pi\in T_z\Sigma$. 
When the constraint equations are solved exactly, 
the RATTLE update is reversible and symplectic. 
Symplecticity guarantees preservation of the symplectic volume form 
$d\Omega_\Sigma$.
Volume preservation and reversibility ensure 
the appropriate detailed-balance condition 
when combined with the usual Metropolis correction.

Thus, explicit evaluation of $|\det E|$
is not required during configuration generation.
The phase of $\det E$, however, 
still needs to be evaluated for measurements
through $\calF$ in Eq.~\eqref{gt_rewt}.

\subsection{Worldvolume Hybrid Monte Carlo (WV-HMC)}
\label{sec:wv-hmc}

The TLT method needs to compute $|\det E|$ at every replica exchange 
although it addresses the ergodicity problem. 
On the other hand, 
the GT-HMC method need not compute $|\det E|$ during configuration generation 
but can suffer from the ergodicity problem.
The WV-HMC method \cite{Fukuma:2020fez} was introduced 
to combine these advantages: 
it need not compute $|\det E|$ during configuration generation 
while addressing the ergodicity problem. 

\begin{figure}[t]
  \centering
  \includegraphics[width=90mm]{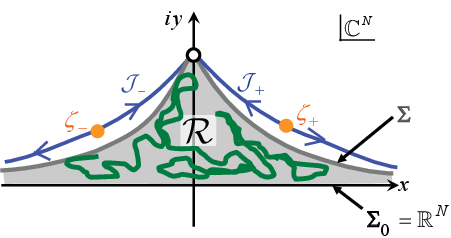}
  \caption{
    Worldvolume $\calR$
    (figure adapted from Ref.~\cite{Fukuma:2023eru}). 
    A Monte Carlo walker on $\Sigma$ 
    can move from one branch to another 
    by passing through a detour in $\calR$. 
  }
  \label{fig:worldvolume}
\end{figure}

We again start from the following expression [see Eq.~\eqref{vev_gt_start}]:
\begin{align}
  \vev{\calO} 
  &= \frac{\int_{\Sigma_t} dz\, e^{-S(z)}\,\calO(z)}
     {\int_{\Sigma_t} dz\, e^{-S(z)}}.
\label{vev_wv_fixed_t}
\end{align}
Since neither the numerator nor the denominator in Eq.~\eqref{vev_wv_fixed_t} 
depends on $t$,
we can take averages over $t$ 
with an arbitrary common weight $e^{-W(t)}$:
\begin{align}
  \vev{\calO}
  = \frac{\int dt \, e^{-W(t)} \int_{\Sigma_t} dz \, e^{-S(z)}\,\calO(z)}
  {\int dt \, e^{-W(t)} \int_{\Sigma_t} dz \, e^{-S(z)}},
\label{vev_wv_start}
\end{align}
which can be viewed 
as a ratio of integrals over the \emph{worldvolume} $\calR$ 
(see Fig.~\ref{fig:worldvolume}) defined by 
\begin{align}
  \calR\equiv \bigcup_{t} \Sigma_t=\{z(t,x) \mid t\in\bbR,\,x\in\bbR^N\}. 
\end{align} 
Monte Carlo sampling over $\calR$ provides detours 
that connect regions originally separated by the potential barriers, 
and can be regarded as a continuum version of the TLT algorithm.  

\begin{figure}[t]
  \centering
  \includegraphics[width=70mm]{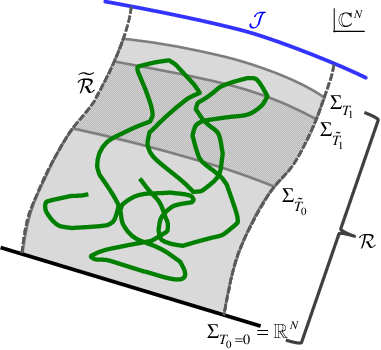}
  \caption{
    Worldvolume $\calR$. 
    Measurements can be restricted to a subregion $\tilde{\calR}$ 
    to improve the signal-to-noise ratio 
    \cite{Fukuma:2020fez,Fukuma:2021aoo}.
  }
  \label{fig:md_flat}
\end{figure}

One can effectively constrain the extent of $\calR$ in the $t$-direction 
to a finite interval $[T_0, T_1]$ (see Fig.~\ref{fig:md_flat}) 
by adjusting the functional form of $W(t)$  
\cite{Fukuma:2023eru}; 
a possible form is discussed in Sect.~\ref{sec:wv_boundary}. 
The lower cutoff $T_0$ is chosen  
such that ergodicity issues are absent on $\Sigma_t$ at $t \sim T_0$. 
The upper cutoff $T_1$ is chosen 
such that oscillatory integrals are sufficiently tamed on $\Sigma_t$ at $t \sim T_1$. 
The latter can be monitored 
by evaluating the average reweighting factor
$\vev{\calF(z)}_{\Sigma_t}$
at various $t$ using GT-HMC \cite{Fukuma:2020fez}. 
After global equilibrium is established across $\calR$, 
we estimate the expectation value $\vev{\mathcal{O}}$ 
using sample averages over a subregion $\tilde\calR$ 
corresponding to a subinterval $[\tilde{T}_0,\tilde{T}_1]$ 
($T_0 \lesssim \tilde{T}_0 < \tilde{T}_1 \lesssim T_1$; 
see Fig.~\ref{fig:md_flat}). 
When necessary, 
this subinterval can exclude the small flow-time region 
that suffers from the sign problem 
as well as the large flow-time region 
that may be difficult to sample due to its complicated geometry. 
The subinterval $[\tilde{T}_0,\tilde{T}_1]$ is determined by the condition
that the estimate of $\vev{\calO}$ remains stable 
within small statistical errors 
against small changes of the subinterval \cite{Fukuma:2020fez}. 

We parametrize $\calR$ 
by the $(N+1)$-dimensional coordinates 
$\hat{x} = (\hat{x}^\mu) = (\hat{x}^0 = t, \hat{x}^a = x^a)$ 
and write  
$d\hat{x} \equiv d\hat{x}^0\wedge d\hat{x}^1\wedge\cdots\wedge d\hat{x}^N
=dt\wedge dx$. 
With the embedding $z=z(\hat{x})=z(t,x)$, 
we define 
\begin{align}
  \hat{E}^i_\mu \equiv \frac{\partial z^i}{\partial \hat{x}^\mu}
\end{align}
and expand the differentials $dz^i(\hat{x})$ as
\begin{align}
  dz^i = \hat{E}^i_\mu\,d\hat{x}^\mu 
  = \xi^i dt + E^i_a dx^a.
\end{align}
The $(N+1)$-form
$dt\wedge dz\equiv dt\wedge dz^1\wedge\cdots\wedge dz^N$
on $\calR$ is then given by 
\begin{align}
  dt \wedge dz = \det E\,dt \wedge dx.
\label{dtdz}
\end{align}
The induced metric on $\calR$ takes the form
\begin{align}
  ds_\calR^2 = \overline{dz^i(\hat{x})} dz^i(\hat{x}) 
  = \hat\gamma_{\mu\nu}(\hat{x})\,d\hat{x}^\mu d\hat{x}^\nu
\end{align}
with 
\begin{align}
  \hat\gamma_{\mu\nu} \equiv \vev{\hat{E}_\mu, \hat{E}_\nu}
  = \re \overline{\hat{E}_\mu^i} \hat{E}_\nu^i,
\end{align}
from which the induced volume element on $\calR$ is given by 
\begin{align}
  |dz|_\calR \equiv \sqrt{\hat\gamma}\, d\hat{x} 
  = \sqrt{\hat\gamma}\,dt \wedge dx
  \quad
  (\hat\gamma\equiv\det(\hat\gamma_{\mu\nu}) ). 
\end{align}

To obtain an explicit form of  $\sqrt{\hat\gamma}$, 
we decompose the vector $\xi(z) = \overline{\partial S(z)} \in T_z\calR$ 
at $z = z(t,x)$ 
into the tangential and normal components 
along $\Sigma_t$ as 
\begin{align}
  \xi^i = \xi_v^i + \xi_n^i
  \quad
  (\xi_v \in T_z \Sigma_t,~\xi_n \in N_z \Sigma_t),
\end{align}
from which we have 
\begin{align}
  dz^i = \xi_n^i dt + E_a^i\,(dx^a + \beta^a dt)
\end{align}
with the shift functions
\begin{align}
  \beta^a \equiv \gamma^{ab}\,\vev{E_b, \xi_v}.
\end{align}
The metric $ds_\calR^2$ can then be expressed in the ADM form, 
\begin{align}
  ds_\calR^2 = \alpha^2\,dt^2 
  + \gamma_{ab}\,(dx^a + \beta^a dt)\,(dx^b + \beta^b dt),
\end{align}
where the lapse function $\alpha$ is given by 
\begin{align}
  \alpha \equiv \sqrt{\vev{\xi_n, \xi_n}}. 
\end{align}
We thus have the relation $\sqrt{\hat\gamma} = \alpha\,\sqrt{\gamma}$, 
and the induced volume element on $\calR$ is given by 
\begin{align}
  |dz|_\calR \
  = \alpha \sqrt{\gamma}\,dt \wedge dx 
  = \alpha\, |\det E|\,dt \wedge dx.
\label{dz_R}
\end{align}
Combining Eqs.~\eqref{dtdz} and \eqref{dz_R}, 
the expression \eqref{vev_wv_start} can be rewritten as 
\begin{align}
  \vev{\calO} 
  = \frac{\int_\calR |dz|_\calR\,e^{-V(z)}\,\calF(z)\,\calO(z)}
    {\int_\calR |dz|_\calR\,e^{-V(z)}\,\calF(z)}
  = \frac{\vev{\calF(z)\,\calO(z)}_\calR}
    {\vev{\calF(z)}_\calR}.
\label{wv_rewt}
\end{align}
Here, $V(z)$ and $\calF(z)$ denote 
the potential and the reweighting factor, respectively,%
\footnote{ 
  Here, $t(z)$ denotes the flow-time coordinate 
  in the parametrization $z=z(t,x)$ of $\calR$.
} 
\begin{align}
  V(z) 
  &\equiv \ReS(z) +  W(t(z)),
\\
  \calF(z) 
  &\equiv \frac{dt \wedge dz}{|dz|_\calR}\,e^{-i\,\ImS(z)} 
  = \alpha^{-1} \frac{\det E}{|\det E|}\,e^{-i\,\ImS(z)}. 
\end{align}
The reweighted average $\vev{\cdots}_\calR$ is defined by 
\begin{align}
  \vev{g(z)}_\calR
  \equiv
  \frac{\int_\calR |dz|_\calR\,e^{-V(z)}\,g(z)}
  {\int_\calR |dz|_\calR\,e^{-V(z)}}.
\label{wv_rewt_avg}
\end{align}

To rewrite the integrals in Eq.~\eqref{wv_rewt_avg} 
as phase-space integrals, 
we introduce an auxiliary momentum $\hat{p} = (\hat{p}_\mu)$ 
with 
$d\hat{p} 
= d\hat{p}_0 \wedge d\hat{p}_1 \wedge \cdots \wedge d\hat{p}_N 
\equiv \prod_\mu d\hat{p}_\mu$. 
We further introduce the cotangent bundle 
$T^\ast \bbR^{N+1} = \{ (\hat{x},\hat{p}) \}$ 
with the symplectic potential 
\begin{align}
  a_\calR \equiv \hat{p}_\mu d\hat{x}^\mu
\end{align}
and the symplectic form 
\begin{align}
  \omega_\calR \equiv da_\calR = d\hat{p}_\mu \wedge d\hat{x}^\mu,
\end{align}
the latter of which gives the symplectic volume form 
(up to irrelevant sign)
\begin{align}
  d\Omega_\calR \equiv \frac{\omega_\calR^{N+1}}{(N+1)!}
  = \pm d\hat{x} \wedge d\hat{p}.
\end{align}
Then, using the identity 
\begin{align}
  \sqrt{\hat\gamma}
  \propto \int d\hat{p}\,e^{-(1/2)\,\hat\gamma^{\mu\nu} \hat{p}_\mu \hat{p}_\nu} 
  \quad
  [(\hat\gamma^{\mu\nu}) \equiv (\hat\gamma_{\mu\nu})^{-1}],
\end{align}
the integral of an arbitrary function $f(z)$ over $\calR$ 
with the induced volume $|dz|_\calR$ 
can be replaced by an integral over $T^\ast \bbR^{N+1}$ as 
\begin{align}
  \int_\calR |dz|_\calR\,f(z) 
  &= \int_{\bbR^{N+1}} d\hat{x}\,\sqrt{\hat\gamma(\hat{x})}\,f(z(\hat{x}))
\nonumber
\\
  &\propto 
  \int_{T^\ast\bbR^{N+1}} d\hat{x} \wedge d\hat{p}\,
  e^{-(1/2)\,\hat\gamma^{\mu\nu}(\hat{x})\, 
  \hat{p}_\mu \hat{p}_\nu}\,f(z(\hat{x}))
\nonumber
\\
  &= \pm \int_{T^\ast\bbR^{N+1}} d\Omega_\calR\,
  e^{-(1/2)\,\hat\gamma^{\mu\nu}(\hat{x})\, 
  \hat{p}_\mu \hat{p}_\nu}\,f(z(\hat{x})), 
\end{align}
and thus the average \eqref{wv_rewt_avg} 
can be rewritten as a ratio of phase-space integrals 
over $T^\ast \bbR^{N+1}$,
\begin{align}
  \vev{g(z)}_\calR
  = \frac{\int_{T^\ast \bbR^{N+1}} d\Omega_\calR\,
    e^{-(1/2)\,\hat\gamma^{\mu\nu}(\hat{x})\,\hat{p}_\mu \hat{p}_\nu 
    - V(z(\hat{x}))}\,g(z(\hat{x}))}
    {\int_{T^\ast \bbR^{N+1}} d\Omega_\calR\,
    e^{-(1/2)\,\hat\gamma^{\mu\nu}(\hat{x})\,\hat{p}_\mu \hat{p}_\nu 
    - V(z(\hat{x}))}}.
\label{vev_calR_flat}
\end{align}

As in the previous subsection on GT-HMC, 
to rewrite the expression \eqref{vev_calR_flat} 
in terms of quantities in the target space $\bbC^N$, 
we introduce the lifted momentum 
$\pi = (\pi^i) \in T_z \calR$ by 
\begin{align}
  \pi^i = \hat{E}^i_\mu\, \hat{p}^\mu
  \quad
  (\hat{p}^\mu \equiv \hat\gamma^{\mu\nu} \hat{p}_\nu),
\end{align}
and consider the tangent bundle as the phase space, 
\begin{align}
  T\calR = \{ (z,\pi)\,|\,z \in \calR,\,\pi \in T_z \calR \}.
\end{align}
Then, one can show the identities 
\begin{align}
  a_\calR &= \vev{\pi, dz} = \re \overline{\pi^i} dz^i,
\\
  \omega_\calR &= d \vev{\pi, dz} = \re \overline{d\pi^i} \wedge dz^i,
\\
  \hat\gamma^{\mu\nu} \hat{p}_\mu \hat{p}_\nu &= \vev{\pi,\pi},
\end{align}
from which the expression \eqref{vev_calR_flat} can be rewritten 
with integrals over $T\calR$ as 
\begin{align}
  \vev{g(z)}_\calR
  = \frac{\int_{T\calR} d\Omega_\calR\,e^{-H(z,\pi)}\,g(z)}
  {\int_{T\calR} d\Omega_\calR\,e^{-H(z,\pi)}},
\label{vev_wv}
\end{align}
where the Hamiltonian $H(z,\pi)$ is now given by 
\begin{align}
  H(z,\pi) = \frac{1}{2}\,\vev{\pi,\pi} + V(z). 
\end{align}

We can now apply HMC 
using molecular dynamics on $T\calR$ 
\cite{Fukuma:2020fez,Fukuma:2023eru}, 
in parallel with GT-HMC on $T\Sigma$ 
\cite{Alexandru:2019,Fukuma:2019uot,Fukuma:2023eru}. 
The momentum $\pi\in T_z\calR$ is refreshed according to 
the Gaussian weight $e^{-\vev{\pi,\pi}/2}$, 
and molecular dynamics (MD) is implemented by the RATTLE algorithm 
\cite{Andersen:1983,Leimkuhler:1994}, 
which constrains $z$ and $\pi$ 
to satisfy $z\in\calR$ and $\pi\in T_z\calR$. 
When the constraint equations are solved exactly, 
the RATTLE update is reversible and symplectic,  
which ensures the appropriate detailed-balance condition 
when combined with the usual Metropolis correction. 

Thus, as in GT-HMC, explicit evaluation of $|\det E|$ 
is not required during configuration generation. 
The phase of $\det E$, however, 
still needs to be evaluated for measurements 
through the reweighting factor $\calF$ in Eq.~\eqref{wv_rewt}. 

We now compare WV-HMC and GT-HMC. 
A major difference is in the capability for addressing the ergodicity issue, 
although both are free from the need to compute $|\det E|$ 
in configuration generation. 
GT-HMC, however, still has its own advantage 
because it is a fast estimator for observables at a fixed flow time, 
which can be used to determine the maximum flow time $T_1$ in WV-HMC. 
As for computational costs, 
it is difficult to predict which method is faster 
in generating a configuration, 
because this depends on the model and parameters under consideration. 
On the one hand, the computational cost of WV-HMC per MD step 
may be smaller than that of GT-HMC at fixed flow time $T_1$, 
because WV-HMC generally samples smaller flow times $t \lesssim T_1$, 
requiring less computational time for flow integration. 
On the other hand, we often observe 
that WV-HMC requires a smaller MD step size than GT-HMC, 
because a large step size may move the trial configuration 
too far from the worldvolume 
for the RATTLE constraint equations to be solved reliably.

\section{Complex analysis on complexified groups}
\label{sec:GbbC}

In this section, we summarize 
the essential aspects of complex analysis on complexified groups 
to provide the mathematical framework for the subsequent sections.

\subsection{Complexification of a compact group}
\label{sec:GbbC_complexification}

We begin by recalling the definition of the Haar measure $(dU_0)$ 
on a connected compact Lie group $G=\{U_0\}$ of dimension $\dimG = N$.%
\footnote{ 
  We work with a faithful unitary representation of $G$. 
  Its Lie algebra can thus be realized by anti-hermitian matrices. 
  Elements in $G$ are generally written with a subscript 0 as $U_0$ 
  to distinguish them from those (written as $U$) 
  in the complexified group $G^\bbC$ defined below. 
} 
Let $\LieG$ be the Lie algebra of $G$ 
with a basis $\{T_a\}$ $(a=1,\ldots,N)$, 
where the basis elements are chosen to be anti-hermitian $(T_a^\dagger = -T_a)$
and normalized such that $\tr T_a^\dagger T_b = -\tr T_a T_b = \delta_{ab}$. 
We introduce the right-invariant Maurer-Cartan 1-form on $G$ by 
\begin{align}
  \theta_0 \equiv dU_0\,U_0^{-1} = T_a \,\theta_0^a
  \quad
  (\text{$\theta_0^a$: real 1-form}),
\end{align}
which satisfies the Maurer-Cartan equation, 
\begin{align}
  d \theta_0 = \theta_0 \wedge \theta_0,
  \quad \text{or in component form,} \quad
  d \theta_0^a = \frac{1}{2}\,C_{bc}{}^a\,\theta_0^b \wedge \theta_0^c,
\label{Maurer-Cartan0}
\end{align}
where $C_{ab}{}^c \in \bbR$ are the structure constants defined by 
$[T_a,T_b] = C_{ab}{}^c\,T_c$.%
\footnote{
  Note that 
  $C_{abc} \equiv -C_{ab}{}^c = \tr [T_a,T_b]\,T_c$ is 
  completely antisymmetric with respect to its indices.
} 
We then define the Riemannian metric on $G$ by 
\begin{align}
  ds_0^2 \equiv \tr \theta_0^\dagger \theta_0 
  \,(= -\tr \theta_0 \theta_0)
  = (\theta_0^a)^2. 
\end{align}
The last equality identifies $\{\theta_0^a\}$ 
as the vielbein forms (orthonormal coframe) of the metric. 
The Haar measure $(dU_0)$ is defined  
as the volume form associated with this metric,%
\footnote{ 
  We ignore the overall normalization of the Haar measure, 
  because it is irrelevant to the following discussion. 
} 
\begin{align}
  (dU_0) \equiv \theta_0^1\wedge\cdots\wedge\theta_0^N.
\label{MC0}
\end{align}
The metric is bi-invariant (i.e., both left- and right-invariant) 
and inversion invariant, 
and the corresponding Haar volume form satisfies 
$(d(g_L U_0 g_R^{-1})) = (dU_0)$ $(\forall g_L, g_R \in G)$
and $(d(U_0^{-1})) = (-1)^N (dU_0)$.%
\footnote{ 
  The sign $(-1)^N$ is irrelevant 
  when the volume form is used as an integration measure. 
} 

The complexification $G^\bbC$ of $G$ is defined as follows: 
We first complexify the Lie algebra 
$\LieG = \bigoplus_a \bbR\, T_a$ 
to $\LieGC \equiv \bigoplus_a \bbC\, T_a$ as a vector space, 
and introduce the Lie bracket 
as the unique complex-bilinear extension of the original commutator:
\begin{align}
  [X+i Y, X'+i Y'] \equiv 
  ([X,X'] - [Y,Y']) + i\, ([X,Y']+ [Y,X'])
  \quad (X,Y,X',Y' \in \LieG).
\end{align}
The complexified group $G^\bbC = \{ U \}$ is then defined 
as the group generated by $\LieGC$, 
consisting of finite products of exponentials: 
\begin{align}
  G^\bbC \equiv \bigl\{
  U = e^{Z} e^{Z'}\cdots e^{Z''} \,|\,
  Z,Z',\ldots,Z'' \in \LieGC \bigr\}.
\end{align}
$\LieGC$ is identified as the (complex) Lie algebra of $G^\bbC$.%
\footnote{ 
  For $G = SU(n)$ and its Lie algebra $\LieG = \mathfrak{su}(n)$, 
  their complexifications are given 
  by $G^\bbC = SL(n,\bbC)$ and $\LieGC = \mathfrak{sl}(n,\bbC)$. 
} 
We introduce the Maurer-Cartan 1-form on $G^\bbC$ by 
\begin{align}
  \theta \equiv dU U^{-1} = T_a \, \theta^a
  \quad
  (\text{$\theta^a$: complex 1-form}),
\end{align}
which satisfies the Maurer-Cartan equation, 
\begin{align}
  d \theta = \theta \wedge \theta,
  \quad \text{or in component form,} \quad
  d \theta^a = \frac{1}{2}\,C_{bc}{}^a\,\theta^b \wedge \theta^c.
\label{Maurer-Cartan}
\end{align}
We then define the metric on $G^\bbC$ by 
\begin{align}
  ds^2 \equiv \tr \theta^\dagger \theta
  = \overline{\theta^a}\, \theta^a,
\end{align}
which represents the squared infinitesimal distance 
between $U$ and $U+dU$ in $G^\bbC$. 
This metric is still bi-invariant under the compact subgroup $G$, 
but is only right-invariant 
under the full complex group $G^\bbC$. 
We identify $T_U G^\bbC$ with $\LieGC$ 
by representing tangent vectors as $\delta U\,U^{-1}\in\LieGC$.

\subsection{Derivatives of functions on $G^\bbC$}
\label{sec:GbbC_derivative}

Any smooth function $f$ on $G^\bbC$ can be treated 
as a function of 
$U=(U_{ij})$ and $U^\dagger=(U^\dagger_{ij}=\overline{U_{ji}})$, 
regarded as independent variables for differentiation, 
and is denoted by $f(U,U^\dagger)$. 
We define the Lie-algebra-valued derivatives 
$Df(U,U^\dagger)$ and $\bar{D}f(U,U^\dagger) \in \LieGC$ 
through the variation: 
\begin{align}
  \delta f(U,U^\dagger)
  &\equiv \tr \bigl[(\delta U U^{-1})\,Df
    + (\delta U U^{-1})^\dagger \bar{D}f \bigr]
\nonumber
\\
  &= (\delta U U^{-1})^a\, D_a f
    + \overline{(\delta U U^{-1})^a\,} \bar{D}_a f,
\end{align}
where we adopt the conventions%
\footnote{
  Unless otherwise specified, 
  indices are raised and lowered using the rule \(A^a=-A_a\), 
  reflecting the normalization $\tr(T_a T_b)=-\delta_{ab}$. 
} 
\begin{align}
  \delta U U^{-1} &= T_a\,(\delta U U^{-1})^a,
  \qquad
  Df = T_a D^a f = -T_a D_a f,
\nonumber
\\
  (\delta U U^{-1})^\dagger &= T^\dagger_a\,\overline{(\delta U U^{-1})^a},
  \qquad
  \bar{D}f = T_a^\dagger\,\bar{D}^af = -T_a^\dagger \bar{D}_a f.
\end{align} 
Note that 
for a real-valued function $f(U,U^\dagger)$, 
the derivatives satisfy the reality conditions 
$\bar{D}f = (D f)^\dagger$ 
and $\bar{D}_a f = \overline{D_a f}$. 
Higher-order derivatives are defined recursively, 
e.g., $D_a D_b f \equiv D_a (D_b f)$. 

We define an inner product of complex matrices $u$ and $v$ by 
\begin{align}
  \vev{u, v} \equiv \re\tr u^\dagger v 
  \, ( = \vev{v, u} = \vev{u^\dagger,v^\dagger} ).
\end{align} 
Then, for a real-valued function $f(U,U^\dagger)$, 
we have the identity
\begin{align}
  \delta f = 2\,\vev{\,\delta U U^{-1},(Df)^\dagger} ,
\label{delta_f}
\end{align}
which is derived as follows:
\begin{align}
  \delta f 
  &= \tr \bigl[ 
   (\delta U U^{-1})\, Df +  (\delta U U^{-1})^\dagger (Df)^\dagger
  \bigr]
\nonumber
\\
  &= 2\,\re \tr \bigl[ (\delta U U^{-1})\, Df \bigr]
  = 2\,\vev{\,\delta U U^{-1},(Df)^\dagger}. 
\label{deltaf_real}
\end{align}
In terms of the exterior derivative $d$, 
this identity is expressed as 
\begin{align}
  df = 2\,\vev{\theta,(Df)^\dagger}
  \quad
  (\text{$f$: real}).
\label{df}
\end{align}

A function $f = f(U)$ is said to be \emph{holomorphic} 
if it depends holomorphically on the matrix elements $U_{ij}$. 
Such a function satisfies the Cauchy-Riemann condition $\bar{D} f=0$, 
and its variation is given by 
\begin{align}
  \delta f(U) = \tr \bigl[ (\delta U U^{-1})\,Df(U) \bigr]
  = (\delta U U^{-1})^a\,D_a f(U).
\label{deltaf(U)}
\end{align}
Replacing $\delta$ with the exterior derivative $d$, 
we obtain the relation 
\begin{align}
  d f(U) = \tr \theta\,Df(U) = \theta^a D_a f(U)
  \quad
  (\text{$f$: holomorphic}).
\label{df(U)}
\end{align}
For any $\xi \in \LieGC$, 
we have the expansion 
\begin{align}
  f(e^\xi U) &= f(U) + \tr \xi\,Df(U) + O(\xi^2)
\nonumber
\\
  &= f(U) + \xi^a\,D_a f(U) + O(\xi^2).
\end{align}

The derivatives $D_a$ do not commute; 
instead, they satisfy the following commutation relation 
when acting on a holomorphic function $f(U)$:
\begin{align}
  D_a D_b f - D_b D_a f = -C_{ab}{}^c\,D_c f.
\label{DaDb}
\end{align}
A proof is given in Appendix~\ref{sec:DaDb}. 

For a holomorphic function $f(U)$, 
we define its Hessian matrix $H(U)=(H_{ab}(U))$ 
of size $N \times N$ by 
\begin{align}
  H_{ab}(U) \equiv D_b D_a f(U). 
\end{align}
Its action on a vector $v=T_a v^a \in \LieGC$ is defined by 
\begin{align}
  H v \equiv T^a (H_{ab}\, v^b) = T^a v^b (D_b D_a f) = D_v Df, 
\label{Hv}
\end{align}
where we have introduced the notation $D_v \equiv v^a D_a$. 
The transpose $H^T(U)$ of $H(U)$ is given by 
\begin{align}
  (H^T)_{ab}(U) = H_{ba}(U) = D_a D_b f(U), 
\end{align}
and acts on $v$ as  
\begin{align}
  H^T v \equiv T^a (H_{ba}\, v^b) = T^a v^b (D_a D_b f) = D D_v f. 
\label{HTv}
\end{align}
The difference between $H v$ and $H^T v$ can be calculated from the formula 
\eqref{DaDb} to be%
\footnote{ 
  This can be shown as follows: 
  $\text{(lhs)}=T^a v^b (D_a D_b f - D_b D_a f)
  = -T^a v^b C_{ab}{}^c D_c f = T_a v^b (D^c f)\, C_{cb}{}^a
  = [Df,v].$
} 
\begin{align}
  H^T v - H v = [Df, v].
\label{HTv-Hv}
\end{align}

\subsection{Cauchy's theorem for group manifolds}
\label{sec:GbbC_cauchy}

\begin{figure}[tb]
  \centering
  \includegraphics[width=60mm]{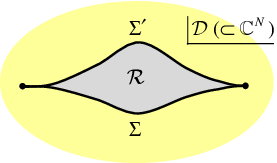}
  \caption{Cauchy's theorem for $\bbC^N$.}
\label{fig:cauchy_flat}
\end{figure}

We begin by recalling the multidimensional version of Cauchy's theorem
for flat space $\bbC^N=\{z=(z^i)\}$.
When the integration surfaces are noncompact,
we assume that the integrals converge
and that no additional boundary contributions arise
under the deformation.
This well-known theorem \cite{Howls:1997,Pham:1967} 
can be stated as follows (see Fig.~\ref{fig:cauchy_flat}):%
\begin{thm}
  Let $\calD$ be a domain in $\bbC^N$ (viewed as $\bbR^{2N}$) 
  and $f(z)$ be a holomorphic function on $\calD$. 
  Then, the integral $I_\Sigma$ of $f(z)$ 
  over a real $N$-dimensional oriented submanifold $\Sigma\subset\calD$,
  \begin{align}
    I_\Sigma = \int_\Sigma dz\,f(z)
    \quad (dz\equiv dz^1\wedge\cdots\wedge dz^N),
  \end{align}
  is invariant under continuous deformations of $\Sigma$ within $\calD$
  that keep its boundary fixed. 
\end{thm}

\begin{proof} 
  Let $\Sigma$ and $\Sigma'$ be 
  real $N$-dimensional oriented submanifolds of $\calD$
  connected by a continuous deformation within $\calD$
  that keeps their common boundary fixed,
  and let $\calR$ be the $(N+1)$-dimensional region 
  swept out by the deformation
  (such that $\partial\calR=\Sigma'-\Sigma$).
  By Stokes' theorem, we have 
  \begin{align}
    I_{\Sigma'} - I_\Sigma
    = \int_{\partial\calR} dz\,f(z)
    = \int_\calR d[dz f(z)]
    = (-1)^N \int_\calR dz \wedge df(z).
  \end{align}
  Since $f(z)$ is holomorphic ($\partial f / \partial \bar{z}^i = 0$), 
  its exterior derivative is simply 
  $df(z) = dz^i\,(\partial f(z)/\partial z^i)$. 
  Consequently, the integrand vanishes: 
  \begin{align}
    dz \wedge df(z)
    = \underbrace{(dz^1\wedge\cdots\wedge dz^N)\wedge dz^i}_{=0}\,
    \frac{\partial f(z)}{\partial z^i}
    = 0,
  \end{align}
  which implies that $I_{\Sigma'} = I_\Sigma$.
\end{proof} 

\begin{figure}[tb]
  \centering
  \includegraphics[width=60mm]{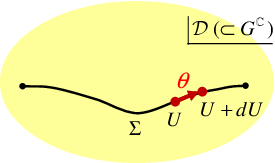}
  \caption{Cauchy's theorem for $G^\bbC$.}
  \label{fig:cauchy_group}
\end{figure}

The extension to complexified groups is straightforward, 
and we present it here to establish the mathematical basis 
for the algorithms developed below. 
The essential difference from the flat case 
is that the holomorphic $N$-form $dz$ 
is replaced by $(dU)$ introduced below: 
\begin{thm}
  Let $\calD$ be a domain in $G^\bbC$ 
  and $f(U)$ be a holomorphic function on $\calD$. 
  Then, the integral $I_\Sigma$ of $f(U)$ 
  over a real $N$-dimensional oriented submanifold $\Sigma\subset\calD$ 
  (see Fig.~\ref{fig:cauchy_group}),
  \begin{align}
    I_\Sigma = \int_\Sigma (dU) \,f(U),
  \end{align}
  is invariant under continuous deformations of $\Sigma$ within $\calD$
  that keep its boundary fixed.
  Here, $(dU)$ denotes the holomorphic $N$-form given by 
  \begin{align}
    (dU) \equiv \theta^1 \wedge \cdots \wedge \theta^N.
  \end{align}
\end{thm}

\begin{proof} 
  Let $\Sigma$ and $\Sigma'$ be 
  real $N$-dimensional oriented submanifolds of $\calD$
  connected by a continuous deformation within $\calD$
  that keeps their common boundary fixed,
  and let $\calR$ be the $(N+1)$-dimensional region 
  swept out by the deformation
  (such that $\partial\calR=\Sigma'-\Sigma$).
  By Stokes' theorem, we have 
  \begin{align}
    I_{\Sigma'} - I_\Sigma
    = \int_{\partial\calR} (dU)\,f(U)
    = \int_\calR d[(dU) f(U)]. 
  \end{align}
  From the Maurer-Cartan equation 
  $d\theta^a=(1/2)\, C_{bc}{}^a\,\theta^b\wedge \theta^c$ 
  [see Eq.~\eqref{Maurer-Cartan}], 
  it follows that the holomorphic $N$-form is closed, 
  $d[(dU)]=0$. 
  Thus, we have 
  \begin{align}
    I_{\Sigma'} - I_\Sigma = (-1)^N \int_\calR (dU) \wedge df(U).
  \end{align} 
  Furthermore, using the expansion 
  $df(U) = \theta^a D_a f(U)$ [see Eq.~\eqref{df(U)}],
  the integrand becomes 
  \begin{align}
    (dU) \wedge df(U) 
    = \underbrace{(\theta^1\wedge\cdots\wedge \theta^N)\wedge \theta^a}_{=0}\,
    D_a f(U)
    = 0,
  \end{align}
  which completes the proof that $I_{\Sigma'} = I_\Sigma$.
\end{proof} 

\section{Outline of GT/WV-HMC for group manifolds}
\label{sec:outline}

In this section, 
we introduce anti-holomorphic gradient flows in the complexified group $G^\bbC$ 
and outline the generalized thimble HMC (GT-HMC) 
and Worldvolume HMC (WV-HMC) algorithms for $G$, 
which are discussed in greater detail in subsequent sections. 
Our main goal here is to show that 
the path integrals in both GT-HMC and WV-HMC can be reformulated 
as phase-space integrals over the tangent bundles of the configuration spaces, 
which naturally carry symplectic structures.

\subsection{Anti-holomorphic gradient flows in $G^\bbC$}
\label{sec:outline_anti-holomorphic_flow}

\begin{figure}[tb]
  \centering
  \includegraphics[width=70mm]{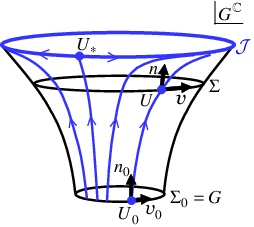}
  \caption{
    Deformation of $\Sigma_0 = G$ 
    into a submanifold $\Sigma$ of $G^\bbC$. 
    The deformed surface $\Sigma$ approaches a Lefschetz thimble $\calJ$, 
    which consists of points flowing out from a critical point $U_\ast$. 
    $v \in T_U \Sigma$ ($n \in N_U \Sigma$) 
    is the tangent (normal) vector at $U\in\Sigma$
    lifted from a tangent (normal) vector
    $v_0 \in T_{U_0} \Sigma_0$ ($n_0 \in N_{U_0} \Sigma_0$) 
    at $U_0\in\Sigma_0$. 
  }
\label{fig:gt_group}
\end{figure}

We consider the expectation value 
of an observable $\calO(U_0)$ 
defined by a path integral 
over a connected compact group $G = \{U_0\}$: 
\begin{align}
  \vev{\calO} \equiv
  \frac{\int_G (dU_0)\,e^{-S(U_0)}\,\calO(U_0)}
  {\int_G (dU_0)\,e^{-S(U_0)}}. 
\label{vev_G}
\end{align}
We complexify $G$ to $G^\bbC = \{U\}$ 
and assume that both $e^{-S(U)}$ and $e^{-S(U)}\,\calO(U)$ 
are holomorphic on $G^\bbC$ 
(which usually holds in cases of physical interest). 
By Cauchy's theorem (Sect.~\ref{sec:GbbC_cauchy}), 
the expression above can be rewritten 
as a ratio of integrals over a new integration surface $\Sigma$ 
that is obtained by a continuous deformation of $G$ 
(see Fig.~\ref{fig:gt_group}): 
\begin{align}
  \vev{\calO}
  =
  \frac{\int_\Sigma (dU)\,e^{-S(U)}\,\calO(U)}
  {\int_\Sigma (dU)\,e^{-S(U)}}. 
\label{vev_Sigma}
\end{align}
Thus, even when the original path integral on $\Sigma_0=G$ 
suffers from a severe sign problem 
due to the highly oscillatory behavior of $e^{-i\,\ImS(U_0)}$, 
this problem can be significantly alleviated 
if $\ImS(U)$ is almost constant on the new integration surface $\Sigma$. 

In this work, 
we generate such a deformation using the anti-holomorphic gradient flow 
\cite{Cristoforetti:2012su}:
\begin{align}
  \dot{U} = \xi(U)\,U
  ~~\text{with}~~
  U|_{t=0} = U_0,
\label{flow_c}
\end{align}
where $\dot{U} = dU/dt$  
and the drift term is given by 
\begin{align}
  \xi(U) \equiv [DS(U)]^\dagger.
\end{align}
This leads to the monotonicity relation
\begin{align}
  [S(U)]^\centerdot = \tr [(\dot{U}\,U^{-1}) DS(U)]
  =\tr \bigl[ (DS(U))^\dagger\,(DS(U)) \bigr] \geq 0, 
\end{align} 
which shows that 
the real part $\ReS(U)$ always increases along the flow 
(except at critical points where $DS(U)$ vanishes), 
while the imaginary part $\ImS(U)$ remains constant. 
The flow is in the direction of steepest descent 
of $|e^{-S(U)}| = e^{-\ReS(U)}$.%
\footnote{ 
  In fact, holomorphicity $\bar{D} S(U) = 0$ means $D \overline{S(U)} = 0$. 
  Thus, the flow equation \eqref{flow_c} can be rewritten as 
  $\dot{U} U^{-1} = [DS(U)]^\dagger = [DS(U) + D\overline{S(U)}]^\dagger
  = 2 [D \ReS(U)]^\dagger$. 
  The vector $[D \ReS(U)]^\dagger$ points in the steepest ascent direction 
  of $\ReS(U)$, and hence in the steepest descent direction of $|e^{-S(U)}|$, 
  because 
  $\ReS(e^{\epsilon v} U) - \ReS(U) 
  = 2\epsilon\,\vev{v, [D \ReS(U)]^\dagger} + O(\epsilon^2)$ 
  for $v \in T_U G^\bbC$ 
  [see Eq.~\eqref{deltaf_real}]. 
} 
The Lefschetz thimble $\calJ$ associated with a critical point $U_\ast$ 
is defined as the set of points flowing out of $U_\ast$ 
(see Fig.~\ref{fig:gt_group}).  
Since $\ImS(U)$ is invariant along the flow, 
it is constant over $\calJ$; 
$\ImS(U) = \ImS(U_\ast)$ $(U\in\calJ)$. 
Thus, the oscillatory behavior of the integrand 
is expected to be significantly mitigated 
if we deform the integration surface 
with a sufficiently large flow time $t$ 
so that the deformed surface comes to the vicinity 
of the relevant Lefschetz thimbles. 

As we will see in later sections, 
in order to sample $\Sigma$, 
we need to lift a tangent vector $v_0\in T_{U_0}\Sigma_0$ ($\Sigma_0=G$) 
to a tangent vector $v\in T_U\Sigma$ 
(see Fig.~\ref{fig:gt_group}).%
\footnote{ 
  $T_U \Sigma$ denotes the tangent space at $U$, 
  and $N_U \Sigma$ the normal space at $U$, 
  $N_U \Sigma = \{ n\in T_U G^\bbC\,|\,
  \vev{n,v} = 0\,\,(\forall v \in T_U \Sigma) \}$. 
  Note that $v_0^\dagger = -v_0$ for $v_0 \in T_{U_0}\Sigma_0$ 
  while $n_0^\dagger = +n_0$ for $n_0 \in N_{U_0}\Sigma_0$. 
} 
The transformation rule is obtained 
by considering the flow of an adjacent configuration 
of the form $e^{\epsilon v} U$,
\begin{align}
  (e^{\epsilon v} U)^\centerdot
  = [ DS(e^{\epsilon v} U) ]^\dagger\,e^{\epsilon v} U, 
\end{align}
and extracting terms linear in $\epsilon$, 
which leads to \cite{Cristoforetti:2012su} 
\begin{align}
  \dot{v} 
  = (H(U)\, v)^\dagger + [\xi(U),v] 
  ~~\text{with}~~
  v|_{t=0} = v_0. 
\label{flow_t}
\end{align}
Here, $H(U)$ is the Hessian matrix of $S(U)$, 
acting on $v$ as $Hv = D_v DS = T^a (H_{ab} v^b)$ with $H_{ab} = D_b D_a S$  
[see Eq.~\eqref{Hv}]. 
We can also introduce the flow of a normal vector $n=n_t\in N_U\Sigma_t$ 
such that it remains orthogonal to $v=v_t\in T_U\Sigma_t$ 
at any flow time $t$: 
\begin{align}
  \dot{n} &= -(H^T(U)\, n)^\dagger - [\xi^\dagger(U), n]
  ~~\text{with}~~
  n|_{t=0} = n_0,
\label{flow_n}
\end{align} 
where $H^T(U)$ is the transpose of $H(U)$, 
acting on $n$ as $H^T n = D D_n S = T^a (H_{ba} n^b)$ 
[see Eq.~\eqref{HTv}]. 
One can show that if $\vev{n_0,v_0}=0$, 
then $\vev{n,v}=0$ for all $t$. 
Indeed,  
\begin{align}
  \vev{n,v}^\centerdot = (\re \tr n^\dagger v)^\centerdot = 0,
\end{align}
because
\begin{align}
  (\tr n^\dagger v)^\centerdot
  &= \tr (\dot{n}^\dagger v + n^\dagger \dot{v} )
\nonumber
\\
  &= \tr\bigl[
  (-H^T n + [\xi,n^\dagger])\,v
  + n^\dagger\,( (Hv)^\dagger + [\xi,v])
  \bigr]
\nonumber
\\
  &= \tr \bigl[ -(H^T n)\,v + n^\dagger\,(Hv)^\dagger
  + [\xi,n^\dagger v]
  \bigr]
\nonumber
\\
  &= -2 i\,\im \tr n\,(Hv)\,\in i\,\bbR.
\end{align}

\begin{figure}[tb]
  \centering
  \includegraphics[width=85mm]{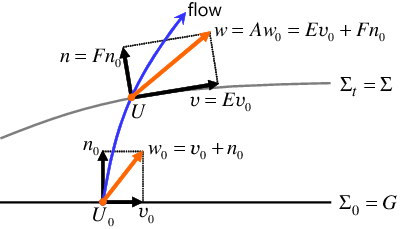}
  \caption{
    $\bbR$-linear map 
    $A:\,w_0=v_0+n_0 \mapsto w = E v_0 + F n_0$.}
  \label{fig:multEF_group}
\end{figure}

The flow equations \eqref{flow_t} and \eqref{flow_n}
define $\bbR$-linear maps 
$E: T_{U_0}\Sigma_0 \to T_U\Sigma$ 
and 
$F: N_{U_0}\Sigma_0 \to N_U\Sigma$ 
(see Fig.~\ref{fig:multEF_group}), 
which together define the $\bbR$-linear map 
$A:\,T_{U_0} G^\bbC \to T_U G^\bbC$ 
by 
\begin{align}
  A:\,T_{U_0} G^\bbC \ni w_0 = v_0 + n_0
  ~\mapsto~ w = v + n = E v_0 + F n_0
  \in  T_U G^\bbC.
\label{multEF}
\end{align}
%

\subsection{Outline of GT-HMC for group manifolds}
\label{sec:outline_gt}

In this subsection, 
we outline GT-HMC for group manifolds. 
The presentation is designed to be 
in parallel with that in Sect.~\ref{sec:gt-hmc} 
to demonstrate the similarity and difference 
between flat space and group manifolds. 

We start from the following expression [Eq.~\eqref{vev_Sigma}]: 
\begin{align}
  \vev{\calO}
  =
  \frac{\int_\Sigma (dU)_\Sigma\,e^{-S(U)}\,\calO(U)}
  {\int_\Sigma (dU)_\Sigma\,e^{-S(U)}}, 
\label{vev_Sigma2}
\end{align}
where $\Sigma\equiv\Sigma_t$ denotes the deformed surface 
at fixed flow time $t$. 
We simply write $U = U(t,U_0)$ as $U = U(U_0)$. 
To stress that geometric quantities below 
are defined on $\Sigma$, 
we denote the Maurer-Cartan form on $\Sigma$ 
and the corresponding holomorphic $N$-form 
as $\theta_\Sigma \equiv \theta|_\Sigma = T_a\,\theta_\Sigma^a$ 
and $(dU)_\Sigma \equiv (dU)|_\Sigma 
= \theta_\Sigma^1 \wedge \cdots \wedge \theta_\Sigma^N$, 
respectively. 
The one-forms $\theta_\Sigma^a$ are linear in $\theta_0^a$ 
and take the form 
\begin{align}
  \theta_\Sigma^a = E^a_b\,\theta_0^b,
\end{align}
from which we obtain the relation  
\begin{align}
  (dU)_\Sigma = \det E \, (dU_0). 
\label{dU_Sigma}
\end{align}
The Jacobian matrix $E = (E^a_b)$ can be computed 
by integrating the flow equations 
for the vectors $E_a \equiv T_b \, E^b_a \in T_U \Sigma$ 
[see Eq.~\eqref{flow_t}],
\begin{align}
  \dot{E}_a = (H E_a)^\dagger + [\xi,E_a]
  ~~\text{with}~~
  E_a|_{t=0} = T_a.
\end{align}
Since $\theta_\Sigma$ can be expressed as 
\begin{align}
  \theta_\Sigma = E_a\, \theta_0^a,
\end{align}
the induced metric on $\Sigma$ is given by 
\begin{align}
  ds_\Sigma^2 = \tr \theta_\Sigma^\dagger \theta_\Sigma
  = \gamma_{ab}(U_0)\,\theta_0^a\, \theta_0^b
\end{align}
with
\begin{align}
  \gamma_{ab} \equiv \vev{E_a, E_b}
  = \re \tr E_a^\dagger E_b,
\end{align}
which represents the squared distance between $U,\,U+dU \in \Sigma$. 
This metric defines the induced volume element on $\Sigma$ as 
\begin{align}
  |dU|_\Sigma = \sqrt{\gamma}\,(dU_0)
  \quad
  (\gamma = \det (\gamma_{ab})).
\end{align}
One can easily confirm 
that $\im \tr E_a^\dagger E_b$ need not vanish in general 
(as opposed to the flat case \cite{Fujii:2013sra}). 
This implies that one cannot write $\gamma_{ab}$ 
as $\tr E_a^\dagger E_b = \overline{E^c_a} E^c_b$, 
and thus $\sqrt{\gamma}$ does not agree with $|\det E|$ generically. 
Consequently, 
the ratio $(dU)_\Sigma / |dU|_\Sigma = \det E/\sqrt{\gamma}$ 
is not a pure phase factor. 

We rewrite the expectation value \eqref{vev_Sigma2}  
as a ratio of reweighted averages on $\Sigma$, 
\begin{align}
  \vev{\calO} 
  =  \frac{\int_\Sigma |dU|_\Sigma\,e^{-\ReS(U)}\,\calF(U)\,\calO(U)} 
  {\int_\Sigma |dU|_\Sigma\,e^{-\ReS(U)}\,\calF(U)}
  = \frac{\vev{\calF(U)\,\calO(U)}_\Sigma}
  {\vev{\calF(U)}_\Sigma},
\label{vev_gt1}
\end{align}
where the reweighted average $\vev{\cdots}_\Sigma$ is given by
\begin{align}
  \vev{g(U)}_\Sigma
  \equiv
  \frac{\int_\Sigma |dU|_\Sigma\,e^{-\ReS(U)}\,g(U)} 
  {\int_\Sigma |dU|_\Sigma\,e^{-\ReS(U)}},
\label{vev_gt1a}
\end{align}
and the reweighting factor $\calF(U)$ takes the form
\begin{align}
  \calF(U) \equiv \frac{(dU)_\Sigma}{|dU|_\Sigma}\,e^{-i\,\ImS(U)}
  = \frac{\det E}{\sqrt{\gamma}}\,e^{-i\,\ImS(U)}.
\label{F_Sigma}
\end{align}

To rewrite the integrals in Eq.~\eqref{vev_gt1a} as phase-space integrals, 
we introduce an auxiliary momentum $p = (p_a)$ 
with $dp = dp_1 \wedge \cdots \wedge dp_N \equiv \prod_a dp_a$. 
We further introduce the cotangent bundle 
$T^\ast G = \{ (U_0,p) \}$ 
with the symplectic potential 
\begin{align}
  a_\Sigma \equiv p_a \theta_0^a
\end{align}
and the symplectic form 
\begin{align}
  \omega_\Sigma \equiv da_\Sigma = dp_a \wedge \theta_0^a
  + \frac{1}{2}\,C_{bc}{}^a\, p_a\, \theta_0^b \wedge \theta_0^c,
\label{omega_Sigma}
\end{align}
the latter of which gives the symplectic volume form 
(up to irrelevant sign)
\begin{align}
  d\Omega_\Sigma \equiv \frac{\omega_\Sigma^N}{N!}
  = \pm \theta_0^1 \wedge \cdots \wedge \theta_0^N \wedge dp
  = \pm (dU_0) \wedge dp. 
\label{dOmega_Sigma}
\end{align}
Note that 
terms proportional to $\theta_0^b \wedge \theta_0^c$ in $\omega_\Sigma$ 
drop out in $\omega_\Sigma^N$. 
Then, using the identity 
\begin{align}
  \sqrt{\gamma}
  \propto 
    \int dp\,e^{-(1/2)\,\gamma^{ab}p_a p_b} 
  \quad
  [(\gamma^{ab}) \equiv (\gamma_{ab})^{-1}],
\end{align}
the integral of an arbitrary function $f(U)$ over $\Sigma$ 
with the induced volume element $|dU|_\Sigma$ 
can be replaced by an integral over $T^\ast G$ as 
\begin{align}
  &\int_\Sigma |dU|_\Sigma\,f(U) 
  = \int_G (dU_0)\,\sqrt{\gamma(U_0)}\,f(U(U_0))
\nonumber
\\
  &\propto
  \int_{T^\ast G} (dU_0) \wedge dp\,
  e^{-(1/2)\,\gamma^{ab}(U_0)\,p_a p_b}\,f(U(U_0))
  = \pm \int_{T^\ast G} d\Omega_\Sigma\,
  e^{-(1/2)\,\gamma^{ab}(U_0)\,p_a p_b}\,f(U(U_0)), 
\end{align}
and thus the average \eqref{vev_gt1a} 
can be rewritten as a ratio of phase-space integrals 
over $T^\ast G$,
\begin{align}
  \vev{g(U)}_\Sigma
  = \frac{\int_{T^\ast G} d\Omega_\Sigma\,
    e^{-(1/2)\,\gamma^{ab}(U_0)\,p_a p_b - \ReS(U(U_0))}\,g(U(U_0))}
    {\int_{T^\ast G} d\Omega_\Sigma\,
    e^{-(1/2)\,\gamma^{ab}(U_0)\,p_a p_b - \ReS(U(U_0))}}.
\label{vev_gt2}
\end{align}

For practical Monte Carlo computations,
it is more convenient to rewrite everything 
in terms of quantities in the target space $G^\bbC$ 
at the cost of projecting them onto $\Sigma$. 
As in the flat case, 
we adopt the RATTLE algorithm \cite{Andersen:1983,Leimkuhler:1994} 
for projection. 
There, we introduce the lifted momentum 
(or more precisely ``velocity'') 
$\pi = T_a \pi^a\in T_U \Sigma$ by 
\begin{align}
  \pi = E_a\, p^a
  \quad
  (p^a \equiv \gamma^{ab} p_b),
\end{align}
and consider the tangent bundle as the phase space,
\begin{align}
  T\Sigma = \{ (U,\pi)\,|\,U \in \Sigma,\,\pi \in T_U \Sigma \}.
\end{align}
Then, one can show the identities 
\begin{align}
  a_\Sigma &= \vev{\pi, \theta_\Sigma} = \re \tr \pi^\dagger \theta_\Sigma,
\\
  \omega_\Sigma &= d\vev{\pi, \theta_\Sigma}
  = \re \tr ( d\pi^\dagger \wedge \theta_\Sigma 
    + \pi^\dagger\, \theta_\Sigma \wedge \theta_\Sigma), 
\\
  \gamma^{ab}\,p_a p_b &= \vev{\pi,\pi},
\end{align}
from which the expression \eqref{vev_gt2} can be rewritten 
with integrals over $T\Sigma$ as
\begin{align}
  \vev{g(U)}_\Sigma
  = \frac{\int_{T\Sigma} d\Omega_\Sigma\,e^{-H(U,\pi)}\,g(U)}
  {\int_{T\Sigma} d\Omega_\Sigma\,e^{-H(U,\pi)}},
\label{vev_gt3}
\end{align}
where the Hamiltonian $H(U,\pi)$ is given by 
\begin{align}
  H(U,\pi) = \frac{1}{2}\,\vev{\pi,\pi} + \ReS(U). 
\label{H_Sigma}
\end{align}

We can now apply HMC using molecular dynamics on $T\Sigma$. 
The momentum $\pi \in T_U\Sigma$ 
is refreshed according to the Gaussian weight 
$e^{-\vev{\pi,\pi}/2}$, 
and molecular dynamics is implemented by the RATTLE algorithm 
\cite{Andersen:1983,Leimkuhler:1994}, 
which constrains $U$ and $\pi$ to satisfy 
$U\in\Sigma$ and $\pi\in T_U\Sigma$. 
When the constraint equations are solved exactly, 
the RATTLE update is reversible and symplectic. 
Symplecticity guarantees preservation of the symplectic volume form 
$d\Omega_\Sigma$.
Volume preservation and reversibility 
ensure the appropriate detailed-balance condition 
when combined with the usual Metropolis correction.

Thus, explicit evaluation of $|\det E|$
is not required during configuration generation.
However, the factor $\det E/\sqrt{\gamma}$
must still be evaluated for measurements
through $\calF(U)$ in Eq.~\eqref{vev_gt1}.
An explicit algorithm for generating configurations $(U,\pi) \in T \Sigma$ 
from the distribution $\propto e^{-H(U,\pi)}$ 
is presented in Sect.~\ref{sec:gt} 
after we establish a general theory of HMC 
on constrained submanifolds of $G^\bbC$.

\subsection{Outline of WV-HMC for group manifolds}
\label{sec:outline_wv}

In this subsection, 
we outline WV-HMC for group manifolds. 
The presentation is designed to be 
in parallel with that in Sect.~\ref{sec:wv-hmc} 
to demonstrate the similarity and difference 
between flat space and group manifolds. 

\begin{figure}[tb]
  \centering
  \includegraphics[width=75mm]{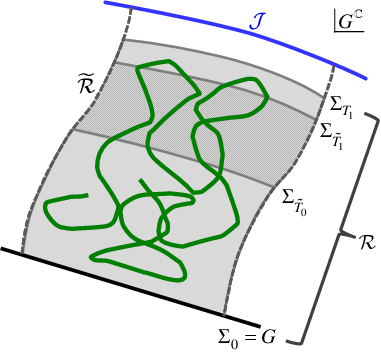}
  \caption{
    WV-HMC over the worldvolume $\calR$. 
    The worldvolume can be restricted to a finite region 
    $T_0 \lesssim t \lesssim T_1$ 
    by using the function $W(t)$. 
    Measurements are performed for configurations in a subregion 
    $\tilde{\calR}$ 
    consisting of configurations with $t \in [\tilde{T}_0, \tilde{T}_1]$. 
    We set $T_0 = 0$ in the figure.
  }
\label{fig:worldvolume_group2}
\end{figure}

As reviewed in Sect.~\ref{sec:picard-lefschetz_wv-hmc}, 
the GT-HMC algorithm often introduces ergodicity issues 
when the flow time is taken to be sufficiently large 
to reduce the oscillatory behavior of $e^{-i\, \ImS(U)}$. 
This trade-off between reducing the sign problem 
and maintaining ergodicity can be resolved 
by employing the Worldvolume HMC (WV-HMC) method \cite{Fukuma:2020fez}, 
in which the path integral over a single deformed surface 
$\Sigma$ [Eq.~\eqref{vev_Sigma}] 
is replaced by an integral 
over a continuous union of deformed surfaces 
(the worldvolume; see Fig.~\ref{fig:worldvolume_group2}). 

Although the method was originally introduced for the flat case 
(see Sect.~\ref{sec:wv-hmc}), 
it can be generalized to group manifolds as follows. 
We first note that 
when we set the deformed surface to $\Sigma=\Sigma_t$ 
(the deformed surface at flow time $t$), 
both the numerator and the denominator of Eq.~\eqref{vev_Sigma} 
are independent of $t$ due to Cauchy's theorem. 
Thus, we can take averages over $t$ separately 
with an arbitrary common weight $e^{-W(t)}$ 
as in Ref.~\cite{Fukuma:2020fez}:
\begin{align}
  \vev{\calO}
  =
  \frac{\int_{\Sigma_t} (dU)_{\Sigma_t}\,e^{-S(U)}\,\calO(U)}
  {\int_{\Sigma_t} (dU)_{\Sigma_t}\,e^{-S(U)}}
  =
  \frac{\int dt\,e^{-W(t)}\,\int_{\Sigma_t} (dU)_{\Sigma_t}\,e^{-S(U)}\,\calO(U)}
  {\int dt\,e^{-W(t)}\,\int_{\Sigma_t} (dU)_{\Sigma_t}\,e^{-S(U)}}.
\label{vev_R}
\end{align}
This can be regarded as a ratio of integrals 
over the \emph{worldvolume} $\calR$ defined by 
\begin{align}
  \calR \equiv \bigcup_t \Sigma_t 
  = \{ U(t,U_0) \in G^\bbC \,|\, t \in \bbR,\, U_0 \in G \},
\end{align}
where $U(t,U_0)$ denotes the configuration reached at flow time $t$ 
starting from the initial configuration $U_0$.

Using the parametrization $U = U(t,U_0) \in \calR$,
we introduce $\hat\theta_0^\mu$ $(\mu = 0,1,\ldots,N)$ as
\begin{align}
  \hat{\theta}_0^0 \equiv dt, 
  \qquad
  \hat{\theta}_0^a \equiv \theta_0^a \quad (a=1,\ldots,N).
\end{align}
The Maurer-Cartan form on $\calR$, 
$\theta_\calR \equiv \theta|_\calR = dU U^{-1}|_\calR = T_a \theta_\calR^a$, 
can be expanded as 
\begin{align}
  \theta_\calR
  = (\dot{U} U^{-1})\,dt + \theta_{\Sigma_t}
  = \xi\,dt + E_a\,\theta_0^a
  \equiv  \hat{E}_\mu\,\hat{\theta}_0^\mu,
\end{align}
where $\hat{E}_\mu = T_a \hat{E}^a_\mu \in T_U \calR$ $(\mu = 0,1,\ldots,N)$ 
are given by 
\begin{align}
  \hat{E}_0 = \xi, 
  \qquad
  \hat{E}_a = E_a = T_b\,E^b_a \quad (a=1,\ldots,N).
\end{align}
Due to Eq.~\eqref{dU_Sigma}, we have 
\begin{align}
  dt \wedge (dU)_{\Sigma_t} = \det E \, dt \wedge (dU_0).
\label{dtdU}
\end{align}
The induced metric on $\calR$ takes the form 
\begin{align}
  ds_\calR^2 = \tr \theta_\calR^\dagger \theta_\calR
  = \hat\gamma_{\mu\nu}(t,U_0)\,\hat\theta_0^\mu\,\hat\theta_0^\nu
\end{align}
with
\begin{align}
  \hat\gamma_{\mu\nu} \equiv \vev{\hat{E}_\mu,\hat{E}_\nu}
  = \re \tr \hat{E}_\mu^\dagger \hat{E}_\nu,
\end{align}
from which the induced volume element on $\calR$ is given by 
\begin{align}
  |dU|_\calR \equiv \sqrt{\hat\gamma}\,\prod_\mu \hat\theta_0^\mu
  = \sqrt{\hat\gamma}\,dt \wedge (dU_0)
  \quad
  (\hat\gamma \equiv \det (\hat\gamma_{\mu\nu})).
\end{align}

\begin{figure}[tb]
  \centering
  \includegraphics[width=60mm]{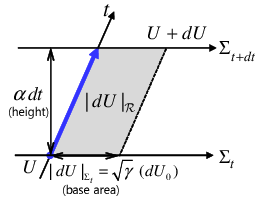}
  \caption{
    Induced volume element $|dU|_\calR$ on worldvolume $\calR$.
  }
\label{fig:volume_element_group}
\end{figure}

To obtain an explicit form of $\sqrt{\hat\gamma}$, 
we decompose the vector $\xi(U) = DS(U)^\dagger \in T_U \calR$ 
at $U = U(t,U_0)$ 
into its tangential and normal components along $\Sigma_t$ as 
\begin{align}
  \xi = \xi_v + \xi_n
  \quad
  (\xi_v = T_a\, \xi_v^a\in T_U \Sigma_t,~
   \xi_n = T_a\, \xi_n^a \in N_U \Sigma_t),
\end{align}
from which we have 
\begin{align}
  \theta_\calR = \xi_n\, dt + ( \theta_{\Sigma_t} + \xi_v\, dt)
  = \xi_n\,dt + E_a\,(\theta_0^a + \beta^a dt)
\end{align}
with the shift functions
\begin{align}
  \beta^a \equiv \gamma^{ab}\,\vev{E_b,\xi_v}.
\end{align}
The metric $ds_\calR^2$ can then be expressed in the ADM form, 
\begin{align}
  ds_\calR^2 
  = \alpha^2\, dt^2 + 
    \gamma_{ab}\,(\theta_0^a + \beta^a dt) (\theta_0^b + \beta^b dt),
\end{align}
where the lapse function $\alpha$ is given by 
\begin{align}
  \alpha = \sqrt{\vev{\xi_n, \xi_n}}.
\end{align}
In what follows, we assume that $\xi_n\neq 0$ 
(and thus $\alpha > 0$) 
except on a set of measure zero. 
We thus have the relation $\sqrt{\hat\gamma} = \alpha\,\sqrt{\gamma}$, 
and the induced volume element on $\calR$ is given by 
(see Fig.~\ref{fig:volume_element_group})
\begin{align}
  |dU|_\calR 
  = \alpha \sqrt{\gamma}\,dt \wedge (dU_0). 
\label{dU_R}
\end{align}
Combining Eqs.~\eqref{dtdU} and \eqref{dU_R}, 
the expression \eqref{vev_R} can be rewritten as 
\begin{align}
  \vev{\calO}
  = \frac{\int_\calR |dU|_\calR\,e^{-V(U)}\,\calF(U)\,\calO(U)}
  {\int_\calR |dU|_\calR\,e^{-V(U)}\,\calF(U)}
  = \frac{ \vev{\calF(U)\,\calO(U)}_\calR }
  { \vev{\calF(U)}_\calR }.
\label{vev_wv1}
\end{align}
Here, $V(U)$ and $\calF(U)$ denote the potential 
and the reweighting factor, respectively:%
\footnote{ 
  The real-valued function $t(U)$ returns the flow time $t$ 
  for configuration $U=U(t,U_0)$. 
} 
\begin{align}
  V(U) 
  &\equiv \ReS(U) + W(t(U)),
\label{V_calR}
  \\
  \calF(U) &\equiv
  \frac{dt \wedge (dU)_{\Sigma_t}}{|dU|_\calR}\,e^{-i\,\ImS(U)}
  =
  \alpha^{-1}\, \frac{\det E}{\sqrt{\gamma}}\, e^{-i\,\ImS(U)}. 
\label{F_calR}
\end{align}
The reweighted average $\vev{\cdots}_\calR$ is defined by
\begin{align}
  \vev{g(U)}_\calR
  \equiv 
  \frac{ \int_\calR |dU|_\calR\,e^{-V(U)}\,g(U) }
  { \int_\calR |dU|_\calR\,e^{-V(U)} }.
\label{vev_wv2}
\end{align}

To rewrite the integrals in Eq.~\eqref{vev_wv2} 
as phase space integrals, 
we introduce an auxiliary momentum $\hat{p} = (\hat{p}_\mu)$ 
with 
$d\hat{p} 
= d\hat{p}_0 \wedge d\hat{p}_1 \wedge \cdots \wedge d\hat{p}_N 
\equiv \prod_\mu d\hat{p}_\mu$. 
We further introduce the cotangent bundle 
$T^\ast (\bbR \times G) = \{ ((t,U_0),\hat{p}) \}$ 
with the symplectic potential 
\begin{align}
  a_\calR \equiv \hat{p}_\mu \hat\theta_0^\mu
\end{align}
and the symplectic form 
\begin{align}
  \omega_\calR 
  &\equiv da_\calR
  = d(\hat{p}_\mu\,\hat\theta_0^\mu)
  = d(\hat{p}_0\, dt) + d( \hat{p}_a\,\theta_0^a)
\nonumber
\\
  &= d\hat{p}_0 \wedge dt 
  + d\hat{p}_a \wedge \theta_0^a
  + \frac{1}{2}\,C_{bc}{}^a\,\hat{p}_a\,\theta_0^b \wedge \theta_0^c
\nonumber
\\
  &= d\hat{p}_\mu \wedge \hat\theta_0^\mu
  + \frac{1}{2}\,C_{bc}{}^a\,\hat{p}_a\,\theta_0^b \wedge \theta_0^c,
\label{omega_calR}
\end{align} 
the latter of which gives the symplectic volume form 
(up to irrelevant sign) 
\begin{align}
  d\Omega_\calR \equiv \frac{\omega_\calR^{N+1}}{(N+1)!}
  = \pm \biggl(\prod_{\mu=0}^N \hat\theta_0^\mu\biggr) \wedge
    \biggl(\prod_{\mu=0}^N d\hat p_\mu\biggr).
\label{dOmega_calR}
\end{align}
Note that terms proportional to $\theta_0^b \wedge \theta_0^c$ 
in $\omega_\calR$ 
drop out in $\omega_\calR^{N+1}$. 
Then, using the identity 
\begin{align}
  \sqrt{\hat\gamma}
  \propto \int d\hat{p}\,e^{-(1/2)\,\hat\gamma^{\mu\nu} \hat{p}_\mu \hat{p}_\nu} 
  \quad
  [(\hat\gamma^{\mu\nu}) \equiv (\hat\gamma_{\mu\nu})^{-1}],
\end{align}
the integral of an arbitrary function $f(U)$ over $\calR$ 
with the induced volume $|dU|_\calR$ 
can be replaced by an integral over $T^\ast (\bbR \times G)$ as 
\begin{align}
  \int_\calR |dU|_\calR\,f(U) 
  &= \int_{\bbR \times G} \biggl( \prod_\mu \hat\theta_0^\mu \biggr)\,
  \sqrt{\hat\gamma(t,U_0)}\,f(U(t,U_0))
\nonumber
\\
  &\propto 
  \int_{T^\ast (\bbR \times G)} 
  \biggl( \prod_\mu \hat\theta_0^\mu \biggr) \wedge d\hat{p}\,
  e^{-(1/2)\,\hat\gamma^{\mu\nu}(t,U_0)\, 
  \hat{p}_\mu \hat{p}_\nu}\,f(U(t,U_0))
\nonumber
\\
  &= \pm \int_{T^\ast (\bbR \times G)} d\Omega_\calR\,
  e^{-(1/2)\,\hat\gamma^{\mu\nu}(t,U_0)\, 
  \hat{p}_\mu \hat{p}_\nu}\,f(U(t,U_0)), 
\end{align}
and thus the average \eqref{vev_wv2} 
can be rewritten as a ratio of phase-space integrals 
over $T^\ast (\bbR \times G)$,
\begin{align}
  \vev{g(U)}_\calR
  = \frac{\int_{T^\ast (\bbR \times G)} d\Omega_\calR\,
    e^{-(1/2)\,\hat\gamma^{\mu\nu}(t,U_0)\,\hat{p}_\mu \hat{p}_\nu 
    - V(U(t,U_0))}\,g(U(t,U_0))}
    {\int_{T^\ast (\bbR \times G)} d\Omega_\calR\,
    e^{-(1/2)\,\hat\gamma^{\mu\nu}(t,U_0)\,\hat{p}_\mu \hat{p}_\nu 
    - V(U(t,U_0))}}.
\label{vev_calR_G}
\end{align}

As in the previous subsection on GT-HMC, 
to rewrite the expression \eqref{vev_calR_G} 
in terms of quantities in the target space $G^\bbC$, 
we introduce the lifted momentum 
$\pi = T_a \pi^a \in T_U \calR$ by 
\begin{align}
  \pi = \hat{E}_\mu\, \hat{p}^\mu
  \quad
  (\hat{p}^\mu \equiv \hat\gamma^{\mu\nu} \hat{p}_\nu),
\end{align}
and consider the tangent bundle as the phase space, 
\begin{align}
  T\calR = \{ (U,\pi)\,|\,U \in \calR,\,\pi \in T_U \calR \}.
\end{align}
Then, one can show the identities 
\begin{align}
  a_\calR &= \vev{\pi, \theta_\calR} 
  = \re \tr \pi^\dagger\, \theta_\calR,
\\
  \omega_\calR &= d \vev{\pi, \theta_\calR} 
  = \re \tr ( d \pi^\dagger \wedge \theta_\calR 
  + \pi^\dagger\, \theta_\calR \wedge \theta_\calR),
\\
  \hat\gamma^{\mu\nu} \hat{p}_\mu \hat{p}_\nu &= \vev{\pi,\pi},
\end{align}
from which the expression \eqref{vev_calR_G} can be rewritten 
with integrals over $T\calR$ as 
\begin{align}
  \vev{g(U)}_\calR
  = \frac{\int_{T\calR} d\Omega_\calR\,e^{-H(U,\pi)}\,g(U)}
  {\int_{T\calR} d\Omega_\calR\,e^{-H(U,\pi)}},
\label{vev_wv3}
\end{align}
where the Hamiltonian $H(U,\pi)$ is now given by 
\begin{align}
  H(U,\pi) = \frac{1}{2}\,\vev{\pi,\pi} + V(U). 
\label{H_calR}
\end{align}

We can now apply HMC using molecular dynamics on $T\calR$, 
in parallel with GT-HMC on $T\Sigma$.
The momentum $\pi\in T_U\calR$
is refreshed according to the Gaussian weight
$e^{-\vev{\pi,\pi}/2}$,
and molecular dynamics is implemented
by the RATTLE algorithm \cite{Andersen:1983,Leimkuhler:1994},
which constrains $U$ and $\pi$ to satisfy
$U\in\calR$ and $\pi\in T_U\calR$.
When the constraint equations are solved exactly,
the RATTLE update is reversible and symplectic,
which ensures the appropriate detailed-balance condition 
when combined with the usual Metropolis correction.

Thus, as in GT-HMC, 
explicit evaluation of $|\det E|$
is not required during configuration generation.
However, the same factor $\det E/\sqrt{\gamma}$
must still be evaluated for measurements
through the reweighting factor $\calF(U)$ in Eq.~\eqref{vev_wv1}.
An explicit algorithm for generating configurations $(U,\pi) \in T \calR$ 
from the distribution $\propto e^{-H(U,\pi)}$ 
is presented in Sect.~\ref{sec:wv} 
after we establish a general theory of HMC 
on constrained submanifolds of $G^\bbC$.

\section{Unconstrained molecular dynamics on $G^\bbC$}
\label{sec:unconstrained}

Although our main interest lies in molecular dynamics 
on a constrained submanifold of $G^\bbC$ 
in the context of GT/WV-HMC, 
we begin by considering unconstrained molecular dynamics on $G^\bbC$ 
to clarify the mathematical structure 
that remains unchanged by the introduction of constraints. 
For a review of the HMC algorithm for a compact group $G$, 
presented in a manner consistent with this section, 
see Appendix~\ref{sec:G}.

\subsection{Continuum version (Hamiltonian dynamics)}
\label{sec:unconstrained_continuum}

Our phase space is given by the tangent bundle over $G^\bbC$, 
\begin{align}
  T G^\bbC = \{(U,\pi) \,|\,U \in G^\bbC,\,\pi\in T_U G^\bbC\}.
\end{align} 
We introduce the symplectic potential $a$ by 
\begin{align}
  a \equiv \vev{\pi,\theta}
  = \re \tr \pi^\dagger \theta
  = \frac{1}{2}\,\tr (\pi^\dagger \theta + \pi \theta^\dagger),
\end{align}
with which the symplectic 2-form $\omega$ 
and the volume $4N$-form $d\Omega$ are obtained as 
(recall that $\dim_\bbR G^\bbC = 2N$)
\begin{align}
  \omega = da,
  \qquad
  d\Omega = \frac{1}{(2N)!}\,\omega^{2N}. 
\end{align}
We restrict our attention to Hamiltonians of the form%
\footnote{ 
  A more precise expression will be 
  $H(U,U^\dagger,\pi,\pi^\dagger) 
  = K(\pi,\pi^\dagger) + V(U,U^\dagger)$ 
  because $U$ and $U^\dagger$ (and also $\pi$ and $\pi^\dagger$) 
  are regarded as independent variables 
  under differentiation.  
  A similar comment applies to any non-holomorphic functions on $T G^\bbC$. 
} 
\begin{align}
  H(U,\pi) = K(\pi) + V(U)
  \text{~~with~~}
  K(\pi) = \frac{1}{2}\,\vev{\pi,\pi}. 
\end{align}
To \emph{define} consistent dynamics 
with a given Hamiltonian $H(U,\pi)$ 
and the symplectic potential $a$, 
we consider a trajectory $(U(s),\pi(s))$ 
parametrized by a continuous parameter $s$, 
and define it as a stationary path of the action of the first-order form, 
\begin{align}
  I[U(s),\pi(s)]
  = \int ds\,\bigl[
    \bigvev{\pi,\circdot{U}U^{-1}} - H(U,\pi)
  \bigr],
\label{action_GC}
\end{align}
where $\circdot{U}\equiv dU/ds$.
The variation of the action gives%
\footnote{ 
  We have neglected boundary terms 
  using the identity 
  \begin{align}
    \tr \pi^\dagger \delta(\circdot{U} U^{-1})
    = \tr \delta U U^{-1} (-\circdot{\pi}{}^\dagger 
    + [\circdot{U} U^{-1},\pi^\dagger])
      + \text{(total derivatives)}.
  \nonumber
  \end{align} 
} 
\begin{align}
  \delta I[U,\pi]
  &= \frac{1}{2}\, \int ds\,\tr\Bigl\{
    \delta\pi^\dagger\,[\circdot{U} U^{-1} - \pi]
    + \delta\pi\,[(\circdot{U} U^{-1})^\dagger - \pi^\dagger]
\nonumber
\\
  &~~~ + (\delta U U^{-1})\,[-\circdot{\pi}{}^\dagger 
       + [\circdot{U} U^{-1},\pi^\dagger]
       -2\,DV]
\nonumber
\\
  &~~~ + (\delta U U^{-1})^\dagger\,[-\circdot{\pi} 
       + [\pi,(\circdot{U} U^{-1})^\dagger]
       -2\,(DV)^\dagger ]
  \Bigr\}.
\end{align}
Setting $\delta I=0$, we obtain
\begin{align}
  \circdot{U} 
  &= \pi\,U,
\label{md_cont1}
\\
  \circdot{\pi} 
  &= -2\,[DV(U)]^\dagger
  + [\pi,(\circdot{U} U^{-1})^\dagger], 
\nonumber
\\
  &= -2\,[DV(U)]^\dagger
  + [\pi,\pi^\dagger], 
\label{md_cont2}
\end{align}
and 
\begin{align}
  \circdot{U}{}^\dagger 
  &= (\pi\,U)^\dagger,
  \label{md_cont3}
  \\
  \circdot{\pi}{}^\dagger 
  &= -2\,DV(U)
  + [\circdot{U} U^{-1}, \pi^\dagger], 
\nonumber
\\
  &= -2\,DV(U)
  + [\pi,\pi^\dagger]. 
  \label{md_cont4}
\end{align}
Note the appearance of the term $[\pi,\pi^\dagger]$, 
which is not present in Hamiltonian dynamics for $G$ 
(where $\pi^\dagger = -\pi$).

Hamilton's equations \eqref{md_cont1}--\eqref{md_cont4} 
satisfy the following three key properties:

\noindent
\underline{(a) Reversibility}: 
When a trajectory $(U(s),\pi(s))$ $(0 \leq s \leq 1)$ 
is a solution to Hamilton's equations, 
its time-reversal, 
\begin{align}
  U(s) &\to \tilde{U}(s) \equiv U(1-s),
\\
  \pi(s) &\to \tilde\pi(s) \equiv -\pi(1-s),
\end{align}
is also a solution. 
Indeed, 
we have
\begin{align}
  \circdot{\tilde{U}}(s) = -\circdot{U}(1-s)
  = -\pi(1-s)\,U(1-s) = \tilde\pi(s)\,\tilde{U}(s),
\end{align}
and 
\begin{align}
  \circdot{\tilde\pi}(s) &= \circdot{\pi}(1-s) 
\nonumber
\\
  &= -2\,[ DV(U(1-s)) ]^\dagger 
  + [\pi(1-s),\pi^\dagger(1-s)] 
\nonumber
\\
  &= -2\,[ DV(\tilde{U}(s)) ]^\dagger 
  + [\tilde\pi(s),\tilde\pi^\dagger(s)].
\end{align}

\noindent
\underline{(b) Symplecticity}: 
\begin{align}
  \circdot{\omega} = 0.
\end{align}
Indeed, explicit calculations show%
\footnote{ 
  We have used the identities 
  $2 \vev{(DV)^\dagger,\theta} = dV$
  [Eq.~\eqref{df}] 
  and $\circdot{\theta} = (dU U^{-1})^{\circdot{}} = d\pi - [\theta,\pi]$.
\label{a_dot}
} 
\begin{align}
  \circdot{a}
  = \vev{\pi,\theta}^\circ
  = \vev{\circdot{\pi},\theta} + \vev{\pi,\circdot{\theta}}
  = d\Bigl( \frac{1}{2}\,\tr \pi^\dagger \pi - V \Bigr),
\end{align}
and thus we have
\begin{align}
  \circdot{\omega} = d \circdot{a} = 0.
\end{align}

\noindent
\underline{(c) Energy conservation}: 
\begin{align}
  \circdot{H} = 0.
\end{align}

When $G$ is $SU(n)$ or $U(n)$, 
the Poisson brackets take the following forms:%
\footnote{ 
  This can be obtained in parallel to the derivation 
  given in Appendix~\ref{sec:G_symplectic}. 
  The $-(1/n)$ terms are not present when $G=U(n)$. 
} 
\begin{align}
  \{U_{ij}, U_{kl}\} &= 0,
\label{U_U}
\\
  \{U_{ij}, U^\dagger_{kl}\} &= 0,
\label{U_UH}
\\
  \{U_{ij}, \pi_{kl}\} &= 0,
\label{U_pi}
\\
  \{\pi_{ij}, U_{kl}\} &= 0,
\label{pi_U}
\\
  \{U_{ij}, \pi^\dagger_{kl}\} &= 
  2\,\Bigl(\delta_{il}\,U_{kj} 
    - \frac{1}{n}\,U_{ij}\,\delta_{kl}\Bigr),
\label{U_piH}
\\
  \{\pi^\dagger_{ij}, U_{kl}\} &= 
  2\,\Bigl(-U_{il}\,\delta_{kj} 
    + \frac{1}{n}\,\delta_{ij}\,U_{kl}\Bigr),
\label{piH_U}
\\
  \{U^\dagger_{ij}, \pi_{kl}\} &= 
  2\,\Bigl(U^\dagger_{il}\,\delta_{kj} 
    - \frac{1}{n}\,U^\dagger_{ij}\,\delta_{kl}\Bigr),
\label{UH_pi}
\\
  \{ \pi_{ij}, U^\dagger_{kl}\} &= 
  2\,\Bigl( -\delta_{il}\,U^\dagger_{kj} 
    + \frac{1}{n}\,\delta_{ij}\,U^\dagger_{kl}\Bigr),
\label{pi_UH}
\\
  \{U^\dagger_{ij}, \pi^\dagger_{kl}\} &= 0,
\label{UH_piH}
\\
  \{ \pi^\dagger_{ij}, U^\dagger_{kl}\} &= 0,
\label{piH_UH}
\\
  \{\pi_{ij}, \pi_{kl}\} &= 
  2\,(-\delta_{il}\,\pi_{kj} + \pi_{il}\,\delta_{kj} ),
\label{pi_pi}
\\
  \{\pi^\dagger_{ij}, \pi^\dagger_{kl}\} &= 
  2\,(\delta_{il}\,\pi^\dagger_{kj} - \pi^\dagger_{il}\,\delta_{kj} ),
\label{piH_piH}
\\
  \{\pi_{ij}, \pi^\dagger_{kl}\} &= 0.
\label{pi_piH}
\end{align}
Using the Poisson brackets, 
one can easily show the identities 
\begin{align}
  \{U, K\} &= \pi\,U,
\label{U_K}
\\
  \{U, V\} &= 0,
\label{U_V}
\\
  \{\pi, K\} &= [\pi,\pi^\dagger],
\label{pi_K}
\\
  \{\pi, V\} &= -2\,(DV)^\dagger,
\label{pi_V}
\\
  \{U^\dagger, K\} &= U^\dagger\,\pi^\dagger,
\label{UH_K}
\\
  \{U^\dagger, V\} &= 0,
\label{UH_V}
\\
  \{\pi^\dagger, K\} &= [\pi,\pi^\dagger],
\label{piH_K}
\\
  \{\pi^\dagger, V\} &= -2\,DV,
\label{piH_V}
\end{align}
which reproduce Hamilton's equations \eqref{md_cont1}--\eqref{md_cont4} 
in the form $\circdot{\calO} = \{\calO, H\}$ with
\begin{align}
  \{U, H\} &= \pi\,U,
  \qquad~~~\,
  \{\pi, H\} =-2\,(DV)^\dagger + [\pi,\pi^\dagger],
\\
  \{U^\dagger, H\} &= U^\dagger \pi^\dagger,
  \qquad
  \{\pi^\dagger, H\} =-2\,DV + [\pi,\pi^\dagger]. 
\label{pi_H_V}
\end{align}

We now make an important observation  
that will play a crucial role in the discussions to follow. 
From Eqs.~\eqref{pi_K} and \eqref{piH_K}, 
we have 
\begin{align}
  \{\pi, K\} &= [\pi,\pi^\dagger] = [\pi-\pi^\dagger, \pi],
\\
  \{\pi^\dagger, K\} &= [\pi,\pi^\dagger] = [\pi-\pi^\dagger, \pi^\dagger].
\end{align}
This implies that
the Poisson bracket $\{\ast,K\}$ acts as the commutator
$[\pi-\pi^\dagger,\ast]$
on arbitrary matrices $f$ built from $\pi$ and $\pi^\dagger$:
\begin{align}
  \{f(\pi,\pi^\dagger), K\} = 
  [\pi-\pi^\dagger, f(\pi,\pi^\dagger)] .
\label{f_K}
\end{align}

\subsection{Discrete version (molecular dynamics)}
\label{sec:unconstrained_discrete}

We discretize the evolution operator $T_H = e^{\epsilon\,\{\ast,H\}}$ 
with step size $\Delta s = \epsilon$ 
in the symmetric form:
\begin{align}
  T \equiv T_{V/2}\,T_K\,T_{V/2},
\label{T_def}
\end{align}
whose deviation from $T_H$ is $O(\epsilon^3)$, 
$T = T_H + O(\epsilon^3)$.%
\footnote{
  For notational simplicity, 
  we denote $e^{\epsilon\, \{\ast,f\}}$ 
  by $T_f$ (rather than $T_{\epsilon f}$).
  The relation $T = T_H + O(\epsilon^3)$ 
  follows directly from the identity 
  $e^{(\epsilon/2) A}\,e^{\epsilon B}\,e^{(\epsilon/2) A} 
  = e^{\epsilon (A+B)} + O(\epsilon^3)$, 
  which holds for linear operators $A$ and $B$ of any type.
} 

Straightforward computation yields the following identities 
(see Appendix~\ref{sec:T_V/2_U-T_K_pi} for a proof):
\begin{align}
  T_{V/2}\, U &= U,
\label{T_V/2_U}
\\
  T_{V/2}\, \pi &= \pi - \epsilon\,[DV(U)]^\dagger
  ~(\,\equiv \pi_{1/2}),
\label{T_V/2_pi}
\\
  T_K\, U &= e^{\epsilon (\pi-\pi^\dagger)}\,e^{\epsilon \pi^\dagger}\,U,
\label{T_K_U}
\\
  T_K\, \pi &= e^{\epsilon (\pi-\pi^\dagger)}\,\pi\,e^{-\epsilon (\pi-\pi^\dagger)}.
\label{T_K_pi}
\end{align}
We thus find that the transformed variables 
$U' \equiv T U$ and $\pi' \equiv T \pi$ 
are given by 
\begin{align}
  U' 
  &= T_{V/2}\,T_K\,T_{V/2}\,U
  = T_{V/2}\,T_K\,U
  \nonumber
  \\
  &= T_{V/2}\, \bigl\{
    e^{\epsilon\,(\pi-\pi^\dagger)}\,
    e^{\epsilon\,\pi^\dagger}\,U
  \bigr\}
  \nonumber
\\
  &= e^{\epsilon\,(\pi_{1/2}-\pi_{1/2}^\dagger)}\,
  e^{\epsilon\,\pi_{1/2}^\dagger}\,U,
\end{align}
and 
\begin{align}
  \pi' 
  &= T_{V/2}\,T_K\,T_{V/2}\,\pi
  = T_{V/2}\,T_K\,\bigl( \pi - \epsilon\,[DV(U)]^\dagger \bigr)
  \nonumber
  \\
  &= T_{V/2}\,\bigl\{
  e^{\epsilon\,(\pi-\pi^\dagger)}\,\pi\,e^{-\epsilon\,(\pi-\pi^\dagger)}
  - \epsilon\,\bigl[
  DV( e^{\epsilon\,(\pi-\pi^\dagger)}\,e^{\epsilon\,\pi^\dagger}\,U)
  \bigr]^\dagger
  \bigr\}
  \nonumber
  \\
  &= 
  e^{\epsilon\,(\pi_{1/2}-\pi_{1/2}^\dagger)}\,
  \pi_{1/2}\,e^{-\epsilon\,(\pi_{1/2}-\pi_{1/2}^\dagger)}
  - \epsilon\,[
  DV(U')
  ]^\dagger.
\end{align}
This gives a unit of molecular dynamics (MD) on $T G^\bbC$: 
\begin{align}
  \pi_{1/2} &= \pi - \epsilon\,[DV(U)]^\dagger,
\label{md1}
\\
  U' &= e^{\epsilon\,(\pi_{1/2} - \pi_{1/2}^\dagger)}\,
  e^{\epsilon\,\pi_{1/2}^\dagger}\,U,
\label{md2}
\\
  \pi' &= e^{\epsilon\,(\pi_{1/2} - \pi_{1/2}^\dagger)}\,\pi_{1/2}\,
  e^{-\epsilon\,(\pi_{1/2} - \pi_{1/2}^\dagger)}
  - \epsilon\,[ DV(U') ]^\dagger. 
\label{md3}
\end{align}

The MD step \eqref{md1}--\eqref{md3} 
satisfies the following three properties, 
just as the standard leapfrog algorithm does in flat space:

\noindent
\underline{(a) Exact reversibility}:
The transformation is invariant under time reversal,
\begin{align}
  U &\to \tilde{U} \equiv U',
\label{rev1}
\\
  \pi &\to \tilde\pi \equiv -\pi',
\label{rev2}
\\
  \pi_{1/2} &\to \tilde\pi_{1/2}
  \equiv -e^{\epsilon\,(\pi_{1/2} - \pi_{1/2}^\dagger)}\,\pi_{1/2}\,
    e^{-\epsilon\,(\pi_{1/2} - \pi_{1/2}^\dagger)},
\label{rev3}
\\
  U'&\to \tilde{U}' \equiv U,
\label{rev4}
\\
  \pi' &\to \tilde\pi' \equiv -\pi.
\label{rev5}
\end{align}
 
\noindent
\underline{(b) Symplecticity}: 
The symplectic form $\omega$ 
and thus the symplectic volume form $d\Omega = \omega^{2N}/(2N)!$ 
are preserved,
\begin{align}
  \omega' = \omega,
  \qquad
  d\Omega' = d\Omega.
\label{symp}
\end{align}

\noindent
\underline{(c) Approximate energy conservation}:
\begin{align}
  H(U',\pi') = H(U,\pi) + O(\epsilon^3).
\label{ene_consv}
\end{align}

Among these, 
only the first property is nontrivial. 
Indeed, property (b) follows directly 
from the fact that the transformation $T$ 
is composed of canonical transformations of the form 
$T_f = e^{\epsilon\,\{\ast,f\}}$.%
\footnote{ 
  This property can also be verified explicitly 
  from the identity 
  \begin{align}
    a' = a + \frac{\epsilon}{2}\,d\bigl[
    \tr \pi_{1/2}^\dag \pi_{1/2}
    - V(U) - V(U')
    \bigr],
  \nonumber
  \end{align}
  which is proved in Appendix~\ref{sec:symplecticity} 
  for a more general case. 
} 
Property (c) holds because 
$T = T_H + O(\epsilon^3)$ 
and $T H = T_H H + O(\epsilon^3) = H + O(\epsilon^3)$. 
Property (a) is proved in Appendix~\ref{sec:rev}.

\section{Constrained molecular dynamics on $G^\bbC$}
\label{sec:constrained}

In this section, 
we develop a general framework for molecular dynamics (MD) 
on a real $m$-dimensional submanifold $\calS$ of $G^\bbC$ ($m \leq 2N$). 
Specifically, we set $\calS = \Sigma$ 
with $m = N$ for GT-HMC (Sect.~\ref{sec:gt}),
and $\calS = \calR$ with $m = N+1$ for WV-HMC (Sect.~\ref{sec:wv}).

\subsection{Continuum version (Hamiltonian dynamics)}
\label{sec:constrained_continuum}

We assume that 
the submanifold $\calS$ is defined 
by $2N-m$ independent real-valued functions $\phi^r(U)=0$ 
$(r=1,\ldots,2N-m)$. 
Our phase space is the tangent bundle over $\calS$, 
\begin{align}
  T \calS \equiv \{ (U,\pi) \,|\, U \in \calS,\,\pi \in T_U\calS \},
\end{align}
with symplectic potential $a_\calS$ 
and symplectic form $\omega_\calS = d a_\calS$ defined by 
\begin{align}
  a_\calS 
  &= \vev{\pi,\theta_\calS},
\\
  \omega_\calS 
  &= d \vev{\pi,\theta_\calS},
\end{align}
where $\theta_\calS = (dU U^{-1})|_\calS$ is the Maurer-Cartan form 
on $\calS$. 
The symplectic volume form is given by 
\begin{align}
  d\Omega_\calS \equiv \frac{\omega_\calS^m}{m!}.
\end{align}

To define consistent Hamiltonian dynamics on $T \calS$ 
with a Hamiltonian of the form 
\begin{align}
  H(U,\pi) = \frac{1}{2}\,\vev{\pi,\pi} + V(U),
\end{align} 
we consider the first-order action 
\begin{align}
  I[U,\pi,\lambda] = 
  \int ds\,\Bigl[ \vev{\pi, \circdot{U}U^{-1}} - H(U,\pi)
  - \lambda_r\, \phi^r(U) \Bigr],
\end{align}
where $\circdot{U}\equiv dU/ds$ 
and $\lambda_r\in\bbR$ are Lagrange multipliers. 
The resulting equations of motion are 
\begin{align}
  \circdot{U} &= \pi\,U,
\label{gt_md1}
\\
  \circdot{\pi} &= - 2\, [D V(U)]^\dagger
  + [\pi, \pi^\dagger]
  - 2\,\lambda_r\, [D\phi^r(U)]^\dagger,
\label{gt_md2}
\end{align}
subject to the constraints%
\footnote{ 
  The second constraint follows from 
  $0 = \circdot{\phi}{}^r (U) = 2\,\vev{\circdot{U} U^{-1},(D\phi^r)^\dagger}
  = 2\,\vev{\pi, (D\phi^r)^\dagger}$ 
  [see Eq.~\eqref{delta_f}].
} 
\begin{align}
  \phi^r(U) &= 0,
\label{gt_md3}
\\
  \vev{\pi, (D\phi^r)^\dagger} &= 0.
\label{gt_md4}
\end{align}
It is easy to verify that 
the symplectic potential $a_\calS$ changes as 
$\circdot{a}_\calS 
= d[(1/2)\vev{\pi,\pi} - V(U) - \lambda_r \phi^r(U)]$,%
\footnote{ 
  See footnote \ref{a_dot}.
} 
from which it follows that 
$\circdot{\omega}_\calS = d\circdot{a}_\calS = 0$. 
Furthermore, we define
\begin{align}
  \lambda_\textrm{cont} \equiv \lambda_r\,(D \phi^r)^\dagger,
\label{lambda_cont}
\end{align}
which lies in the normal space $N_U\calS$.%
\footnote{
  Indeed, for any vector $v\in T_U\calS$, 
  we have 
  \begin{align}
    \vev{\lambda_\textrm{cont}, v}
    = \lambda_r\,\vev{(D\phi^r)^\dagger,v}
    = (\lambda_r/2)\,\tr [v\,D\phi^r + v^\dagger (D\phi^r)^\dagger ]
    = (\lambda_r/2)\,\lim_{\epsilon\to 0}(1/\epsilon)[\phi^r(e^{\epsilon v}U)-\phi^r(U)]
    = 0.
  \nonumber
  \end{align}
} 
It then follows that $\circdot{H} = -2\,\vev{\pi,\lambda_\textrm{cont}} = 0$.

\subsection{Discrete version (molecular dynamics)}
\label{sec:constrained_discrete}

A discretized version of Eqs.~\eqref{gt_md1}--\eqref{gt_md4} 
with step size $\Delta s = \epsilon$ is obtained 
by modifying Eqs.~\eqref{md1}--\eqref{md3} 
using the RATTLE algorithm \cite{Andersen:1983,Leimkuhler:1994}
as follows:
\begin{align}
  \pi_{1/2} &= \pi - \epsilon\,[DV(U)]^\dagger
  - \epsilon\,\lambda,
\label{rattle1}
  \\
  U' &= e^{\epsilon\,(\pi_{1/2} - \pi_{1/2}^\dagger)}\,
  e^{\epsilon\,\pi_{1/2}^\dagger}\,U,
\label{rattle2}
\\
  \pi' &= e^{\epsilon\,(\pi_{1/2} - \pi_{1/2}^\dagger)}\,\pi_{1/2}\,
  e^{-\epsilon\,(\pi_{1/2} - \pi_{1/2}^\dagger)}
  - \epsilon\,[DV(U')]^\dagger
  - \epsilon\,\lambda'.
\label{rattle3}
\end{align}
Here, the Lagrange multipliers 
$\lambda \in N_U \calS$ and $\lambda' \in N_{U'} \calS$ 
are determined, respectively, 
such that $U' \in \calS$ and $\pi' \in T_{U'} \calS$. 
If one introduces the orthogonal projector 
$\Pi_\calS$ ($\Pi'_\calS$) 
from $T_U G^\bbC$ ($T_{U'} G^\bbC$) 
onto $T_U \calS$ ($T_{U'} \calS$), 
then the determination of $\lambda'$ is replaced 
by the projection: 
\begin{align}
  \pi' = \Pi'_\calS\, \tilde\pi',
\end{align}
where 
\begin{align}
  \tilde\pi' \equiv e^{\epsilon\,(\pi_{1/2} - \pi_{1/2}^\dagger)}\,\pi_{1/2}\,
  e^{-\epsilon\,(\pi_{1/2} - \pi_{1/2}^\dagger)}
  - \epsilon\,[DV(U')]^\dagger.
\end{align}

The RATTLE update \eqref{rattle1}--\eqref{rattle3} 
satisfies the following three properties, 
just as in the unconstrained case:

\noindent
\underline{(a) Exact reversibility}:
The transformation is invariant under time reversal,
\begin{align}
  U &\to \tilde{U} \equiv U',
\label{rattle_rev1}
\\
  \pi &\to \tilde\pi \equiv -\pi',
\label{rattle_rev2}
\\
  \pi_{1/2} &\to \tilde\pi_{1/2}
  \equiv -e^{\epsilon\,(\pi_{1/2} - \pi_{1/2}^\dagger)}\,\pi_{1/2}\,
  e^{-\epsilon\,(\pi_{1/2} - \pi_{1/2}^\dagger)},
\label{rattle_rev3}
\\
  U'&\to \tilde{U}' \equiv U,
\label{rattle_rev4}
\\
  \pi' &\to \tilde\pi' \equiv -\pi,
\label{rattle_rev5}
\\
  \lambda &\to \tilde\lambda \equiv \lambda',
\label{rattle_rev6}
\\
  \lambda' &\to \tilde\lambda' \equiv \lambda.
\label{rattle_rev7}
\end{align}

\noindent
\underline{(b) Symplecticity}: 
The symplectic form $\omega_\calS$ 
and thus the symplectic volume form $d\Omega_\calS = \omega_\calS^m/m!$ 
are preserved,
\begin{align}
  \omega'_\calS = \omega_\calS,
  \qquad
  d\Omega'_\calS = d\Omega_\calS.
\end{align}

\noindent
\underline{(c) Approximate energy conservation}:
\begin{align}
  H(U',\pi') = H(U,\pi) + O(\epsilon^3).
\label{ene_consv2}
\end{align}

Property (a) can be proved in a way similar to the unconstrained case. 
A simple proof of Property (c) [Eq.~\eqref{ene_consv2}] 
is given in Appendix~\ref{sec:ene_consv2}.
Property (b) is not obvious now, 
because the RATTLE update is not manifestly a composition of transformations 
of the form $T_f = e^{\epsilon\,\{\ast,f\}}$. 
One way to verify this property is to use the identity 
\begin{align}
  a_\calS' = a_\calS + \frac{\epsilon}{2}\,d\bigl[
  \tr \pi_{1/2}^\dagger \pi_{1/2}
  - V(U) - V(U')
  \bigr],
\label{rattle_symplecticity}
\end{align}
which is shown in Appendix~\ref{sec:symplecticity}.

\subsection{HMC on a submanifold $\calS$ of $G^\bbC$}
\label{sec:constrained_calS}

To simplify expressions, 
we introduce the notation $\Omega^T \equiv (U,-\pi)$ 
for $\Omega = (U,\pi)$. 
Our aim is to numerically evaluate the reweighted average $\vev{\cdots}_\calS$ 
that is expressed as an integral over the tangent bundle $T \calS$,
\begin{align}
  \vev{g(U)}_\calS
  = \frac{\int_{T\calS} d\Omega_\calS\,e^{-H(U,\pi)}\,g(U)}
  {\int_{T\calS} d\Omega_\calS\,e^{-H(U,\pi)}}
  = \frac{\int_{T\calS} d\Omega_\calS\,e^{-H(\Omega)}\,g(U)}
  {\int_{T\calS} d\Omega_\calS\,e^{-H(\Omega)}}.
\end{align}
This can be estimated by a sample average 
over configurations $\{\Omega = (U,\pi)\}$ 
drawn from the distribution $\propto e^{-H(\Omega)}$.

In the remainder of this paper
(except in Appendices~\ref{sec:ene_consv2} and~\ref{sec:symplecticity}),  
we rescale the Lagrange multipliers as 
$\lambda \to \lambda / \epsilon^2$ 
and 
$\lambda' \to \lambda' / \epsilon^2$
to simplify later expressions, 
which means that $\lambda$ and $\lambda'$ are $O(\epsilon^2)$ quantities. 
We also introduce an intermediate variable 
\begin{align}
  Z \equiv \epsilon \pi - \epsilon^2\,[DV(U)]^\dagger,
\label{Z_def1}
\end{align}
so that 
\begin{align}
  \pi_{1/2} = (Z-\lambda)/\epsilon. 
\label{Z_def2}
\end{align}
A single MD step 
$T:\,\Omega = (U,\pi) \mapsto \Omega' = (U',\pi')$ 
is then given as follows  
[see Eqs.~\eqref{rattle1}--\eqref{rattle3}, \eqref{Z_def1} and \eqref{Z_def2} 
with $\lambda \to \lambda/\epsilon^2$]: 
\begin{align}
  Z &= \epsilon \pi - \epsilon^2\,[DV(U)]^\dagger,
\label{rattle1a}
\\
  U' &= e^{(Z - \lambda) - (Z - \lambda)^\dagger}\,
  e^{(Z - \lambda)^\dagger}\,U,
\label{rattle2a}
\\
  \tilde\pi' &= e^{(Z - \lambda) - (Z - \lambda)^\dagger}\,
  \frac{1}{\epsilon}\,(Z - \lambda)\,
  e^{-(Z - \lambda) + (Z - \lambda)^\dagger}
  - \epsilon\,[DV(U')]^\dagger,
\label{rattle3a}
\\
  \pi' &= \Pi'_\calS\, \tilde\pi',
\label{rattle4a}
\end{align}
where $\lambda \in N_U \calS$ is determined 
so that $U' \in \calS$.

To construct a Markov chain 
with equilibrium distribution proportional to $e^{-H(\Omega)}$ 
with $H$ satisfying $H(\Omega^T) = H(\Omega)$, 
we define a set of stochastic processes 
$P_{(k)}(\Omega'\,|\,\Omega)$ $(k=1,\ldots,K)$
such that 
each satisfies the detailed balance condition 
of molecular dynamics, 
\begin{align}
  P_{(k)}(\Omega'\,|\,\Omega)\,e^{-H(\Omega)}
  = P_{(k)}(\Omega^T\,|\,\Omega^{\prime\,T})\,e^{-H(\Omega^{\prime\,T})}.
\label{detailed_balance_MD}
\end{align}
We also assume that the product $P \equiv P_{(K)}\cdots P_{(1)}$ 
has a proper ergodicity property 
that ensures unique convergence. 
Under these conditions, 
it can be shown that $e^{-H(\Omega)}$ 
is the unique eigenvector of $P$ with unit eigenvalue 
(see Appendix~\ref{sec:hmc_flat}), 
\begin{align}
  \int_{T\calS} d\Omega'_\calS\,
  P(\Omega\,|\,\Omega')\,e^{-H(\Omega')} = e^{-H(\Omega)},
\end{align}
and thus $e^{-H(\Omega)} / \int_{T\calS} d\Omega_\calS\,e^{-H(\Omega)}$ 
is the equilibrium distribution reached by the Markov chain 
defined by $P$. 

In the present case, 
we consider the following two stochastic processes $P_{(1)}$ and $P_{(2)}$. 
Although neither $P_{(1)}$ nor $P_{(2)}$ is ergodic, 
their product $P \equiv P_{(2)}\,P_{(1)}$ is expected to be ergodic as usual, 
guaranteeing convergence to the correct equilibrium.

\noindent
\underline{(1) Heat bath for $\pi$}:
\begin{align}
  P_{(1)}(\Omega'\,|\,\Omega) 
  = \frac{1}{(2\pi)^{m/2}}\,e^{-(1/2)\,\vev{\pi',\pi'}}\,
  \delta_\calS(U',\, U),
\end{align}
where $\delta_\calS(U',\,U)$ is the delta function on $\calS$.%
\footnote{ 
  When $U \in \calS$ is parametrized as $U = U(t,U_0)$ 
  as in GT/WV-HMC, 
  the delta function is proportional to  
  $\delta(U'_0,\, U_0)$ for GT-HMC ($\calS = \Sigma$), 
  and to $\delta(t'-t)\,\delta(U'_0,\, U_0)$ for WV-HMC ($\calS = \calR$), 
  where $\delta(U'_0,\, U_0)$ is the bi-invariant delta function 
  associated with the Haar measure on $G$. 
  Jacobian factors can be ignored 
  because they do not matter in the argument for detailed balance 
  under the condition $U'=U$.
} 
$\pi'$ can be generated by drawing $\tilde\pi \in T_U G^\bbC \simeq \LieGC$ 
from the Gaussian distribution $e^{-(1/2)\,\vev{\tilde\pi,\tilde\pi}}$ 
(see Appendix~\ref{sec:Gaussian})
and projecting it onto $T_U \calS$ 
as
\begin{align}
  \pi' = \Pi_\calS\, \tilde\pi.
\end{align}
This process satisfies the detailed balance condition \eqref{detailed_balance_MD}, 
because 
\begin{align}
  P_{(1)}(\Omega'\,|\,\Omega)\,e^{-H(\Omega)}
  &= \frac{1}{(2\pi)^{m/2}}\,e^{-(1/2)\,(\vev{\pi,\pi} + \vev{\pi',\pi'})}\,
  \delta_\calS(U',U)\,e^{ -V(U) }
\nonumber
\\
  &= \frac{1}{(2\pi)^{m/2}}\,e^{-(1/2)\,(\vev{-\pi,-\pi} + \vev{-\pi',-\pi'})}\,
  \delta_\calS(U,U')\,e^{ -V(U') }
\nonumber
\\
  &= P_{(1)}(\Omega^T\,|\,\Omega^{\prime\,T})\,e^{-H(\Omega^{\prime\,T})}.
\end{align}

\noindent
\underline{(2) MD followed by Metropolis test}:

\noindent
We introduce a reversible and volume-preserving map 
$f: \Omega \mapsto f(\Omega)$, 
constructed by repeating the MD update  
\eqref{rattle1a}--\eqref{rattle4a} 
$N_\textrm{MD}$ times: 
$f = T^{N_\textrm{MD}}$. 
We update $\Omega$ to $\Omega' = f(\Omega)$
with probability 
$A(\Omega) \equiv \min(1,e^{-H(f(\Omega)) + H(\Omega)})$. 
Otherwise, we update $\Omega$ to $\Omega' = \Omega^T$.%
\footnote{ 
  See, e.g., Ref.~\cite{Horowitz:1991rr}.
} 
The transition probability can be compactly summarized as 
\begin{align}
  P_{(2)}(\Omega'\,|\,\Omega) 
  = A(\Omega)\,\delta_{T\calS}(\Omega',f(\Omega))
    + [1 - A(\Omega)]\,\delta_{T\calS}(\Omega',\Omega^T),
\end{align}
where $\delta_{T\calS}$ is the symplectic delta function 
with respect to the symplectic volume form $d\Omega_\calS$. 
Since the momentum is fully refreshed before each MD trajectory 
and our observables depend only on $U$, 
the momentum flip upon rejection 
need not be performed in practice.
As in Appendix~\ref{sec:hmc_flat}, 
one can prove that 
$P_{(2)}$ satisfies the normalization condition 
\begin{align}
  \int_{T\calS} d\Omega'_\calS\,P_{(2)}(\Omega'|\Omega) = 1
\end{align}
and the detailed balance condition 
\begin{align}
  P_{(2)}(\Omega'|\Omega)\,e^{-H(\Omega)}
  = P_{(2)}(\Omega^T|\Omega^{\prime\,T})\,e^{-H(\Omega^{\prime\,T})}. 
\end{align}

Since our observables depend only on $U$, 
the algorithm can be viewed as a stochastic process 
on $U$ alone: 
\begin{itemize}
  \item
  \underline{Step 1 (momentum refresh)}:\\
  Given $U \in \calS$, 
  generate $\tilde\pi \in T_U G^\bbC \simeq \LieGC$ 
  from the Gaussian distribution 
  $\propto e^{-(1/2) \vev{\tilde\pi,\tilde\pi}}$,
  and project it to get $\pi = \Pi_\calS\, \tilde\pi \in T_U \calS$.
  
  \item
  \underline{Step 2 (molecular dynamics)}:\\
  Evolve $(U,\pi) \to (U',\pi')$ 
  by repeatedly performing the RATTLE update \eqref{rattle1a}--\eqref{rattle4a}.
  
  \item
  \underline{Step 3 (Metropolis test)}:\\
  Accept the proposed $U'$ 
  with probability $\min \bigl(1, e^{-[H(U',\pi') - H(U,\pi)]} \bigr)$.
  
\end{itemize}

The only remaining ingredients in the whole algorithm 
are the construction of the projector $\Pi_\calS$
and the determination of the Lagrange multiplier $\lambda \in N_U \calS$, 
both of which depend on the details of the geometry of $\calS$. 
We explain the case $\calS = \Sigma$ (GT-HMC) in Sect.~\ref{sec:gt} 
and the case $\calS = \calR$ (WV-HMC) in Sect.~\ref{sec:wv}.

\section{Algorithm of GT-HMC for group manifolds}
\label{sec:gt}

In GT-HMC, 
we consider a submanifold $\calS = \Sigma \equiv \Sigma_t$, 
which is the deformed surface at fixed flow time $t$.

\subsection{Projector in GT-HMC}
\label{sec:gt_projector}

We first detail an algorithm to project a vector $w\in T_U G^\bbC$ 
onto the tangent space $T_U\Sigma$ and the normal space $N_U\Sigma$: 
\begin{align}
  w =  w_v + w_n \quad
  \bigl(w_v \equiv \Pi_\Sigma\, w \in T_U\Sigma,
  ~ w_n \equiv (1 - \Pi_\Sigma)\, w\in N_U\Sigma\bigr). 
\label{gt_projector}
\end{align}
By using the $\bbR$-linear map 
$A:\,T_{U_0} G^\bbC \ni w_0 = v_0 + n_0  
\mapsto w = v + n = E v_0 + F n_0 \in T_U G^\bbC$ 
[Eq.~\eqref{multEF}], 
the decomposition \eqref{gt_projector} of a given $w\in T_U G^\bbC$ 
can be carried out as in Algorithm~\ref{alg:gt_decomposition}.
Note that, if an iterative method (such as BiCGStab) is used 
to solve $A w_0 = w$ in Step~1, 
then Steps~2 and~3 are no longer necessary 
as in the flat case \cite{Alexandru:2017lqr}. 
This is because, 
during the iteration, 
one repeatedly computes $E \tilde{v}_0$ and $F \tilde{n}_0$ 
for a candidate solution $\tilde{w}_0 = \tilde{v}_0 + \tilde{n}_0$, 
so that $v = E v_0$ and $n = F n_0$ are already obtained 
upon convergence.
%
\begin{algorithm}[tb]
  \caption{
    Orthogonal decomposition of $w\in T_U G^\bbC$ 
    into $w_v\in T_U\Sigma$ and $w_n\in N_U\Sigma$
  }
  \label{alg:gt_decomposition}
  
  \begin{algorithmic}[1]
    \State%
    Solve the linear system $A w_0 = w$ for $w_0$ [Eq.~\eqref{multEF}]

    \State%
    Decompose $w_0$ 
    into $v_0 = (1/2)(w_0 - w_0^\dagger)$ 
    and $n_0 = (1/2)(w_0 + w_0^\dagger)$

    \State%
    Compute $v = E v_0$ and $n = F n_0$ 
    by integrating Eqs.~\eqref{flow_t} and \eqref{flow_n}, 
    and set $w_v = v$ and $w_n = n$.%
  \end{algorithmic}
\end{algorithm}

\subsection{Constrained molecular dynamics on $T\Sigma$}
\label{sec:gt_md}

\begin{figure}[tb]
  \centering
  \includegraphics[width=75mm]{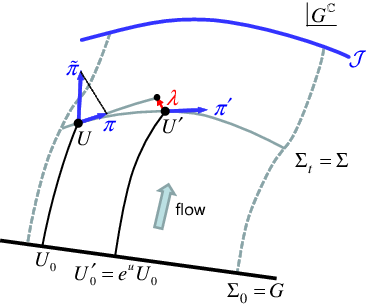}
  \caption{
    RATTLE in GT-HMC: $(U,\pi)\to (U',\pi')$.
    The initial momentum $\pi$ is obtained 
    by projecting $\tilde\pi\in T_U G^\bbC$ onto $T_U\Sigma$,
    where $\tilde\pi$ is isotropically generated with the Gaussian distribution 
    $\propto e^{-(1/2)\vev{\tilde\pi, \tilde\pi}}$ 
    (see Appendix~\ref{sec:Gaussian}). 
  }
\label{fig:gt_rattle_group}
\end{figure}

We begin by recalling the quantity we aim to evaluate [Eq.~\eqref{vev_gt1}]:
\begin{align}
	\vev{\calO} 
	&= \frac{\vev{\calF(U)\,\calO(U)}_\Sigma}
	{\vev{\calF(U)}_\Sigma}.
\end{align}
Here, the reweighted average on $\Sigma$ is given by 
[Eq.~\eqref{vev_gt3}]
\begin{align}
	\vev{ g(U) }_\Sigma 
  &= \frac{\int_{T\Sigma}\, d\Omega_\Sigma\,e^{-H(U,\pi)}\,g(U)}
  {\int_{T\Sigma}\, d\Omega_\Sigma\,e^{-H(U,\pi)}},
\end{align}
where [Eqs.~\eqref{dOmega_Sigma}, \eqref{omega_Sigma}, \eqref{H_Sigma}]
\begin{align}
  d\Omega_\Sigma 
  &= \frac{\omega_\Sigma^N}{N!}
  ~~\text{with}~~
  \omega_\Sigma = d \vev{\pi, \theta_\Sigma},
\\
  H(U,\pi)
  &= \frac{1}{2}\,\vev{\pi,\pi} + \ReS(U).
\end{align}
The reweighting factor $\calF(U)$ takes the form [Eq.~\eqref{F_Sigma}]
\begin{align}
  \calF(U) 
  = \frac{\det E}{\sqrt{\gamma}}\,e^{-i\,\ImS(U)}.
\end{align}
Since $[DV(U)]^\dagger = (1/2)\,[DS(U)]^\dagger = (1/2)\,\xi(U)$ 
for $V(U) = \ReS(U)$, 
a single MD step [Eqs.~\eqref{rattle1}--\eqref{rattle3}] 
is given by 
\begin{align}
  Z &= \epsilon \pi - \frac{\epsilon^2}{2}\,\xi(U),
\label{gt_rattle1a}
\\
  U' &= e^{(Z - \lambda) - (Z - \lambda)^\dagger}\,
  e^{(Z - \lambda)^\dagger}\,U,
\label{gt_rattle2a}
\\
  \tilde\pi' &= e^{(Z - \lambda) - (Z - \lambda)^\dagger}\,
  \frac{1}{\epsilon}\,(Z - \lambda)\,
  e^{-(Z - \lambda) + (Z - \lambda)^\dagger}
  - \frac{\epsilon}{2}\,\xi(U'),
\label{gt_rattle3a}
\\
  \pi' &= \Pi'_\Sigma\, \tilde\pi'.
\label{gt_rattle4a}
\end{align}
Here, the Lagrange multiplier $\lambda \in N_U \Sigma$ 
is determined such that $U' \in \Sigma$, 
which is equivalent 
to finding a pair $(u \in T_{U_0} \Sigma_0,\,\lambda \in N_U \Sigma)$ 
satisfying the following equation 
for a given $U = U(t, U_0)$ (see Fig.~\ref{fig:gt_rattle_group}):
\begin{align}
  e^{(Z-\lambda) - (Z-\lambda)^\dagger}\,e^{(Z-\lambda)^\dagger}
  U(t, U_0)
  = U(t, e^u U_0). 
\label{gt_existence}
\end{align}

We solve Eq.~\eqref{gt_existence} by using Newton's method 
based on the premise that an update 
$u \leftarrow u + \Delta u$ 
and $\lambda \leftarrow \lambda + \Delta\lambda$ 
can be efficiently performed by solving the equation 
\begin{align}
  e^{-\Delta \lambda}\,e^{(Z-\lambda) - (Z-\lambda)^\dagger}\,
  e^{(Z-\lambda)^\dagger}\,U(t,U_0)
  = U(t, e^{\Delta u} e^u U_0).
\end{align}
We linearize this 
by writing $e^{-\Delta\lambda} \approx 1 - \Delta\lambda$ 
and $U(t, e^{\Delta u} e^u U_0)\approx 
(1 + E_\textrm{new} \Delta u)\,U(t,e^u U_0)$, 
where $E_\textrm{new}$ is the $\bbR$-linear map  
$E_\textrm{new}:\,T_{e^u U_0}\Sigma_0 \to T_{U_\textrm{new}} \Sigma$ 
with $U_\textrm{new} \equiv U(t,e^u U_0)$.
The algorithm then proceeds as follows:

\noindent
\underline{Step 1}: 
Initialize $u=0$ and $\lambda=0$.

\noindent
\underline{Step 2}: 
Solve the linearized equation 
\begin{align}
  E_\textrm{new} \Delta u 
  + \Delta\lambda\,e^{(Z-\lambda) - (Z-\lambda)^\dagger}\,
  e^{(Z-\lambda)^\dagger}\,U U_\textrm{new}^{-1}
  =
  e^{(Z-\lambda) - (Z-\lambda)^\dagger}\,
  e^{(Z-\lambda)^\dagger}\,U U_\textrm{new}^{-1} - 1,
\label{gt_newton1}
\end{align}
where $U = U(t,U_0)$ and $U_\textrm{new} = U(t, e^u U_0)$. 

\noindent
\underline{Step 3}: 
Update 
$u \leftarrow u + \Delta u$ 
and $\lambda \leftarrow \lambda + \Delta\lambda$. 

\noindent
\underline{Step 4}: 
Repeat Steps~2 and~3 
until the Frobenius norm of the right-hand side of Eq.~\eqref{gt_newton1} 
becomes smaller than a prescribed tolerance. 

Note that, 
once $(u,\lambda)$ is sufficiently close to the solution, 
the map $E_\textrm{new}$ is nearly equal to $E$ 
and the right coefficient of $\Delta\lambda$ becomes close to unity. 
We therefore employ the simplified Newton method 
of the following form \cite{Fukuma:2023eru}:%
\footnote{ 
  This is a group-manifold extension 
  of the simplified GT-HMC algorithm \cite{Fukuma:2023eru} 
  that uses the fixed-point method 
  introduced in Ref.~\cite{Fujii:2013sra}, 
  the latter performing HMC updates directly on a single Lefschetz thimble 
  in flat space $\bbC^N$.
} 
\begin{align}
  E\,\Delta u + \Delta\lambda = B,
\label{gt_newton2}
\end{align}
where $E$ is the original $\bbR$-linear map, 
$E:\,T_{U_0}\Sigma_0 \to T_U \Sigma$, 
and the right-hand side $B$ is given by%
\footnote{ 
  $B$ is not necessarily traceless even when $G = SU(n)$ 
  although the solution ($u,\,\lambda$) must be traceless 
  for that case. 
  A simple algorithm is that 
  (1) we solve Eq.~\eqref{gt_newton2} 
  with respect to $\Delta u$ and $\Delta \lambda$ 
  without imposing the traceless condition, 
  (2) we make them traceless,  
  and (3) we update $u$ and $\lambda$ 
  as $u \leftarrow u + \Delta u$ 
  and $\lambda \leftarrow \lambda + \Delta\lambda$. 
  Here, the $\bbR$-linear map $A$ in Eq.~\eqref{multEF} 
  is extended to traceful matrices by using 
  the same flow equations (4.7) and (4.8). 
  Explicit numerical tests in Sect.~\ref{sec:1site} show that 
  the repetition of processes (1)--(3) converges quickly 
  for the one-site model. 
  The same prescription also applies to the WV-HMC algorithm. 
  One should not make $B$ traceless 
  before solving Eq.~\eqref{gt_newton2} 
  based on the property $\det (B+1) = 1$. 
  In fact, $\det(B+1) = 1$ does not guarantee $B \sim 0$ 
  for a finite step size 
  when $G$ has nontrivial center elements, 
  so that iterations may converge to an incorrect solution. 
\label{fn:traceful_B}
} 
\begin{align}
  B = 
  e^{(Z-\lambda) - (Z-\lambda)^\dagger}\,
  e^{(Z-\lambda)^\dagger}\,U U_\textrm{new}^{-1} - 1.
\label{gt_newton3}
\end{align}
The linear equation \eqref{gt_newton2} can be efficiently solved 
using the projection introduced in Sect.~\ref{sec:gt_projector}. 
Indeed, 
from the orthogonal decomposition 
\begin{align}
  B = B_v + B_n 
  = E B_{0,v} + F B_{0,n}
\end{align}
with 
$B_v\in T_U\Sigma$, $B_n\in N_U\Sigma$,
and $B_{0,v}\in T_{U_0}\Sigma_0$ and $B_{0,n}\in N_{U_0}\Sigma_0$,
we write 
\begin{align}
  B = E B_{0,v} + B_n. 
\end{align}
Comparing this with the left-hand side of 
the simplified Newton equation \eqref{gt_newton2}, 
we obtain 
\begin{align}
  \Delta u = B_{0,v},
  \qquad
  \Delta \lambda = B_n.
\end{align}

A single RATTLE step from $(U,\pi)$ to $(U',\pi')$ on $T\Sigma$  
is summarized in Algorithm~\ref{alg:gt_rattle}. 
%
\begin{algorithm}[tb]
  \caption{RATTLE $(U,\pi)\to(U',\pi')$ in GT-HMC
  with $U = U(t,U_0)$ and $\pi \in T_U \Sigma$}  
  \label{alg:gt_rattle}
  
  \begin{algorithmic}[1]
    \State%
    Compute $\xi(U) = [DS(U)]^\dagger$ 
    and $Z = \epsilon \pi - (\epsilon^2/2)\,\xi(U)$

    \State%
    Set $u = 0$ and $\lambda = 0$
    \For{$k=0,1,\ldots$}%
    \State%
    Compute $U_\textrm{new}=U(t,e^u U_0)$ 
    and $B$ [Eq.~\eqref{gt_newton3}]
    \If{$\| B \|$ is sufficiently small} 
    \State \textbf{break}
    \EndIf
    \State%
    Decompose $B = E B_{0,v} + B_n$ 
    by using Algorithm~\ref{alg:gt_decomposition} 
    \State%
    Update: $u \leftarrow u + B_{0,v}$, $\lambda \leftarrow \lambda + B_n$
    \EndFor
    \State%
    Set $U' = U_\textrm{new}$ and 
    $\tilde\pi'
    = e^{(Z-\lambda) - (Z-\lambda)^\dagger}\,((Z-\lambda)/\epsilon)\,
    e^{-(Z-\lambda) + (Z-\lambda)^\dagger}
    - (\epsilon/2)\,\xi(U')$
    \State%
    Decompose $\tilde\pi' \in T_{U'}G^\bbC$ 
    as $\tilde\pi' = \tilde\pi'_v + \tilde\pi'_n$ 
    by using Algorithm~\ref{alg:gt_decomposition},
    and set $\pi' = \tilde\pi'_v$
  \end{algorithmic}
\end{algorithm}

\subsection{Summary of GT-HMC}
\label{sec:gt_summary}

We summarize the GT-HMC algorithm in Algorithm~\ref{alg:gt_hmc}, 
which uses the simplified RATTLE integrator of Algorithm~\ref{alg:gt_rattle}.

%
\begin{algorithm}[tb]
  \caption{GT-HMC}
  \label{alg:gt_hmc}
  
  \begin{algorithmic}[1]
    \State%
    Given $U\in \Sigma$, 
    generate $\tilde\pi\in T_U G^\bbC \simeq \LieGC$ 
    from the Gaussian distribution 
    $\propto e^{-(1/2)\,\vev{\tilde\pi, \tilde\pi} }$
    \State%
    Decompose $\tilde\pi$ as 
    $\tilde\pi = \tilde\pi_v + \tilde\pi_n$ 
    by using Algorithm~\ref{alg:gt_decomposition}, 
    and set $\pi = \tilde\pi_v$
    \State%
    Apply a sequence of MD steps using Algorithm~\ref{alg:gt_rattle} 
    to evolve $(U,\pi) \to (U',\pi')$ 
    \State%
    Compute $\Delta H\equiv H(U',\pi') - H(U,\pi)$, 
    and accept $U'$ as the new configuration 
    with probability $\min(1,e^{-\Delta H})$. 
    Otherwise, retain $U$ as the next configuration
  \end{algorithmic}
\end{algorithm}

\section{Algorithm of WV-HMC for group manifolds}
\label{sec:wv}

In WV-HMC, 
we consider a submanifold $\calS = \calR$, 
which represents the worldvolume of interest. 
The presentation below is structured 
in parallel with the previous section, 
in order to emphasize both the similarities and the differences 
between GT-HMC and WV-HMC.

\subsection{Projector in WV-HMC}
\label{sec:wv_projector}

We first detail an algorithm to project a vector $w\in T_U G^\bbC$ 
at $U \in \Sigma_t \subset \calR$
onto the tangent space $T_U\calR$ and the normal space $N_U\calR$: 
\begin{align}
  w = w_\parallel + w_\perp \quad
  \bigl(w_\parallel = \Pi_\calR\, w 
  \in T_U\calR,~ 
  w_\perp = (1 - \Pi_\calR)\, w \in N_U\calR\bigr). 
  \label{wv_projector1}
\end{align}
This decomposition can be carried out 
by making use of the tangent/normal decomposition at $\Sigma_t$ 
[see Eq.~\eqref{gt_projector} and Algorithm~\ref{alg:gt_decomposition}], 
and is summarized in Algorithm~\ref{alg:wv_decomposition}.%
\footnote{ 
  One can easily show that 
  $\vev{\xi_n,w_\perp}=0$
  and $\vev{w_\parallel, w_\perp} = 0$, 
  confirming orthogonality. 
} 
%
\begin{algorithm}[thb]
  \caption{Orthogonal decomposition of $w\in T_U G^\bbC$ 
    into $w_\parallel \in T_U\calR$ 
    and $w_\perp \in N_U\calR$}
  \label{alg:wv_decomposition}
  
  \begin{algorithmic}[1]
    \State%
    Compute $\xi(U) = [DS(U)]^\dagger$

    \State%
    Decompose $w$ and $\xi$ as 
    $w = w_v + w_n$ and $\xi = \xi_v + \xi_n$ 
    ($w_v,\,\xi_v \in T_U \Sigma_t$, $w_n,\,\xi_n \in N_U \Sigma_t$)
    by using Algorithm~\ref{alg:gt_decomposition}

    \State%
    Compute $c = \vev{\xi_n, w_n} / \vev{\xi_n, \xi_n}
    \,(= \vev{\xi, w_n} / \vev{\xi_n, \xi_n})$

    \State%
    Set $w_\parallel = w_v + c\,\xi_n$ and 
    $w_\perp = w_n - c\,\xi_n$
  \end{algorithmic}
\end{algorithm}

\subsection{Constrained molecular dynamics on $T\calR$}
\label{sec:wv_md}

\begin{figure}[tb]
  \centering
  \includegraphics[width=75mm]{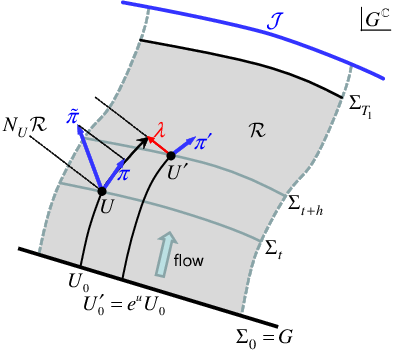}
  \caption{
    RATTLE in WV-HMC: $(U,\pi)\to (U',\pi')$. 
    The initial momentum $\pi$ is obtained 
    by projecting $\tilde\pi\in T_U G^\bbC$ onto $T_U\calR$,
    where $\tilde\pi$ is isotropically generated with the Gaussian distribution 
    $\propto e^{-(1/2)\vev{\tilde\pi, \tilde\pi}}$ 
    (see Appendix~\ref{sec:Gaussian}). 
  }
  \label{fig:wv_rattle_group}
\end{figure}

We begin by recalling the quantity we aim to evaluate 
[Eq.~\eqref{vev_wv1}]:
\begin{align}
  \vev{\calO} 
  &= \frac{\vev{\calF(U)\,\calO(U)}_\calR}
  {\vev{\calF(U)}_\calR}.
\end{align}
Here, the reweighted average on $\calR$ is given by [Eq.~\eqref{vev_wv3}]
\begin{align}
  \vev{ g(U) }_\calR 
  &= \frac{\int_{T \calR}\, d\Omega_\calR \,e^{-H(U,\pi)}\,g(U)}
  {\int_{T \calR}\, d\Omega_\calR \,e^{-H(U,\pi)}},
\end{align}
where [Eqs.~\eqref{dOmega_calR}, \eqref{omega_calR}, 
\eqref{H_calR}, \eqref{V_calR}]
\begin{align}
  d\Omega_\calR
  &= \frac{\omega_\calR^{N+1}}{(N+1)!}
  ~~\text{with}~~
  \omega_\calR
  = d \vev{\pi, \theta_\calR},
\\
  H(U,\pi)
  &= \frac{1}{2}\,\vev{\pi,\pi} + V(U)
  ~~\text{with}~~
  V(U) = \ReS(U) + W(t(U)).
\end{align}
The reweighting factor $\calF(U)$ takes the form [Eq.~\eqref{F_calR}]
\begin{align}
  \calF(U) 
  = \alpha^{-1}\,\frac{\det E}{\sqrt{\gamma}}\,e^{-i\,\ImS(U)}.
\end{align}
A single MD step is given by 
\begin{align}
  Z &= \epsilon \pi - \epsilon^2\,[DV(U)]^\dagger,
\label{wv_rattle1}
\\
  U' &= e^{(Z - \lambda) - (Z - \lambda)^\dagger}\,
  e^{(Z - \lambda)^\dagger}\,U,
\label{wv_rattle2}
\\
  \tilde\pi' &= e^{(Z - \lambda) - (Z - \lambda)^\dagger}\,
  \frac{1}{\epsilon}\,(Z - \lambda)\,
  e^{-(Z - \lambda) + (Z - \lambda)^\dagger}
  - \epsilon\,[DV(U')]^\dagger,
\label{wv_rattle3}
\\
  \pi' &= \Pi'_\calR\, \tilde\pi'.
\label{wv_rattle4}
\end{align}
Here, the gradient of the potential, $[DV(U)]^\dagger$, 
can be set to the following form: 
\begin{align}
  [DV(U)]^\dagger = 
  \frac{1}{2}\,\Bigl[ \xi + \frac{W'(t)}
  {\vev{\xi_n,\xi_n}}\,\xi_n \Bigr].
\label{wv_force}
\end{align}
A proof is given in Appendix~\ref{sec:wv_force}.
The Lagrange multiplier $\lambda \in N_U \calR$ 
is determined such that $U' \in \calR$, 
which is equivalent 
to finding a triplet 
$(h \in \bbR,\,u \in T_{U_0} \Sigma_0,\,\lambda \in N_U \calR)$ 
satisfying the following equation 
for a given $U = U(t, U_0)$ (see Fig.~\ref{fig:wv_rattle_group}):
\begin{align}
  e^{(Z-\lambda) - (Z-\lambda)^\dagger}\,e^{(Z-\lambda)^\dagger}
  U(t, U_0)
  = U(t+h, e^u U_0). 
\label{wv_newton}
\end{align}

We solve Eq.~\eqref{wv_newton} using Newton's method 
based on the premise that an update 
$h \leftarrow h + \Delta h$, 
$u \leftarrow u + \Delta u$ 
and $\lambda \leftarrow \lambda + \Delta\lambda$ 
can be efficiently performed by solving the equation 
\begin{align}
  e^{-\Delta \lambda}\,e^{(Z-\lambda) - (Z-\lambda)^\dagger}\,
  e^{(Z-\lambda)^\dagger}\,U(t,U_0)
  = U(t+h+\Delta h, e^{\Delta u} e^u U_0).
\end{align}
After linearization as in Sect.~\ref{sec:gt_md}, 
the algorithm proceeds as follows:

\noindent
\underline{Step 1}: 
Initialize $h=0$, $u=0$ and $\lambda=0$.

\noindent
\underline{Step 2}: 
Solve the linearized equation 
\begin{align}
  \xi_\textrm{new}\,\Delta h
  + E_\textrm{new} \Delta u 
  + \Delta\lambda\,e^{(Z-\lambda) - (Z-\lambda)^\dagger}\,
  e^{(Z-\lambda)^\dagger}\,U U_\textrm{new}^{-1}
  =
  e^{(Z-\lambda) - (Z-\lambda)^\dagger}\,
  e^{(Z-\lambda)^\dagger}\,U U_\textrm{new}^{-1} - 1.
\label{wv_newton1}
\end{align}
Here, $U = U(t,U_0)$, $U_\textrm{new} = U(t+h, e^u U_0)$, 
$\xi_\textrm{new} = \xi(U_\textrm{new})$, 
and $E$, $E_\textrm{new}$ are the $\bbR$-linear maps, 
$E:\,T_{U_0}\Sigma_0 \to T_U \Sigma_t$ and 
$E_\textrm{new}:\,T_{e^u U_0}\Sigma_0 \to T_{U_\textrm{new}} \Sigma_{t+h}$, 
respectively 
[see Eq.~\eqref{multEF}].

\noindent
\underline{Step 3}: 
Update 
$h \leftarrow h + \Delta h$, 
$u \leftarrow u + \Delta u$, 
$\lambda \leftarrow \lambda + \Delta\lambda$.

\noindent
\underline{Step 4}: 
Repeat Steps~2 and~3 
until the Frobenius norm of the right-hand side of Eq.~\eqref{wv_newton1} 
becomes smaller than a prescribed tolerance. 

When $(h,u,\lambda)$ is sufficiently close to the solution, 
we have
$\xi_\textrm{new} \approx \xi$ and $E_\textrm{new} \approx E$, 
and the right coefficient of $\Delta\lambda$ is nearly unity. 
We therefore employ the simplified Newton method 
of the following form \cite{Fukuma:2023eru}:%
\footnote{ 
  This is a group-manifold extension 
  of the simplified WV-HMC algorithm \cite{Fukuma:2023eru} 
  that uses the fixed-point method 
  introduced in Ref.~\cite{Fujii:2013sra}, 
  the latter performing HMC updates directly on a single Lefschetz thimble 
  in flat space $\bbC^N$.
} 
\begin{align}
  \xi\,\Delta h + E\,\Delta u + \Delta\lambda = B,
\label{wv_newton2}
\end{align}
where $B$ is given by%
\footnote{
  See footnote~\ref{fn:traceful_B}.
} 
\begin{align}
  B = e^{(Z-\lambda) - (Z-\lambda)^\dagger}\,
  e^{(Z-\lambda)^\dagger}\,U U_\textrm{new}^{-1} - 1.
\label{wv_newton3}
\end{align}
The linear equation \eqref{wv_newton2} can be solved 
using the projection introduced 
in Sects.~\ref{sec:gt_projector} and \ref{sec:wv_projector}. 
To see this, 
we note that 
from the orthogonal decomposition 
(see Algorithm~\ref{alg:gt_decomposition})
\begin{align}
  \xi = E\, \xi_{0,v} + \xi_n, 
\end{align}
the left-hand side of Eq.~\eqref{wv_newton2} 
can be rewritten as 
\begin{align}
  E\,(\xi_{0,v}\,\Delta h + \Delta u) + \xi_n\,\Delta h 
  + \Delta\lambda. 
\end{align}
Comparing this with the decomposition (see Algorithm~\ref{alg:wv_decomposition})
\begin{align}
  B&= B_\parallel + B_\perp
  = [ B_v + c_B\, \xi_n ] + [ B_n - c_B\,\xi_n ] 
\nonumber
\\
  &= E B_{0,v} + c_B\, \xi_n + [ B_n - c_B\,\xi_n ]
\\
  &\Bigl( c_B \equiv \frac{\vev{B_n,\xi_n}}{\vev{\xi_n,\xi_n}} 
  = \frac{\vev{B,\xi_n}}{\vev{\xi_n,\xi_n}}\Bigr), 
\end{align}
we obtain the following equations: 
\begin{align}
  \xi_{0,v}\,\Delta h + \Delta u &= B_{0,v},
\\
  \Delta h &= c_B,
\\
  \Delta \lambda &= B_n - c_B\,\xi_n,
\end{align}
which can be solved explicitly as 
\begin{align}
  \Delta h &= c_B,
\\
  \Delta u &= B_{0,v} - c_B\,\xi_{0,v},
\\
  \Delta \lambda &= B_n - c_B\,\xi_n.
\end{align}

A single RATTLE step from $(U,\pi)$ to $(U',\pi')$ on $T\calR$  
is summarized in Algorithm~\ref{alg:wv_rattle}. 
%
\begin{algorithm}[tb]
  \caption{RATTLE $(U,\pi)\to(U',\pi')$ in WV-HMC
    with $U = U(t,U_0)$ and $\pi \in T_U \calR$} 
  \label{alg:wv_rattle}
  
  \begin{algorithmic}[1]
    \State%
    Compute $\xi(U) = [DS(U)]^\dagger$ 
    and decompose it as $\xi ~= E\, \xi_{0,v} + \xi_n$ 
    by using Algorithm~\ref{alg:gt_decomposition}

    \State%
    Compute $Z = \epsilon \pi - \epsilon^2\,[DV(U)]^\dagger
    = \epsilon \pi - (\epsilon^2/2)\,[\xi + (W'(t)/\vev{\xi_n,\xi_n})\,\xi_n]$
    [Eq.~\eqref{wv_force}]

    \State%
    Set $h = 0$, $u = 0$, and $\lambda = 0$

    \For{$k=0,1,\ldots$}%
    \State%
    Compute $U_\textrm{new}=U(t+h,e^u U_0)$ 
    and $B$ [Eq.~\eqref{wv_newton3}]
    \If{$\| B \|$ is sufficiently small} 
    \State \textbf{break}
    \EndIf
    \State%
    Decompose $B = E B_{0,v} + B_n$ 
    by using Algorithm~\ref{alg:gt_decomposition}

    \State%
    Compute $c_B = \vev{B,\xi_n} / \vev{\xi_n,\xi_n}$

    \State%
    Set $\Delta h = c_B$, 
    $\Delta u = B_{0,v} - c_B\, \xi_{0,v}$,
    $\Delta \lambda = B_n - c_B\, \xi_n$ 

    \State%
    Update: $h \leftarrow h + \Delta h$, 
    $u \leftarrow u + \Delta u$, 
    $\lambda \leftarrow \lambda + \Delta \lambda$
    \EndFor
    \State%
    Set $U' = U_\textrm{new}$ and 
    $\tilde\pi'
    = e^{(Z-\lambda) - (Z-\lambda)^\dagger}\,((Z-\lambda)/\epsilon)\,
    e^{-(Z-\lambda) + (Z-\lambda)^\dagger}
    - \epsilon\,[DV(U')]^\dagger$
    \State%
    Decompose $\tilde\pi' \in T_{U'} G^\bbC$ 
    as $\tilde\pi' = \tilde\pi'_\parallel + \tilde\pi'_\perp$ 
    by using Algorithm~\ref{alg:wv_decomposition}, 
    and set $\pi' = \tilde\pi'_\parallel$
  \end{algorithmic}
\end{algorithm}

\subsection{Treatment of the boundary}
\label{sec:wv_boundary}

The temporal extent of the worldvolume can be restricted 
to a finite region $T_0\lesssim t\lesssim T_1$ 
by adjusting the function $W(t)$ in the potential $V(U)$. 
A possible form of the function (modified from Ref.~\cite{Fukuma:2023eru}) 
is given by 
\begin{align}
  W(t) =
  \begin{cases}
    -\,\gamma(t-T_0)
    - c_0\,(t-T_0)\,\bigl(e^{(t-T_0)^2/2d_0^2} - 1\bigr)
    & (t < T_0) \\
    -\,\gamma(t-T_0)
    & (T_0 \leq t \leq T_1) \\
    -\,\gamma(t-T_0)
    + c_1\,(t-T_1)\,\bigl(e^{(t-T_1)^2/2d_1^2} - 1\bigr)
    & (t > T_1).
  \end{cases}
\label{W(t)}
\end{align}
The exponential terms introduce soft reflective boundaries 
at $t = T_0$ (lower boundary) and $t = T_1$ (upper boundary) 
with penetration depths $d_0$ and $d_1$, respectively. 
The wall terms start at cubic order in the distance from the boundaries, 
so that $W(t)$ is twice continuously differentiable, 
guaranteeing an $O(\epsilon^3)$ scaling of $\Delta H$ 
for a single MD step with step size $\Delta s = \epsilon$. 
The coefficients $c_0$ and $c_1$ control the wall heights
at $t=T_0-d_0$ and $t=T_1+d_1$ 
with gradients $-\gamma - c_0\,(2 e^{1/2} - 1)$ 
and $-\gamma + c_1\,(2 e^{1/2} - 1)$, respectively. 
However, due to the finite MD step size, 
some configurations may penetrate deeply into the wall regions, 
which yields a strong repulsive contribution  
$-W'(t)\,[Dt(U)]^\dagger$ 
to the force term $-[DV(U)]^\dagger$, 
significantly degrading numerical stability. 
Moreover, with finite $\Delta s = \epsilon$, 
configurations may overshoot the boundary 
and enter regions that cannot be reached by flows from $\Sigma_0$. 
To avoid this issue, 
while preserving 
(a) exact reversibility, 
(b) exact volume preservation, 
and (c) approximate energy conservation of the same order of accuracy, 
we adopt the simple and effective solution 
introduced in Ref.~\cite{Fukuma:2020fez}: 
reversing the trajectory 
whenever a configuration over-penetrates the boundary. 
The procedure is summarized in Algorithm~\ref{alg:wv_boundary}.  
%
\begin{algorithm}[tb]
  \caption{MD step $(U,\pi)\to(U',\pi')$ in WV-HMC with boundary}
  \label{alg:wv_boundary}
  
  \begin{algorithmic}[1]
    \State%
    Given $(U,\pi)$ with $U=U(t,U_0)$, 
    compute a trial MD step 
    $(U,\pi) \to (\tilde U,\tilde\pi)$ 
    with $\tilde U=U(\tilde t,\tilde{U}_0)$
    by using Algorithm~\ref{alg:wv_rattle}
    \If{$T_0-d_0 < \tilde t < T_1+d_1$}%
    \State%
    Set $U' = \tilde U$ and $\pi' = \tilde\pi$
    \Else%
    \State%
    Set $U' = U$ and $\pi' = -\pi$
    \EndIf
  \end{algorithmic}
\end{algorithm}

\subsection{Summary of WV-HMC}
\label{sec:wv_summary}

We summarize the WV-HMC algorithm in Algorithm~\ref{alg:wv_hmc}, 
which uses the boundary-treated MD step 
of Algorithm~\ref{alg:wv_boundary}.
%
\begin{algorithm}[tb]
  \caption{WV-HMC}
  \label{alg:wv_hmc}
  
  \begin{algorithmic}[1]
    \State%
    Given $U\in\calR$, 
    generate $\tilde\pi\in T_U G^\bbC \simeq \LieGC$ 
    from the Gaussian distribution 
    $\propto e^{-(1/2) \vev{\tilde\pi,\tilde\pi}}$
    \State%
    Decompose $\tilde\pi$ as
    $\tilde\pi = \tilde\pi_\parallel + \tilde\pi_\perp$ 
    by using Algorithm~\ref{alg:wv_decomposition}, 
    and set $\pi = \tilde\pi_\parallel$
    \State%
    Apply a sequence of MD steps 
    using Algorithm~\ref{alg:wv_boundary}
    to evolve $(U,\pi) \to (U',\pi')$ 
    \State%
    Compute $\Delta H\equiv H(U',\pi') - H(U,\pi)$, 
    and accept $U'$ as the new configuration 
    with probability $\min(1,e^{-\Delta H})$. 
    Otherwise, retain $U$ as the next configuration
  \end{algorithmic}
\end{algorithm}

\section{Numerical tests: one-site model}
\label{sec:1site}

We validate the correctness of the proposed algorithm 
by performing numerical tests on the one-site model.

\subsection{The model}
\label{sec:1site_model}

The one-site model for a compact group $G=SU(n)$ 
is defined by the action 
\begin{align}
  S(U) 
  \equiv \beta e(U) ,
\end{align}
where $e(U)$ is the energy ``density''
\begin{align}
  e(U) 
  \equiv -\frac{1}{2n}\,\tr (U + U^{-1}).
\end{align}
We take $\beta$ to be purely imaginary, 
which makes the action complex-valued. 
Note that in this case, 
the Boltzmann weight becomes a pure phase factor of constant modulus, 
so that naive reweighting based on the Haar measure 
can suffer from severe phase cancellations. 
Along the anti-holomorphic gradient flow \eqref{flow_c} 
on $G^\bbC = SL(n,\bbC)$, 
the real part of the action becomes nonconstant, 
so that $e^{-\ReS(U)}$ provides a nonuniform positive weight 
on the deformed surfaces.

To make the algebraic structure more transparent,
we generalize the model to the form%
\footnote{ 
  A topological-term-like contribution can be introduced 
  by setting $a = \beta/(2n) + \theta/(4\pi)$ 
  and $b = \beta/(2n) - \theta/(4\pi)$, 
  which can render the action complex-valued 
  even when both $\beta$ and $\theta$ are real. 
  This contribution vanishes identically for $SU(2)$, 
  but is generally nonzero for $SU(n)$ with $n\geq 3$ 
  or when $G$ includes a $U(1)$ factor.
} 
\begin{align}
  S(U) =  -a\,\tr U - b\,\tr U^{-1},
\end{align}
where the original model corresponds to 
$a = b = \beta/(2n)$. 
We perform two validation tests 
to verify the correctness of our formalism. 
The first is to confirm 
that the energy difference $\Delta H$ scales as $\epsilon^3$ 
for a single MD step of step size $\Delta s = \epsilon$. 
This serves as a test of the algorithmic correctness of HMC. 
The second is to verify that analytical results are correctly reproduced. 
This checks the correctness of the constructed observables, 
including the reweighting factor $\calF(U)$.

In the following numerical tests, 
we consider the energy density $e(U)$ as the observable. 
The expectation value $\vev{e}$ for $G=SU(n)$ 
with respect to the normalized Haar measure $(dU)$ 
can be computed analytically from the partition function 
\begin{align}
  Z_G \equiv 
  \int_G (dU)\,e^{a\, \tr U + b\, \tr U^{-1}}
  = \sum_{q\in\bbZ}\, (b/a)^{n q /2}\,
  \det [ I_{q+j-i}(2\sqrt{ab}) ]_{i,j=1,\ldots,n}, 
\label{formula}
\end{align}
where $I_k(z)$ denotes the modified Bessel function of the first kind. 
For complex $a$ and $b$, the expression is understood 
by analytic continuation from positive real $a$ and $b$.
The expectation value is then given by 
$\vev{e} = -(\partial/\partial\beta)\,\ln Z_G$ 
after setting $a = b = \beta / (2n)$.

\subsection{Flow equations}
\label{sec:1site_flow_eqs}

The variation of the action is given by 
($\calP$ denotes the traceless projector) 
\begin{align}
  \delta S(U) 
  &= -\tr [\delta U U^{-1} (a\,U - b\, U^{-1})]
  = - \tr [\delta U U^{-1} \calP (a\,U - b\, U^{-1})].
\end{align}
Comparing this with $\delta S(U) = \tr [\delta U U^{-1} DS(U)]$, 
we obtain 
\begin{align}
  DS(U) = -\calP (a\,U - b\, U^{-1}),
\end{align}
and therefore 
\begin{align}
   \xi(U) = [DS(U)]^\dagger = -[\calP (a\,U - b\, U^{-1})]^\dagger.
\end{align}
This defines the flow of a configuration,
\begin{align}
   \dot{U} = \xi(U)\,U
   ~~\text{with}~~
   U|_{t=0} = U_0.
\label{flow_c_1site}
\end{align}
For a tangent vector $v \in T_U \Sigma$, 
we obtain%
\footnote{ 
  The symbol $|_{v^1}$ indicates the linear part with respect to $v$.
} 
\begin{align}
  H v 
  &= D_v DS(U) = DS(e^v U)|_{v^1}
  = -\calP (a\, e^v U - b\,U^{-1}\,e^{-v}) |_{v^1}
\nonumber
\\
  &= -\calP (a v U + b\,U^{-1} v),
\end{align}
which defines the flow of a tangent vector, 
\begin{align}
  \dot{v} = (H v)^\dagger + [\xi, v]
   ~~\text{with}~~
  v|_{t=0} = v_0.
\label{flow_t_1site}
\end{align}
For a normal vector $n \in N_U \Sigma$, 
we have%
\footnote{
  The traceless projector $\calP$ is omitted in the last expression, 
  because $n$ is traceless 
  [$n \in N_U \Sigma \subset T_U G^\bbC \simeq \LieGC = \mathfrak{sl}(n,\bbC)$].
} 
\begin{align}
  D D_n S(U) = D\,\tr [n DS(U)]
  = - D\,\tr [n (a\, U - b\, U^{-1}) ].
\end{align}
To evaluate the right-hand side,
we consider the variation 
\begin{align}
  \delta\, \tr [n (a\, U - b\, U^{-1})]
  &= \tr [\delta U U^{-1} (a\, U n + b\, n U^{-1}) ]
  = \tr [\delta U U^{-1} \calP (a\, U n + b\, n U^{-1}) ]. 
\end{align}
Comparing this with 
$\tr\bigl[ \delta U U^{-1}\, D\,\tr [n (a\, U - b\, U^{-1})] \bigr]$, 
we have 
$D\,\tr [n (a\, U - b\, U^{-1})]
= \calP (a\, U n + b\, n U^{-1})$,
and thus obtain 
\begin{align}
  H^T n = D D_n S(U) = -\calP( a\, U n + b\, n U^{-1}),
\label{HTn}
\end{align}
which defines the flow of a normal vector, 
\begin{align}
  \dot{n} = -(H^T n)^\dagger - [\xi^\dagger, n]
   ~~\text{with}~~
  n|_{t=0} = n_0.
\label{flow_n_1site}
\end{align}
The expression \eqref{HTn} can be obtained more easily 
by using the formula \eqref{HTv-Hv} as 
\begin{align}
  H^T n 
  &= H n + [DS, n] 
  = -\calP \bigl( a\, n U + b\, U^{-1} n)
  - [\calP (a\, U - b U^{-1}), n]
\nonumber
\\
  &= -\calP\,\bigl(
  a\, n U + b\, U^{-1} n + [ a\, U - b\, U^{-1}, n] \bigr)
\nonumber
\\
  &= -\calP (a U n + b\, n U^{-1}),
\end{align}
where we have used the identity 
$[\calP X, Y] = [X,Y] = \calP [X,Y]$ 
for matrices $X$ and $Y$. 
One can easily confirm  
that $\vev{n,v}^\centerdot = 0$ 
by using the explicit forms of the flow equations.
We integrate the flow equations 
\eqref{flow_c_1site}, \eqref{flow_t_1site} and \eqref{flow_n_1site}
using an adaptive version 
of the Runge-Kutta-Munthe-Kaas algorithm \cite{RKMK1,RKMK2}.

\subsection{$G=SU(2)$ with a purely imaginary coupling constant} 
\label{sec:1site_SU2}

In the $SU(2)$ case with $a = b = \beta/4$,  
the infinite series in Eq.~\eqref{formula} 
can be summed analytically, yielding 
$Z_{SU(2)}(\beta) = (2/\beta)\,I_1(\beta)$, 
from which we obtain $\vev{e} = -I_2(\beta)/I_1(\beta)$. 
One can show that $\re\vev{e} = 0$
for $\beta \in i\,\bbR$.
Indeed, the partition function $Z_{SU(2)}(\beta)$ 
is an even function of $\beta$ 
because the change of variables $U \to -U$ 
corresponds to $\beta \to -\beta$,  
and thus $\vev{e}$ is an odd function of $\beta$. 
Also, $Z_{SU(2)}(\beta)$ is real-valued when $\beta$ is real.
These imply that $\vev{e}$ must be purely imaginary when $\beta \in i\, \bbR$.

\begin{figure}[htb]
  \centering
  \includegraphics[width=75mm]{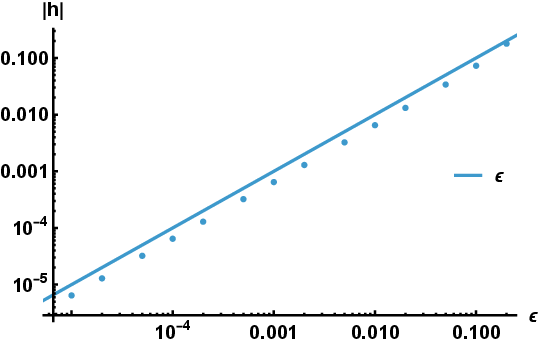}
  \hspace{5mm}
  \includegraphics[width=75mm]{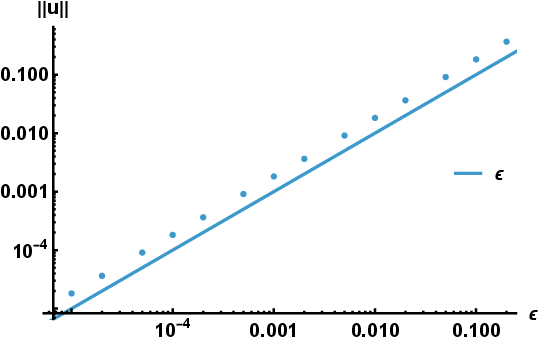}\\
  \vspace{10pt}
  \includegraphics[width=75mm]{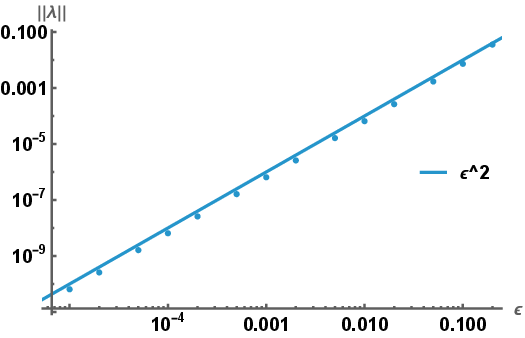}
  \caption{Scaling of $(h,u,\lambda)$  
     as a function of $\epsilon$ 
    in the $SU(2)$ one-site model with $\beta = i$. 
    The Frobenius norm  
    $\|A\| \equiv (\tr A^\dagger A )^{1/2}$ is used. 
  }
  \label{fig:scaling_SU(2)_1}
\end{figure}

We set the parameters in $W(t)$ [Eq.~\eqref{W(t)}] to 
$T_0 = 0$, $T_1 = 0.5$, $\gamma = 0$, 
$c_0 = c_1 = 10$, and $d_0 = d_1 = 0.1$. 
We expect that 
the solution $(h,u,\lambda)$ to Eq.~\eqref{wv_newton} 
scales with the molecular dynamics step size $\Delta s = \epsilon$ 
as 
\begin{align}
  h = O(\epsilon), \qquad 
  u = O(\epsilon), \qquad 
  \lambda = O(\epsilon^2). 
\end{align}
Figure~\ref{fig:scaling_SU(2)_1} shows 
$|h|$, $\|u\|$ and $\|\lambda\|$ 
for various values of $\epsilon$ 
at $\beta = i$ and flow time $t = 0.5$, 
with $U_0$ and $\pi$ chosen randomly and kept fixed. 
The results confirm the expected scaling behavior.
This scaling can be exploited 
to construct a good initial guess 
that enables the simplified Newton method 
for iteratively solving Eq.~\eqref{wv_newton} 
to converge to the correct solution. 
A simple algorithm proceeds as follows:  

\noindent
\underline{Step 1}: 
Scale down $\epsilon$ to $\tilde\epsilon = f \epsilon$ 
with a small scaling factor $f\,(\ll 1)$, 
so that $\tilde Z \equiv Z|_{\epsilon\to\tilde\epsilon}$ 
in Eq.~\eqref{wv_rattle1} becomes small. 

\noindent
\underline{Step 2}: 
Solve Eq.~\eqref{wv_newton} with $Z$ replaced by $\tilde Z$, 
and obtain the solution $(\tilde h,\, \tilde u,\,\tilde\lambda)$. 

\noindent
\underline{Step 3}: 
Set the initial guess for the original equation as  
$h_\textrm{init} \equiv \tilde h / f$, $u_\textrm{init} \equiv \tilde u / f$, 
$\lambda_\textrm{init} \equiv \tilde\lambda / f^2$. 

\noindent
The same trick also applies to GT-HMC.

\begin{figure}[htb]
    \centering
    \includegraphics[width=75mm]{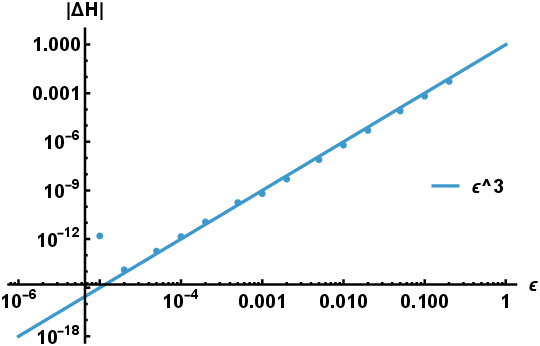}
    \caption{
        Energy difference $\Delta H$ of a single MD step 
        versus step size $\epsilon$ 
        for the $SU(2)$ one-site model with $\beta = i$. 
        Absolute and relative tolerances in recursive procedures 
        are set to $10^{-10}$.
    }
    \label{fig:scaling_SU(2)_2}
\end{figure}

Figure~\ref{fig:scaling_SU(2)_2} shows that 
the energy difference $\Delta H$ 
during a single MD step scales as $\epsilon^3$, 
where the parameters are set as above, 
and molecular dynamics starts at flow time $t=0.5$.
This result strongly supports the algorithmic correctness 
of the proposed method. 

\begin{figure}[htb]
  \centering
  \includegraphics[width=75mm]
  {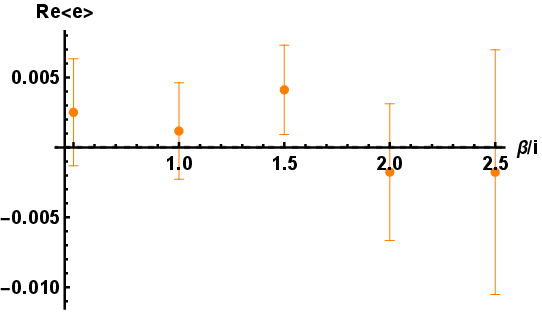}
  \hspace{5mm}
  \includegraphics[width=75mm]
  {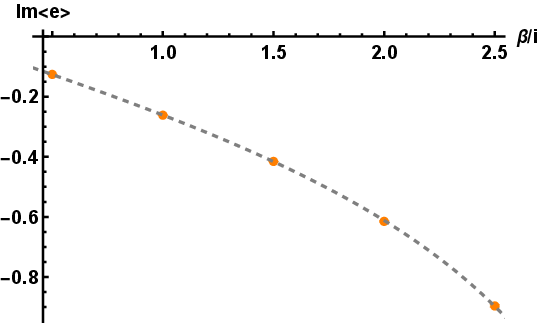}
  \caption{
    Real and imaginary parts of $\vev{e}$ 
    in the $SU(2)$ one-site model for $\beta \in i\,\bbR$. 
    The dashed lines represent the analytical results.
  }
  \label{fig:energy_density_SU(2)}
\end{figure}

To estimate the observable, 
we set the molecular dynamics step size 
to $\Delta s = \epsilon = 0.01$ 
with $N_\textrm{MD} = 50$ steps per trajectory. 
For each of ten independent runs with different random seeds, 
we generate 50,000 configurations using WV-HMC, 
discarding the first 10,000 configurations for thermalization. 
We then combine the measurements from the ten runs 
to estimate $\vev{e}$. 
We find that the estimates remain stable 
under changes of the analysis window 
to $[0,\,0.4]$ and $[0.1,\,0.5]$, 
and therefore adopt 
$[\tilde T_0,\tilde T_1]=[T_0,T_1]=[0,\,0.5]$ 
to maximize the sample size. 
The 40,000 configurations per run remaining after thermalization 
are distributed over the flow-time range $-0.1 < t < 0.6$, 
of which approximately 70\% fall within the chosen analysis window. 
Statistical errors are estimated 
using a jackknife analysis over the ten independent runs.%
\footnote{ 
  See Ref.~\cite{Fukuma:2021aoo} 
  for an argument to justify 
  the use of standard statistical analysis methods 
  for the sample taken from a subregion $[\tilde{T}_0,\tilde{T}_1]$. 
} 
Figure~\ref{fig:energy_density_SU(2)} shows 
the real and imaginary parts of the energy density $\vev{e}$ 
for various values of $\beta \in i\,\bbR$. 
The results are in good agreement with the analytical values, 
supporting the validity of the constructed observables, 
including the reweighting factor $\calF(U)$. 
%

\subsection{$G=SU(3)$ with a purely imaginary coupling constant} 
\label{sec:1site_SU3}

\begin{figure}[htb]
  \centering
  \includegraphics[width=75mm]
  {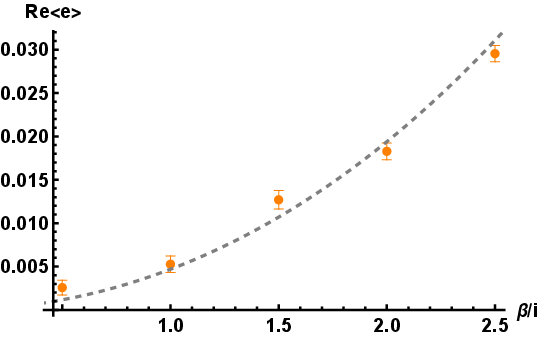}
  \hspace{5mm}
  \includegraphics[width=75mm]
  {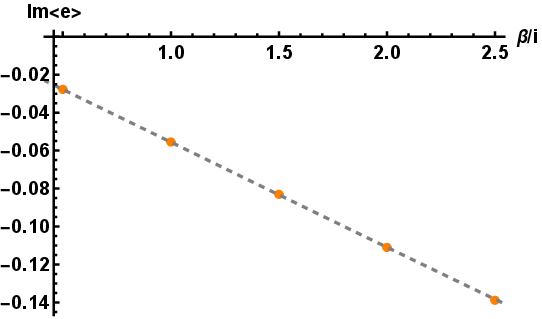}
  \caption{
    Real and imaginary parts of $\vev{e}$ 
    in the $SU(3)$ one-site model for $\beta \in i\,\bbR$. 
    The dashed lines represent the analytical results.
  }
  \label{fig:energy_density_SU(3)}
\end{figure}

Other compact groups can be treated in the same manner. 
Figure~\ref{fig:energy_density_SU(3)} shows the energy density $\vev{e}$ 
for the $SU(3)$ one-site model 
with various values of $\beta \in i\,\bbR$. 
The simulation parameters are set to be the same as those used in the $SU(2)$ case. 
The numerical results are again in good agreement with the analytical results. 
This provides further strong evidence for the validity of the present formalism.
%

\FloatBarrier 
\section{Conclusions and outlook}
\label{sec:conclusion}

After developing a general theory of unconstrained and constrained HMC 
on the complexified group $G^\bbC$, 
we have demonstrated that the WV-HMC algorithm 
(as well as the GT-HMC algorithm) 
can be extended to group manifolds. 
The key ingredient is to introduce a symplectic structure 
on the tangent bundle of the worldvolume $\calR$ 
(or the fixed flowed surface $\Sigma$) 
and to rewrite the path integrals 
as phase-space integrals over the tangent bundle. 
We have validated the correctness of the proposed algorithm 
through numerical simulations of the one-site model. 

A potential issue with the present formalism 
lies in the presence of the factor $\det E/\sqrt{\gamma}$ 
in the reweighting factor $\calF(U)$ [see Eq.~\eqref{F_calR}]. 
This factor is not generally a pure phase, 
and thus may lead to an overlap problem. 
Although no such difficulty was observed in our study of the one-site model,
this potential complication should be kept in mind 
when applying the method to larger systems. 

The present formalism can be directly applied to lattice gauge theories 
without any modification to the algorithmic structure. 
The compact group $G$ becomes a product group 
$G = \prod_{x,\mu} G_{x,\mu}$ 
[e.g., $G_{x,\mu} = SU(n)$ at each link $(x,\mu)$], 
and the corresponding Lie algebra is given by 
$\LieG = \bigoplus_{x,\mu} \LieG_{x,\mu} 
= \bigoplus_{x,\mu,a} \bbR\,(T_{x,\mu})_a$ 
with the commutation relations 
$[(T_{x,\mu})_a, (T_{y,\nu})_b] = \delta_{x y}\,\delta_{\mu \nu}\,
C_{ab}{}^c\,(T_{x,\mu})_c$. 
A study of lattice gauge theories with complex actions 
will be reported in forthcoming publications.


\section*{Acknowledgments}
The author thanks Sinya Aoki, Ken-Ichi Ishikawa, 
Yoshio Kikukawa and Yusuke Namekawa 
for valuable discussions.
This work was partially supported 
by JSPS KAKENHI 
Grant Numbers JP20H01900, JP23H00112, JP23H04506, JP25H01533, JP26K21723; 
by MEXT as ``Program for Promoting Researches on the Supercomputer Fugaku'' 
(Simulation for basic science: approaching the new quantum era, JPMXP1020230411); 
and by SPIRIT2 2025 of Kyoto University.


\appendix

\section{Hamiltonian dynamics on a symplectic manifold}
\label{sec:hamiltonian_dynamics}

In this appendix, 
we review Hamiltonian dynamics on a symplectic manifold, 
focusing on the material 
relevant to the main text.

\subsection{Symplectic manifolds, symplectic forms, and symplectic potentials}
\label{sec:hmc_general}

We begin by considering Hamiltonian dynamics 
on a general symplectic manifold $(\calM,\omega)$. 
Here, $\calM=\{x = (x^i)\}$ $(i=1,\ldots,2N)$ is a real $2N$-dimensional manifold, 
and $\omega=(1/2)\,\omega_{ij}(x)\,dx^{i}\wedge dx^{j}$ 
is a \emph{symplectic 2-form} on $\calM$ 
that satisfies the following conditions:%
\footnote{ 
  We omit the symbol $\wedge$ (wedge) 
  when no confusion is expected. 
} 
\begin{align}
  \text{(i) }& \text{Nondegeneracy: The matrix $(\omega_{ij})$ is invertible, 
    with inverse denoted by $(\omega^{ij})$}. 
\label{nondegeneracy}
\\
  \text{(ii) }& \text{Closedness: 
  The exterior derivative of $\omega$ vanishes, i.e., $d\omega = 0$.} 
\label{closedness}
\end{align}
Condition (i) allows us to define 
the Poisson bracket of smooth functions $f(x)$ and $g(x)$ 
on $\calM$ by%
\footnote{ 
  In particular, we have $\{x^i, x^j\} = \omega^{ij}(x)$.
} 
\begin{align}
  \{f,g\}(x) \equiv \omega^{ij}(x)\,\partial_i f(x)\, \partial_j g(x) 
  = -\{g,f\}(x).
\end{align}
Condition (ii) ensures that 
the Poisson bracket satisfies the Jacobi identity, 
$\{\{f,g\},h\} = \{\{f,h\},g\} + \{f,\{g,h\}\}$. 
Moreover, the closedness of $\omega$ implies 
the (local) existence of a 1-form $a=a_i(x)\, dx^i$, 
called the \emph{symplectic potential}, 
such that  
\begin{align}
  \omega = da, 
\end{align}
which in components reads as $\omega_{ij} = \partial_i a_j - \partial_j a_i$.

\subsection{Finite symplectic transformations and pullback}
\label{sec:pullback}

Given a function $f(x)$, 
we define the associated vector field $v_f=v_f^i(x)\,\partial_i$
and the transformation $T_f$ as 
\begin{align}
  v_f^i(x) &\equiv \omega^{ij}(x)\,\partial_j f(x),
\\
  T_f &\equiv e^{\{\ast,\,f\}}.
\end{align}
We define the action of $T_f$ on an $n$-form 
$A=(1/n!)\,A_{i_1\ldots i_n}(x)\,dx^{i_1}\wedge\cdots\wedge dx^{i_n}$
as 
\begin{align}
  T_f A \equiv \frac{1}{n!}\,(T_f A_{i_1\ldots i_n})(x)\,
  d(T_f x^{i_1})\wedge \cdots \wedge d(T_f x^{i_n}).
\label{T_fA}
\end{align}
The Poisson bracket $\{\ast,f\}$ acts on a function $G(x)$ as
\begin{align}
  \{G,f\} = \omega^{ij}\,\partial_i G\,\partial_j f
  = v_f^i\,\partial_i G = v_f G,
\end{align}
which implies that
\begin{align}
  (T_f G)(x) = (e^{v_f} G)(x).
\end{align}
Thus, the action of $T_f$ on differential forms [Eq.~\eqref{T_fA}] 
can be identified 
with the pullback $(e^{v_f})^\ast$ of the diffeomorphism $e^{v_f}$,
\begin{align}
  T_f A
  = (e^{v_f})^\ast A 
  = e^{\calL_{v_f}} \! A.
\end{align}
Here, $\calL_v$ denotes the Lie derivative 
along a vector field $v=v^i(x)\,\partial_i$, 
which is given 
in terms of the exterior derivative $d=dx^i\wedge \partial_i$ 
and the interior product $i_v$ 
by the Cartan formula%
\footnote{ 
  The interior product acts on an $n$-form 
  as 
  $i_v A = (1/(n-1)!)\,(i_v A)_{i_1\ldots i_{n-1}}(x)\,
  dx^{i_1}\wedge\cdots\wedge dx^{i_{n-1}}$ 
  with $(i_v A)_{i_1\ldots i_{n-1}} 
  = v^i\,A_{i\,i_1\ldots i_{n-1}}$.
} 
\begin{align}
  \calL_v = i_v d + d\,i_v.
\label{identity_calL}
\end{align}

The transformation 
$T_f = e^{\{\ast,f\}} = e^{\calL_{v_f}}$ is a symplectic transformation, 
meaning that 
\begin{align}
  T_f\, \omega = \omega.
\end{align}
To see this, 
note first that 
\begin{align}
  \calL_{v_f} a
  = (i_{v_f} d + d\, i_{v_f})\,a
  = i_{v_f}\, \omega + d(i_{v_f} a).
\end{align}
Since $i_{v_f}\,\omega$ can be rewritten as 
\begin{align}
  i_{v_f}\,\omega = v_f^j\,\omega_{ji}\,dx^i
  = \omega^{jk}\,\partial_k f\,\omega_{ji}\,dx^i
  = -\partial_i f\, dx^i
  = - df,
\end{align}
we have 
\begin{align}
  \calL_{v_f} a = d (i_{v_f} a - f). 
\end{align}
This implies that%
\footnote{
  We have used the identity $\calL_v(dA) = d(\calL_v A)$, 
  which follows from Eq.~\eqref{identity_calL} 
  and the nilpotency $d^2=0$. 
  This identity also implies $T_f (dA) = d\, (T_f A)$ 
  because $T_f = e^{\calL_{v_f}}$.
} 
\begin{align}
  \calL_{v_f} \omega = d (\calL_{v_f} a) = 0,
\label{calL_{v_f}}
\end{align}
thus confirming the invariance of $\omega$ 
under $T_f = e^{\calL_{v_f}}$.

\subsection{Hamiltonian dynamics}
\label{sec:hd_flat}

We consider Hamiltonian dynamics on $\calM$ 
with a Hamiltonian $H(x)$. 
A trajectory in $\calM$ 
is described by a smooth function $x(s)$ 
of a continuous parameter $s$.
We characterize such trajectories 
as those that extremize the first-order action 
\begin{align}
  I[x(s)] \equiv \int ds\, [a_i(x)\,\circdot{x}{}^i - H(x) ],
\end{align} 
where $\circdot{x}{}^i(s) \equiv (d/ds)\,x^i(s)$. 
The resulting equations of motion (Hamilton's equations) are 
given by 
$\omega_{ij}(x)\, \circdot{x}{}^j = \partial_i H(x)$, 
or equivalently, 
\begin{align}
  \circdot{x}{}^i = \omega^{ij}(x)\,\partial_j H(x)
  = \{x^i, H(x)\}.
\end{align}
One can show that 
both the symplectic form and the Hamiltonian are conserved 
along the trajectory, 
\begin{align}
  \circdot{\omega} &= 0,
\label{symplectic_continuum}
\\
  \circdot{H} &= 0.
\label{H-conservation_continuum}
\end{align}
Symplecticity [Eq.~\eqref{symplectic_continuum}] 
follows from the fact 
that the Hamiltonian flow is a symplectic transformation 
[replacing $f(x)$ in Eq.~\eqref{calL_{v_f}} with $H(x)$], 
$\circdot{\omega}=\calL_{v_H}\omega=0$. 
The energy conservation [Eq.~\eqref{H-conservation_continuum}] 
is shown as follows:
\begin{align}
  \circdot{H}
  = \partial_i H\,\circdot{x}{}^i
  = \omega^{ij}\,\partial_i H\,\partial_j H
  = 0.
\end{align}
The finite-time evolution of $x$ 
over an interval $\Delta s = \epsilon$ 
is given by 
\begin{align}
  T_{\epsilon H} x = e^{\epsilon\,\{\ast, H\} }\,x.
\end{align}

\subsection{Molecular dynamics (MD)}
\label{sec:md_flat}

We focus here on the case 
where $\calM$ is equipped with canonical coordinates 
$x=(x^i) = (q^a,p_a)$ $(a=1,\ldots,N)$
for which the symplectic potential and the symplectic form   
are given by $a=p_a\,dq^a$ and $\omega = da = dp_a \wedge dq^a$, 
respectively.%
\footnote{ 
  The Poisson brackets are then given by
  $\{q^a,\,p_b\} = \delta^a_b$, $\{q^a,q^b\} = \{p_a,p_b\} = 0$.
} 
We assume that the Hamiltonian takes a separable form, 
\begin{align}
  H(q,p) = K(p) + V(q)
  \text{~with~}
  K(p) \equiv \frac{1}{2}\,p_a^2.
\end{align}

We discretize the continuum evolution operator 
$T_H = e^{\epsilon\,\{\ast,H\}}$ with step size $\Delta s = \epsilon$ 
in the symmetric form,%
\footnote{
  For simplicity, 
  we denote $e^{\epsilon\, \{\ast,f\}}$ 
  by $T_f$ (not $T_{\epsilon f}$).
} 
\begin{align}
  T \equiv T_{V/2}\,T_K\,T_{V/2},
\label{T_def2}
\end{align}
whose deviation from $T_H$ is $O(\epsilon^3)$,%
\footnote{
  This follows from the identity 
  $e^{\epsilon A/2}\,e^{\epsilon B}\,e^{\epsilon A/2}
  = e^{\epsilon (A+B)} + O(\epsilon^3)$, 
  which holds for linear operators $A$, $B$ of any type.
} 
\begin{align}
  T = T_H + O(\epsilon^3).
\label{TsimTH}
\end{align} 
The following identities are straightforward to verify:
\begin{align}
  T_{V/2}\, q &= q,
\\
  T_{V/2}\, p &= p - \frac{\epsilon}{2}\,\partial V(q)
  ~(\,\equiv p_{1/2}),
\\
  T_K\, q &= q + \epsilon\, p,
\\
  T_K\, p &= p,
\end{align}
from which the update rule $T:\,(q,p) \to (q',p')$ is given by 
\begin{align}
  q' &\equiv T q
  = T_{V/2}\,T_K\,T_{V/2}\, q
\nonumber
\\
  &= T_{V/2}\,T_K\, q
\nonumber
\\
  &= T_{V/2}\, (q+\epsilon\,p)
\nonumber
\\
  &= q + \epsilon\, p_{1/2},
\\
  p' &\equiv T p 
  = T_{V/2}\,T_K\,T_{V/2}\, p
\nonumber
\\
  &= T_{V/2}\,T_K\, \Bigl[p - \frac{\epsilon}{2}\,\partial V(q) \Bigr]
\nonumber
\\
  &= T_{V/2}\,\Bigl[p - \frac{\epsilon}{2}\,\partial V(q+\epsilon\,p) \Bigr]
\nonumber
\\
&= p_{1/2} - \frac{\epsilon}{2}\,\partial V(q+\epsilon\,p_{1/2})
\nonumber
\\
  &= p_{1/2} - \frac{\epsilon}{2}\,\partial V(q').
\end{align}
In practice, 
this update is summarized into the process
\begin{align}
  p_{1/2} &= p - \frac{\epsilon}{2}\,\partial V(q),
\label{lf1}
\\
  q' &= q + \epsilon\,p_{1/2},
\label{lf2}
\\
  p' &= p_{1/2} - \frac{\epsilon}{2}\,\partial V(q'),
\label{lf3}
\end{align}
which forms a unit of molecular dynamics (MD). 
When the above is repeated $N_\textrm{MD}$ times, 
the intermediate steps can be combined 
and the full MD trajectory is generated as 
\begin{align}
  &p_{1/2} = p - \frac{\epsilon}{2}\,\partial V(q)
\nonumber
\\
  &q' = q + \epsilon\,p_{1/2}
\nonumber
\\
  &\text{\texttt{for} $i=1,\ldots,N_\textrm{MD}-1$ \texttt{do}}
\nonumber
\\
  &~~~~~ p_{1/2} \leftarrow p_{1/2} - \epsilon\,\partial V(q')
\nonumber
\\
  &~~~~~ q' \leftarrow q' + \epsilon\,p_{1/2}
\nonumber
\\
  &\text{\texttt{end do}}
\nonumber
\\
  &p' = p_{1/2} - \frac{\epsilon}{2}\,\partial V(q').
\label{lf}
\end{align}

The process \eqref{lf1}--\eqref{lf3} can be shown to satisfy 
the following three key properties: 

\noindent
\underline{(a) Exact reversibility}:
The transformation is invariant under time reversal,
\begin{align}
  q &\to \tilde{q} \equiv q',
\label{lf_rev1}
\\
  p &\to \tilde{p} \equiv -p',
\label{lf_rev2}
\\
  p_{1/2} &\to \tilde{p}_{1/2} \equiv - p_{1/2},
\label{lf_rev3}
\\
  q'&\to \tilde{q}' \equiv q,
\label{lf_rev4}
\\
p'&\to \tilde{p}' \equiv -p.
\label{lf_rev5}
\end{align}

\noindent
\underline{(b) Symplecticity}: 
The symplectic form $\omega$ 
and thus the symplectic volume form 
$d\Omega \equiv \omega^N/N! = \pm dq\,dp
\equiv \pm \prod_a dq^a\,\prod_a dp_a$ 
are preserved,
\begin{align}
  \omega' = \omega,
  \qquad
  d\Omega' = d\Omega.
\end{align}

\noindent
\underline{(c) Approximate energy conservation}:
\begin{align}
  H(q',p') = H(q,p) + O(\epsilon^3).
\end{align}

Property (a) is obvious. 
Property (b) follows directly from the fact 
that the transformation $T$ is 
a composition of symplectic transformations [Eq.~\eqref{T_def2}]. 
It can also be shown explicitly 
by noting that 
\begin{align}
  T_{V/2}\,a &= T_{V/2}\,(p\,dq) 
  = a - \frac{\epsilon}{2}\,dV(q),
  \\  
  T_K\,a &= T_K\,(p\,dq)
  = a + \frac{\epsilon}{2}\,d(p^2),
\end{align}
from which it follows that 
\begin{align}
  T\,a
  = a + \frac{\epsilon}{2}\,d\,\bigl[
  p_{1/2}^2 - V(q) - V(q') \bigr],
\end{align}
and thus 
\begin{align}
  T \omega = T da = d\, (Ta) = da = \omega.
\end{align}
Property (c) follows from 
the identity $T = T_H + O(\epsilon^3)$ [Eq.~\eqref{TsimTH}] 
and the fact that the exact flow conserves energy, 
$T_H H = e^{\epsilon\{\ast,H\}}\,H = H$, 
\begin{align}
  H(q',p') &= (T H)(q,p) = (T_H H)(q,p) + O(\epsilon^3)
\nonumber
\\
  &= H(q,p) + O(\epsilon^3).
\end{align}

\subsection{Hybrid Monte Carlo (HMC)}
\label{sec:hmc_flat}

To simplify expressions, 
we introduce the notation $\Omega^T \equiv (q,-p)$ 
for $\Omega = (q,p)$. 
We now consider the numerical evaluation of a path integral of the form
\begin{align}
  \vev{\calO(q)}
  = \frac{\int_{\bbR^N} dq\,e^{-V(q)}\,\calO(q)}
  {\int_{\bbR^N} dq\,e^{-V(q)}},
\end{align}
which can be rewritten as an integral 
over the phase space $T\bbR^N \equiv \{(q,p)\}$,%
\footnote{ 
  In flat Euclidean space, a covector $p_a$ 
  can naturally be identified with a vector $p^a$ 
  using the standard metric. 
  We regard the tangent bundle (rather than the cotangent bundle) 
  as the phase space, 
  which is generally more convenient when discussing molecular dynamics.
} 
\begin{align}
  \vev{\calO(q)}
  = \frac{\int_{T\bbR^N} d\Omega\,e^{-H(q,p)}\,\calO(q)}
  {\int_{T\bbR^N} d\Omega\,e^{-H(q,p)}}
  = \frac{\int_{T\bbR^N} d\Omega\,e^{-H(\Omega)}\,\calO(q)}
  {\int_{T\bbR^N} d\Omega\,e^{-H(\Omega)}}.
\end{align}
Here, $d\Omega = \omega^N/N! = \pm dq\,dp$ 
is the symplectic volume form, 
and $H(\Omega) = H(q,p) = (1/2)\,p_a^2 + V(q)$ is the Hamiltonian. 
This expectation value can be estimated by a sample average 
over configurations $\{\Omega = (q,p)\}$ 
drawn from the distribution $\propto e^{-H(\Omega)}$. 

To construct a Markov chain 
with equilibrium distribution proportional to $e^{-H(\Omega)}$ 
with $H$ satisfying $H(\Omega^T) = H(\Omega)$ 
(as in the present case), 
we introduce a set of stochastic processes 
$P_{(k)}(\Omega'\,|\,\Omega)$ $(k=1,\ldots,K)$
such that 
each satisfies the detailed balance condition 
of molecular dynamics, 
\begin{align}
  P_{(k)}(\Omega'\,|\,\Omega)\,e^{-H(\Omega)}
  = P_{(k)}(\Omega^T\,|\,\Omega^{\prime\,T})\,e^{-H(\Omega^{\prime\,T})}.
\label{detailed_balance_MD2}
\end{align}
We also assume that the product $P \equiv P_{(K)}\cdots P_{(1)}$ 
has a proper ergodicity property 
that ensures unique convergence. 
Under these conditions, 
it can be shown that $e^{-H(\Omega)}$ 
is the unique eigenvector of $P$ with unit eigenvalue,%
\footnote{ 
  When $K = 2$, for example, 
  this can be shown as follows: 
  \begin{align}
    &\int_{T\bbR^N} d\Omega'\,
    (P_{(2)} P_{(1)})(\Omega|\Omega')\,e^{-H(\Omega')}
    =
    \int_{T\bbR^N} d\Omega' d\Omega''\,
    P_{(2)}(\Omega|\Omega'')\,P_{(1)}(\Omega''|\Omega')\,e^{-H(\Omega')}
  \nonumber
  \\
    &=
    \int_{T\bbR^N} d\Omega' d\Omega''\,
    P_{(2)}(\Omega|\Omega'')\,
    P_{(1)}(\Omega^{\prime\,T}|\Omega^{\prime\prime\,T})\,
    e^{-H(\Omega^{\prime\prime\,T})}
    =
    \int_{T\bbR^N} d\Omega''\,
    P_{(2)}(\Omega|\Omega'')\,
    e^{-H(\Omega^{\prime\prime})}
  \nonumber
  \\
    &=
    \int_{T\bbR^N} d\Omega''\,
    P_{(2)}(\Omega^{\prime\prime\,T}|\Omega^T)\,
    e^{-H(\Omega^T)}
    =
    e^{-H(\Omega)}.
  \nonumber
  \end{align}
} 
\begin{align}
  \int_{T\bbR^N} d\Omega'\,
  P(\Omega\,|\,\Omega')\,e^{-H(\Omega')} = e^{-H(\Omega)},
\end{align}
and thus $e^{-H(\Omega)} / \int_{T\bbR^N} d\Omega\,e^{-H(\Omega)}$ 
is the equilibrium distribution reached by the Markov chain 
defined by $P$. 

In the present case, 
we consider the following two stochastic processes $P_{(1)}$ and $P_{(2)}$. 
Although neither $P_{(1)}$ nor $P_{(2)}$ is ergodic, 
their product $P \equiv P_{(2)}\,P_{(1)}$ is typically ergodic, 
guaranteeing convergence to the correct equilibrium.

\noindent
\underline{(1) Heat bath for $p$}:
\begin{align}
  P_{(1)}(\Omega'\,|\,\Omega) 
  = \frac{1}{(2\pi)^{N/2}}\,e^{-(1/2)\,p_a^{\prime\,2}}\,
  \delta(q' - q).
\end{align}
This process satisfies the detailed balance condition \eqref{detailed_balance_MD2}, 
because 
\begin{align}
  P_{(1)}(\Omega'\,|\,\Omega)\,e^{-H(\Omega)}
  &= \frac{1}{(2\pi)^{N/2}}\,e^{-(1/2)\,(p_a^2 + p_a^{\prime\,2})}\,
  \delta(q' - q)\,e^{ -V(q) }
\nonumber
\\
  &= \frac{1}{(2\pi)^{N/2}}\,e^{-(1/2)\,((-p_a)^2 + (-p'_a)^2)}\,
  \delta(q - q')\,e^{ -V(q') }
\nonumber
\\
  &= P_{(1)}(\Omega^T\,|\,\Omega^{\prime\,T})\,e^{-H(\Omega^{\prime\,T})}.
\end{align}

\noindent
\underline{(2) MD followed by Metropolis test}:

\noindent
We introduce a reversible and volume-preserving map 
$f: \Omega \mapsto f(\Omega)$, 
constructed by repeating the MD update  
$N_\textrm{MD}$ times: 
$f = T^{N_\textrm{MD}}$ 
[cf.\ Eq.~\eqref{lf}]. 
We update $\Omega$ to $\Omega' = f(\Omega)$
with probability 
$A(\Omega) \equiv \min(1,e^{-H(f(\Omega)) + H(\Omega)})$. 
Otherwise, we update $\Omega$ to $\Omega' = \Omega^T$.%
\footnote{ 
  See, e.g., Ref.~\cite{Horowitz:1991rr}.
} 
The transition probability can be compactly summarized as 
\begin{align}
  P_{(2)}(\Omega'\,|\,\Omega) 
  = A(\Omega)\,\delta_{T\bbR^N}(\Omega',f(\Omega))
    + [1 - A(\Omega)]\,\delta_{T\bbR^N}(\Omega',\Omega^T),
\end{align}
where $\delta_{T\bbR^N}(q,p) = \delta(q)\,\delta(p)$ 
is the symplectic delta function 
with respect to the symplectic volume form $d\Omega = \pm dq\,dp$. 
Since the momentum is fully refreshed before each MD trajectory 
and our observables depend only on $q$, 
the momentum flip upon rejection 
need not be performed in practice.

The normalization condition holds obviously, 
\begin{align}
  \int_{T\bbR^N} d\Omega'\,P_{(2)}(\Omega'|\Omega) = 1.
\end{align}
One can also show the detailed balance condition 
\begin{align}
  P_{(2)}(\Omega'|\Omega)\,e^{-H(\Omega)}
  = P_{(2)}(\Omega^T|\Omega^{\prime\,T})\,e^{-H(\Omega^{\prime\,T})} 
\end{align}
as follows. 
We first consider the case 
when the update $\Omega' = f(\Omega)$ is accepted. 
The reversibility states that 
the relation $\Omega' = f(\Omega)$ is equivalent to 
$\Omega^T = f(\Omega^{\prime\,T})$. 
The volume preservation ensures that 
the Jacobian of the map $\Omega \mapsto f(\Omega)$ is unity. 
Thus, the reversibility together with the unit Jacobian 
(and also with the equality 
$\delta_{T\bbR^N}(\Omega_1,\Omega_2) = \delta_{T\bbR^N}(\Omega_1^T,\Omega_2^T)$) 
implies 
\begin{align}
  \delta_{T\bbR^N} \bigl( \Omega',\, f(\Omega) \bigr)
  = \delta_{T\bbR^N} \bigl(
      \Omega,\, f^{-1}(\Omega') \bigr)
  = \delta_{T\bbR^N} \bigl(
      \Omega^T,\, f(\Omega^{\prime\,T}) \bigr),
\end{align}
from which it follows that 
$P_{(2)}$ satisfies the detailed balance condition 
\eqref{detailed_balance_MD2} as 
\begin{align}
  P_{(2)}(\Omega' \,|\, \Omega) \, e^{-H(\Omega)}
  &= A(\Omega) \,\delta_{T\bbR^N}( \Omega',f(\Omega) )\,
  e^{-H(\Omega)}
\nonumber
\\
  &= \min( e^{-H(\Omega)}, e^{-H(\Omega')} )\,
  \delta_{T\bbR^N}( \Omega', f(\Omega) )
\nonumber
\\
  &= \min( e^{-H(\Omega^{\prime\,T})}, e^{-H(\Omega^T)} )\,
  \delta_{T\bbR^N}( \Omega^T, f(\Omega^{\prime\,T}) )
\nonumber
\\
  &= P_{(2)}(\Omega^T\,|\,\Omega^{\prime\,T}) \, e^{-H(\Omega^{\prime\,T})}.
\end{align}
When rejected, 
on the other hand, 
since the relation $\Omega' = \Omega^T$ 
is equivalent to $\Omega = \Omega^{\prime\,T}$, 
the left-hand side of the detailed balance condition 
can be rewritten as follows: 
\begin{align}
  P_{(2)}(\Omega' \,|\, \Omega) \, e^{-H(\Omega)}
  &= [1 - A(\Omega)] \,\delta_{T\bbR^N}(\Omega',\Omega^T)\,e^{-H(\Omega)}
\nonumber
\\
  &= [1 - A(\Omega^{\prime\,T})]\,
     \delta_{T\bbR^N}((\Omega^T)^T,\Omega^{\prime\,T})\,
     e^{-H(\Omega^{\prime\,T})}
\nonumber
\\
  &= P_{(2)}(\Omega^T \,|\, \Omega^{\prime\,T}) \, e^{-H(\Omega^{\prime\,T})}.
\end{align}

When the observables depend only on $q$ (as in the present case),
the combined algorithm $P = P_{(2)} P_{(1)}$
can be viewed as a stochastic process on $q$ alone \cite{Duane:1987de}: 
\begin{itemize}
  \item
  \underline{Step 1 (momentum refresh)}:\\
  Given $q \in \bbR^N$, 
  generate $p = (p_a)$ from the Gaussian distribution 
  $\propto e^{-(1/2)\,p_a^2}$.
  
  \item
  \underline{Step 2 (molecular dynamics)}:\\
  Evolve $(q,p) \to (q',p')$ 
  using the MD algorithm [Eq.~\eqref{lf}].
  
  \item
  \underline{Step 3 (Metropolis test)}:\\
  Accept the proposed $q'$ 
  with probability $\min \bigl(1, e^{-[H(q',p') - H(q,p)]} \bigr)$. 
  
\end{itemize}
This process is represented by the marginal transition kernel 
\begin{align}
  P(q'|q) \equiv \frac{1}{(2\pi)^{N/2}}
  \int dp' dp\,P(q',p' | q,p)\,e^{-(1/2)\,p_a^2}. 
\end{align}
This satisfies the detailed balance condition 
with respect to $e^{-V(q)}$ \cite{Duane:1987de},
\begin{align}
  P(q'|q)\,e^{-V(q)} = P(q|q')\,e^{-V(q')},
\end{align}
which implies that 
the process on $q$ reaches the equilibrium distribution 
$e^{-V(q)} / \int dq\,e^{-V(q)}$.

\section{HMC algorithm for a compact group}
\label{sec:G}

In this appendix, 
we review the Hybrid Monte Carlo (HMC) algorithm 
on a connected compact Lie group $G$ of dimension $N$
\cite{Duane:1987de,Kennedy:1989ae}. 
Although the material presented here is likely familiar to experts, 
we present it in a form that parallels the structure of the main text. 
In this appendix only, 
elements of $G$ are denoted by the symbol $U$ 
(rather than $U_0$, as used in the main text). 
We assume that each group element 
is uniquely represented by a unitary matrix. 
Thus, elements of the Lie algebra $\LieG$ are anti-hermitian matrices. 
We choose a basis $\{T_a\}$ $(a=1,\ldots,N)$ of $\LieG$ 
such that $\tr T_a T_b = -\delta_{ab}$. 
Accordingly, we raise and lower indices using the rule $A^a = -A_a$. 
For a smooth real-valued function $f(U)$ on $G$, 
we define its derivative 
$Df(U) = T_a D^a f(U) = -T_a D_a f(U) \in \LieG$ 
from the variation as 
\begin{align}
  \delta f(U) = \tr \delta U U^{-1} Df(U).
\end{align}
This can also be written as 
$D_a f(U) \equiv (d/dt)\,f(e^{tT_a}U)|_{t=0}$. 
Throughout this appendix, 
we employ the concepts and technical terms 
introduced in Appendix~\ref{sec:hamiltonian_dynamics}.

\subsection{Hamiltonian dynamics on a compact group}
\label{sec:G_hd}

Our phase space is the tangent bundle over $G$, 
\begin{align}
  TG = \{(U,\pi)\,|\,U=(U_{ij}) \in G,\,\pi = (\pi_{ij}) = T_a \pi^a \in T_U G\}.
\end{align} 
To \emph{define} consistent Hamiltonian dynamics on $TG$ 
with a given Hamiltonian of the form
\begin{align}
  H(U,\pi) = \frac{1}{2}\,\tr \pi^\dagger \pi + V(U)
  = -\frac{1}{2}\,\tr \pi^2 + V(U),
\end{align} 
we consider the first-order action 
\begin{align}
  I[U(s),\pi(s)] = \int ds\,\bigl[
    \tr \pi^\dagger \circdot{U} U^{-1} - H(U,\pi)
  \bigr],
\end{align}
where $\circdot{U} \equiv dU/ds$. 
Hamilton's equations are obtained by varying $I[U,\pi]$, 
and are given by 
\begin{align}
  \circdot{U} &= \pi\, U,
\label{G_h1}
\\
  \circdot{\pi} &= DV(U) + [\circdot{U} U^{-1},\pi] = DV(U). 
\label{G_h2}
\end{align}
In components, these are expressed as 
\begin{align}
  \circdot{U}_{ij} &= (\pi\, U)_{ij},
\\
  \circdot{\pi^a} &= D^a V(U) = -D_a V(U).
\end{align}

\subsection{Symplectic structure on $TG$}
\label{sec:G_symplectic}

The first-order action determines the symplectic potential as 
\begin{align}
  a = \tr \pi^\dagger \theta = -\tr \pi\, \theta.
\end{align}
Here, $\theta = dU U^{-1}$ is the (right-invariant) Maurer-Cartan form, 
which satisfies the Maurer-Cartan equation 
\begin{align}
  d \theta = \theta \wedge \theta
  ~~ \Leftrightarrow ~~
  d \theta^a = \frac{1}{2}\,C_{bc}{}^a\,\theta^b \wedge \theta^c,
\end{align}
where $C_{ab}{}^c \in \bbR$ are the structure constants, 
$[T_a, T_b] = C_{ab}{}^c\,T_c$. 
The symplectic potential gives the symplectic form as 
\begin{align}
  \omega 
  &= da
\nonumber
\\
  &= - \tr (d\pi\,\theta + \pi\,\theta \wedge \theta)
\nonumber
\\
  &= \frac{1}{2}\,
  \bigl(
    dU_{ij} ~ d\pi_{ij}
  \bigr)
  \left(
  \begin{array}{cc}
    a_{ij,kl} & \delta_{l i} U^{-1}_{jk} \\
    -\delta_{jk} U^{-1}_{l i} & 0
  \end{array}
  \right)
  \left(
  \begin{array}{c}
    dU_{kl}\\
    d\pi_{kl}
  \end{array}
  \right)
\end{align}
with
\begin{align}
  a_{ij,kl} = 
  -U^{-1}_{jk} (U^{-1} \pi)_{l i}
  + U^{-1}_{l i} (U^{-1} \pi)_{jk}.
\end{align}
For $G=U(n)$, 
the $2\times 2$ block matrix admits an analytical inversion, 
yielding the following Poisson brackets 
(the additional terms in parentheses for $G=SU(n)$ 
will be derived later):
\begin{align}
  \{ U_{ij}, U_{kl} \} &= 0,
\label{G_pb1}
\\
  \{ U_{ij}, \pi_{kl} \} &= -\delta_{il}\, U_{kj}
  ~\Bigl( \,+ \frac{1}{n}\,U_{ij}\,\delta_{kl} \Bigr),
\label{G_pbq}
\\
  \{ \pi_{ij}, U_{kl} \} &= U_{il}\, \delta_{kj}
  ~\Bigl( \,- \frac{1}{n}\,\delta_{ij}\,U_{kl} \Bigr),
\label{G_pb3}
\\
  \{ \pi_{ij}, \pi_{kl} \} &= 
  - \delta_{il}\,\pi_{kj} + \pi_{il}\,\delta_{kj}.
\label{G_pb4}
\end{align}
To derive these Poisson brackets 
in such a manner 
that manifestly preserves the traceless condition on $\pi$ 
when $G = SU(n)$, 
it is preferable to express $\pi$ 
in terms of its Lie algebra components $\pi_a$. 
Then, the symplectic potential becomes 
\begin{align}
  a &= -\pi_a\,\theta^a,
\end{align}
and the symplectic form $\omega = da$ is given by 
\begin{align}
  \omega &= -d\pi_a \wedge \theta^a
  - \frac{1}{2}\,C_{ab}{}^c\,\pi_c\, \theta^a \wedge \theta^b
\nonumber
\\
  &= \frac{1}{2}\,
  \bigl(
    \theta^a ~ d\pi_{a}
  \bigr)
  \left(
  \begin{array}{cc}
    - C_{ab}{}^c\, \pi_c & \delta_a^b \\
    -\delta^a_b & 0
  \end{array}
  \right)
  \left(
  \begin{array}{c}
    \theta^b\\
    d\pi_b
  \end{array}
  \right).
\end{align}
The $2\times 2$ block matrix can be easily inverted to give 
\begin{align}
  \{ U_{ij}, U_{kl} \} &= 0,
\\
  (\{ U, \pi_b \}\, U^{-1} )^a &= -\delta^a_b
  \quad
  \Leftrightarrow
  \quad
  \{ U_{ij}, \pi_a \} = - (T_a U)_{ij},
\\
  (\{ \pi_a, U \}\, U^{-1} )^b &= \delta_a^b
  \quad
  \Leftrightarrow
  \quad
  \{ \pi_a, U_{ij} \} =  (T_a U)_{ij},
\\
  \{ \pi_a, \pi_b \} &= -C_{ab}{}^c\,\pi_c.
\end{align}
Contracting both sides with $T^a_{kl}$ 
and using the identity 
\begin{align}
  T_{a,ij}\,T^a_{kl}
  = \delta_{il}\,\delta_{jk}
  ~\Bigl(\,-\frac{1}{n}\,\delta_{ij}\,\delta_{kl} \Bigr),
\end{align}
we recover the Poisson brackets \eqref{G_pb1}--\eqref{G_pb4}. 

Using the Poisson brackets, 
one obtains the following identities: 
\begin{align}
  \{U, K\} &= \pi\,U,
\qquad
  \{U, V\} = 0,
\\
  \{\pi, K\} &= 0,
\qquad\quad
  \{\pi, V\} = DV,
\end{align}
which reproduce Hamilton's equations \eqref{G_h1} and \eqref{G_h2} 
in the form $\circdot{U} = \{U,H\}$ and $\circdot{\pi} = \{\pi,H\}$.

\subsection{Molecular dynamics (MD) on $G$}
\label{sec:G_md}

With a given Hamiltonian of the form $H(U,\pi) = K(\pi) + V(U)$, 
we introduce the evolution operator $T$ of step size $\Delta s = \epsilon$ 
as 
\begin{align}
  T \equiv T_{V/2}\, T_K\, T_{V/2}
  = T_H + O(\epsilon^3),
\label{G_T}
\end{align}
where the transformation $T_f$ is defined 
for a function $f = f(U,\pi)$ by 
\begin{align}
  T_f \equiv e^{\epsilon\,\{\ast, f\}}.
\end{align}
Expanding the exponentials, 
we find 
\begin{align}
  T_{V/2}\,U
  &= U,
\\
  T_{V/2}\,\pi
  &= \pi + \frac{\epsilon}{2}\,DV(U)
  ~(\,\equiv \pi_{1/2}),
\\
  T_K\,U
  &= e^{\epsilon \pi}\,U,
\\
  T_K\,\pi
  &= \pi.
\end{align}
It follows that 
\begin{align}
  U'
  &\equiv T U
  = T_{V/2}\, T_K\, T_{V/2}\,U
  = T_{V/2}\, e^{\epsilon \pi} U
\nonumber
\\
  &= e^{\epsilon \pi_{1/2}} U,
\\
  \pi'
  &\equiv T \pi
  = T_{V/2}\, T_K\, T_{V/2}\,\pi
  = T_{V/2}\, T_K\, \Bigl[ \pi + \frac{\epsilon}{2}\,DV(U) \Bigr]
\nonumber
\\
  &= T_{V/2}\,\Bigl[ 
  \pi + \frac{\epsilon}{2}\,DV(e^{\epsilon \pi} U) \Bigr]
  = \pi_{1/2} + \frac{\epsilon}{2}\,DV(e^{\epsilon \pi_{1/2}} U) 
\nonumber
\\
  &= \pi_{1/2} + \frac{\epsilon}{2} DV(U'),
\end{align}
which is summarized into the process 
\begin{align}
  \pi_{1/2} &= \pi + \frac{\epsilon}{2}\,DV(U),
\label{G_lf1}
\\
  U' &= e^{\epsilon\,\pi_{1/2}} U,
\label{G_lf2}
\\
  \pi' &= \pi_{1/2} + \frac{\epsilon}{2}\,DV(U').
\label{G_lf3}
\end{align}
When the above is repeated $N_\textrm{MD}$ times, 
the intermediate steps can be combined 
and the full MD trajectory is generated as \cite{Duane:1987de}
\begin{align}
  &\pi_{1/2} = \pi + \frac{\epsilon}{2}\,D V(U)
  \nonumber
  \\
  &U' = e^{\epsilon\,\pi_{1/2}}\,U
  \nonumber
  \\
  &\text{\texttt{for} $i=1,\ldots,N_\textrm{MD}-1$ \texttt{do}}
  \nonumber
  \\
  &~~~~~ \pi_{1/2} \leftarrow \pi_{1/2} + \epsilon\,DV(U')
  \nonumber
  \\
  &~~~~~ U' \leftarrow e^{\epsilon\,\pi_{1/2}}\,U'
  \nonumber
  \\
  &\text{\texttt{end do}}
  \nonumber
  \\
  &\pi' = \pi_{1/2} + \frac{\epsilon}{2}\,DV(U').
\label{lf_G}
\end{align}

The process \eqref{G_lf1}--\eqref{G_lf3} 
satisfies the following three properties: 

\noindent
\underline{(a) Exact reversibility}:
The transformation is invariant under time reversal,
\begin{align}
  U &\to \tilde{U} \equiv U',
\label{G_lf_rev1}
\\
  \pi &\to \tilde{\pi} \equiv -\pi',
\label{G_lf_rev2}
\\
  \pi_{1/2} &\to \tilde{\pi}_{1/2} \equiv - \pi_{1/2},
\label{G_lf_rev3}
\\
  U'&\to \tilde{U}' \equiv U,
\label{G_lf_rev4}
\\
  \pi'&\to \tilde{\pi}' \equiv -\pi.
\label{G_lf_rev5}
\end{align}

\noindent
\underline{(b) Symplecticity}: 
The symplectic form $\omega$ 
and thus the symplectic volume form $d\Omega = \omega^N/N!$ 
are preserved, 
\begin{align}
  \omega' = \omega,
  \qquad
  d\Omega' = d\Omega.
\end{align}

\noindent
\underline{(c) Approximate energy conservation}:
\begin{align}
  H(U',\pi') = H(U,\pi) + O(\epsilon^3).
\end{align}

Properties (a) and (c) can be shown in a similar way to the flat case. 
Property (b) is an immediate consequence of the fact 
that $T$ is a composition of symplectic transformations [Eq.~\eqref{G_T}], 
but can also be shown explicitly by noting that 
\begin{align}
  T_{V/2}\,\theta &= \theta,
  \\
  T_K\,\theta &= (d e^{\epsilon \pi})\,e^{-\epsilon \pi}
  + e^{\epsilon \pi}\,\theta\,e^{-\epsilon \pi},
\end{align}
and thus that%
\footnote{ 
  One can further show that 
  \begin{align}
    Ta = a + \frac{\epsilon}{2}\,d\bigl[ 
    \tr \pi_{1/2}^\dagger \pi_{1/2} - V(U) - V(U') \bigr],
    \nonumber
  \end{align}
  from which it again follows $T\omega = \omega$.
} 
\begin{align}
  T_{V/2}\,a &= - \tr (T_{V/2}\,\pi)\,(T_{V/2}\,\theta)
  = - \tr \Bigl[ \pi + \frac{\epsilon}{2}\,DV(U) \Bigr]\,\theta
\nonumber
\\
  &= a - \frac{\epsilon}{2}\,dV(U),
\\  
  T_K\,a &= - \tr (T_K \pi)\,(T_K\,\theta)
  = - \tr \pi\,\bigl[
    (d e^{\epsilon \pi})\,e^{-\epsilon \pi}
    + e^{\epsilon \pi}\,\theta\,e^{-\epsilon \pi}
    \bigr]
  = a - \tr \pi\,(d e^{\epsilon \pi})\,e^{-\epsilon \pi}
\nonumber
\\
  &= a - \frac{\epsilon}{2}\,d\,[ \tr \pi^2 ],
\end{align}
which leads to 
\begin{align}
  T_{V/2}\,\omega = \omega,
\qquad
  T_K\,\omega = \omega,
\end{align}
and thus, 
\begin{align}
  T \omega = \omega.
\end{align}

\subsection{Hybrid Monte Carlo (HMC) on $G$}
\label{sec:G_hmc}

We aim to compute the expectation values 
of observables $\calO(U)$ with the potential $V(U)$, 
\begin{align}
  \vev{ \calO }
  = \frac{
    \int_G (dU)\,e^{-V(U)}\,\calO(U)
  }{
    \int_G (dU)\,e^{-V(U)}
  }.
\label{G_vev}
\end{align}
Here, $(dU)$ denotes the Haar measure on $G$ 
defined using the Maurer-Cartan form $\theta = dU U^{-1} = T_a \theta^a$ 
as [see Eq.~\eqref{MC0}] 
\begin{align}
  (dU) = \theta^1 \wedge \cdots \wedge \theta^N
  \quad (N=\dimG).
\end{align}
The symplectic form $\omega$ is derived 
from the symplectic potential $a= -\tr\pi\,\theta = -\pi_a \theta^a$ as
\begin{align}
  \omega = da = -\tr [d\pi \wedge \theta + \pi\,\theta\wedge\theta]
  = - d\pi_a \wedge \theta^a 
  - \frac{1}{2}\,C_{ab}{}^c \pi_c\, \theta^a \wedge \theta^b,
\end{align}
and thus, the symplectic volume form $d\Omega$ of $T G$ can be written as follows:
\begin{align}
  d\Omega
  &\equiv \frac{1}{N!}\,\omega^{N}
\nonumber
\\
  &= \frac{1}{N!}\,(-\tr d\pi\wedge\theta - \tr \pi\,\theta\wedge\theta)^N
\nonumber
\\
  &= \frac{1}{N!}\,(-\tr d\pi\wedge\theta)^N
\nonumber
\\
  &= \pm (dU)\,\bigl(\prod_a d\pi_a\bigr).
\label{TG_measure}
\end{align}
Note that terms involving $\theta\wedge\theta$ in $\omega$ vanish 
when taking the $N$-th wedge product $\omega^N$. 
Thus, the expectation value \eqref{G_vev} can be written 
as an integral over $TG$, 
\begin{align}
  \vev{ \calO }
  = \frac{
    \int_{TG} d\Omega\,e^{-H(U,\pi)}\,\calO(U)
  }{
    \int_{TG} d\Omega\,e^{-H(U,\pi)}
  }
\label{G_hmc0}
\end{align}
with $H(U,\pi) = (1/2)\,\tr \pi^\dagger \pi + V(U)
= (1/2)\,\pi_a^2 + V(U)$. 

To simplify expressions, 
we introduce the notation $\Omega^T \equiv (U,-\pi)$ 
for $\Omega = (U,\pi)$. 
As discussed in Sect.~\ref{sec:hmc_flat},
we construct the target distribution $\propto e^{-H(\Omega)} = e^{-H(U,\pi)}$ 
through the following two stochastic processes:

\noindent
\underline{(1) Heat bath for $\pi$}:
\begin{align}
  P_{(1)}(\Omega'\,|\,\Omega) = \frac{1}{(2\pi)^{N/2}}\,
  e^{-(1/2)\,\tr\pi^{\prime\dagger} \pi'}\,
  \delta(U',\,U),
\end{align}
where $\delta(U',\,U)$ is the bi-invariant delta function 
with respect to the Haar measure $(dU)$. 
See Appendix~\ref{sec:Gaussian} 
for an algorithm to generate $\pi \in \LieG$ 
with probability $\propto e^{-(1/2)\,\tr \pi^\dagger \pi}$. 

\noindent
\underline{(2) MD followed by Metropolis test}:

\noindent
Writing $f = T^{N_\textrm{MD}}$ [Eq.~\eqref{lf_G}], 
we update $\Omega$ to $\Omega' = f(\Omega)$
with probability 
$A(\Omega) \equiv \min(1,e^{-H(f(\Omega)) + H(\Omega)})$. 
Otherwise, we update $\Omega$ to $\Omega' = \Omega^T$.%
\footnote{ 
  See, e.g., Ref.~\cite{Horowitz:1991rr}.
} 
The transition probability can be compactly summarized as 
\begin{align}
  P_{(2)}(\Omega'\,|\,\Omega) 
  = A(\Omega)\,\delta_{TG}(\Omega',f(\Omega))
    + [1 - A(\Omega)]\,\delta_{TG}(\Omega',\Omega^T),
\end{align}
where $\delta_{TG}$ 
is the symplectic delta function 
with respect to the symplectic volume form $d\Omega = \pm (dU)(\prod_a d\pi_a)$ 
[see Eq.~\eqref{TG_measure}]. 
Since the momentum is fully refreshed before each MD trajectory 
and our observables depend only on $U$, 
the momentum flip upon rejection 
need not be performed in practice.

The stochastic processes $P_{(k)}$ $(k=1,2)$ 
can be shown to satisfy the detailed balance condition of molecular dynamics 
as in the flat case,
\begin{align}
  P_{(k)}(\Omega'\,|\,\Omega)\,e^{-H(\Omega)}
  = P_{(k)}(\Omega^T\,|\,\Omega^{\prime\,T})\,e^{-H(\Omega^{\prime\,T})}.
\end{align} 
Although neither $P_{(1)}$ nor $P_{(2)}$ is ergodic, 
their product $P \equiv P_{(2)} P_{(1)}$ 
is expected to be ergodic 
to guarantee unique convergence to equilibrium, 
and thus the Markov chain defined by $P$ will give  
$e^{-H(\Omega)} / \int_{TG} d\Omega\,e^{-H(\Omega)}$
as the equilibrium distribution.%

Since the observables depend only on $U$, 
the algorithm can be expressed 
as a stochastic process on $U$ alone \cite{Duane:1987de}: 
\begin{itemize}
  \item
  \underline{Step 1 (momentum refresh)}:\\
  Given $U \in G$, 
  generate $\pi \in T_U G \simeq \LieG$ 
  from the Gaussian distribution 
  $\propto e^{-(1/2)\,\tr\pi^\dagger \pi}$.
  
  \item
  \underline{Step 2 (molecular dynamics)}:\\
  Evolve $(U,\pi) \to (U',\pi')$ using the MD algorithm [Eq.~\eqref{lf_G}].
  
  \item
  \underline{Step 3 (Metropolis test)}:\\
  Accept the proposed $U'$  
  with probability $\min \bigl(1, e^{-[H(U',\pi') - H(U,\pi)]} \bigr)$.
  
\end{itemize}
One can easily verify that 
the marginal transition kernel 
\begin{align}
  P(U' \,|\, U) \equiv 
  \frac{1}{(2\pi)^{N/2}}\,\int 
  \Bigl(\prod_a d\pi'_a\Bigr)\,\Bigl(\prod_a d\pi_a\Bigr)\,
  P(U',\pi'\,|\,U,\pi)\,e^{-(1/2)\,\tr\pi^\dagger\pi}
\end{align}
satisfies the detailed balance condition 
with respect to $e^{-V(U)}$ \cite{Duane:1987de},
\begin{align}
  P(U' \,|\, U)\,e^{-V(U)} = P(U \,|\, U')\,e^{-V(U')},
\end{align}
which implies that 
the process on $U$ reaches the equilibrium distribution 
$e^{-V(U)} / \int (dU)\,e^{-V(U)}$.

\section{Gaussian sampling in $\LieG$ and $\LieGC$}
\label{sec:Gaussian}

Let $\{T_a\}$ $(a = 1,\ldots, N\,(=\dimG))$ be 
a basis of the Lie algebra $\LieG$, 
satisfying 
$T_a^\dagger = -T_a$ and $\tr (T_a T_b) = -\delta_{ab}$. 
We present algorithms 
for generating random elements $X = T_a x^a \in \LieG$ $(x^a \in \bbR)$ 
according to the Gaussian distribution 
$e^{-\vev{X,X}/(2\sigma^2)} = e^{-(x^a)^2/(2\sigma^2)}$, 
\begin{align}
  \vev{x^a} = 0,
  \qquad
  \vev{x^a x^b} = \sigma^2\,\delta^{ab},
\label{Gaussian_LieG}
\end{align}
and also for generating random elements 
$Z = T_a z^a \in \LieGC$ $(z^a \in \bbC)$ 
according to the Gaussian distribution 
$e^{-\vev{Z,Z}/(2\sigma^2)} = e^{- \overline{z^a} z^a/(2\sigma^2)}$, 
\begin{align}
  \vev{z^a} = 0,
  \qquad
  \vev{z^a z^b} = \vev{\overline{z^a} \overline{z^b}} = 0,
  \qquad
  \vev{z^a \overline{z^b} } = 2\sigma^2\,\delta^{ab}.
\label{Gaussian_LieGC}
\end{align}
A direct algorithm is 
to independently generate $x^a$ and $y^a$ ($a=1,\ldots,N$) 
from the Gaussian distributions $\propto e^{-(x^a)^2/(2\sigma^2)}$ 
and $\propto e^{-(y^a)^2/(2\sigma^2)}$, 
respectively, 
and then set $X = T_a x^a$ and $Z = T_a (x^a + i y^a)$. 
In the following, 
we present an algorithm 
that does not use the explicit form of $T_a$. 
Throughout this section, 
we restrict our discussions to the cases 
where $G = SU(n)$ or $G = U(n)$. 
Other classical compact groups can be treated similarly.

\subsection{$G = SU(n)$}
\label{sec:Gaussian_SU(n)}

In this case, 
$N = \dimG = n^2 - 1$.
A random variable $X = T_a x^a \in \mathfrak{su}(n)$ 
satisfying Eq.~\eqref{Gaussian_LieG} 
can be generated by the following two steps:

\noindent
\underline{(1)}: 
Generate two sets of independent Gaussian random variables 
$\xi = (\xi_{ij})$ and $\eta = (\eta_{ij})$,
\begin{align}
  \xi_{ij} \sim e^{-\xi_{ij}^2/(2\sigma^2)},
  \quad
  \eta_{ij} \sim e^{-\eta_{ij}^2/(2\sigma^2)}
  \quad
  (i,j = 1,\ldots,n).
\end{align}

\noindent
\underline{(2)}: Set $X = (X_{ij})$ as 
\begin{align}
  X = \frac{1}{2}\,(\xi - \xi^T) + \frac{i}{2}\,(\eta + \eta^T)
  - \frac{i}{n}\,(\tr\eta)\,1,
\end{align}
where $1$ is the $n\times n$ identity matrix.

To verify that this construction yields the desired distribution, 
note that $x^a$ are given by 
$x^a = -\tr (T_a X) = \tr (T^a X) = T^a_{ji} X_{ij}$. 
Thus, the first moment becomes 
\begin{align}
  \vev{x^a} = T^a_{ji}\,\vev{X_{ij}} 
  = T^a_{ji} \times \vev{\text{(terms linear in $\xi$ and $\eta$)}}= 0.
\end{align}
For the second moment, 
the expectation $\vev{X_{ij} X_{kl}}$ appearing in 
\begin{align}
  \vev{x^a x^b} = T^a_{ji}\,T^b_{lk}\,\vev{X_{ij} X_{kl}}
\end{align}
evaluates to 
\begin{align}
  \vev{X_{ij} X_{kl}}
  &= \frac{1}{4}\,\Bigl[
  \vev{ (\xi_{ij} - \xi_{ji}) (\xi_{kl} - \xi_{lk}) }
  - \Bigvev{
      \Bigl(\eta_{ij} + \eta_{ji} - \frac{2}{n}\,\eta_{pp}\,\delta_{ij}\Bigr) 
      \Bigl(\eta_{kl} + \eta_{lk} - \frac{2}{n}\,\eta_{qq}\,\delta_{kl}\Bigr) }
  \Bigr]
\nonumber
\\
  &= -\sigma^2\,
  \Bigl( \delta_{il}\,\delta_{jk} - \frac{1}{n}\,\delta_{ij}\,\delta_{kl} \Bigr),
\end{align}
so that we have 
\begin{align}
  \vev{x^a x^b} = -\sigma^2\,\tr (T^a T^b) = \sigma^2 \delta^{ab}.
\end{align}

\subsection{$G^\bbC = SU(n)^\bbC = SL(n,\bbC)$}
\label{sec:Gaussian_SL(n,C)}

In this case, 
$N = \dimG = n^2 - 1$.
A random variable $Z = T_a z^a \in \mathfrak{sl}(n,\bbC)$ 
satisfying Eq.~\eqref{Gaussian_LieGC} 
can be generated by the following two steps:

\noindent
\underline{(1)}: Generate two sets of independent Gaussian random variables 
$\xi = (\xi_{ij})$ and $\eta = (\eta_{ij})$,
\begin{align}
  \xi_{ij} \sim e^{-\xi_{ij}^2/(2\sigma^2)},
  \quad
  \eta_{ij} \sim e^{-\eta_{ij}^2/(2\sigma^2)}
  \quad
  (i,j = 1,\ldots,n).
\end{align}

\noindent
\underline{(2)}: Set $Z = (Z_{ij})$ as 
\begin{align}
  Z = \xi + i \eta - \frac{1}{n}\,(\tr\xi + i\, \tr\eta)\,1,
\end{align}
where $1$ is the $n\times n$ identity matrix. 

To verify that this construction yields the desired distribution, 
note that $z^a$ and $\overline{z^b}$ are expressed as 
$z^a = -\tr (T_a Z) = \tr (T^a Z) = T^a_{ji}\,Z_{ij}$ and 
$\overline{z^b} = \tr (T^{b\,\dagger} Z^\dagger) = -\tr(T^b Z^\dagger)
= -T^b_{kl}\,\overline{Z_{kl}}$. 
Thus, the first moment becomes 
\begin{align}
  \vev{z^a} = T^a_{ji}\,\vev{Z_{ij}} 
  = T^a_{ji} \times \vev{\text{(terms linear in $\xi$ and $\eta$)}}= 0.
\end{align}
For the second moment, 
the expectation $\vev{Z_{ij} Z_{kl}}$ appearing in 
\begin{align}
  \vev{z^a z^b} = T^a_{ji}\,T^b_{lk}\,\vev{Z_{ij} Z_{kl}}
\end{align}
evaluates to 
\begin{align}
  \vev{Z_{ij} Z_{kl}}
  &= 
  \Bigvev{
    \Bigl(\xi_{ij} - \frac{1}{n}\,\xi_{pp}\,\delta_{ij}\Bigr) 
    \Bigl(\xi_{kl} - \frac{1}{n}\,\xi_{qq}\,\delta_{kl}\Bigr) }
\nonumber
\\
  &~~~ 
  - \Bigvev{
      \Bigl(\eta_{ij} - \frac{1}{n}\,\eta_{pp}\,\delta_{ij}\Bigr) 
      \Bigl(\eta_{kl} - \frac{1}{n}\,\eta_{qq}\,\delta_{kl}\Bigr) }
\nonumber
\\
  &= 0,
\end{align}
so that we have 
\begin{align}
  \vev{z^a z^b} = 0.
\end{align}
Furthermore, the expectation $\vev{Z_{ij} \overline{Z_{kl}}}$ appearing in 
\begin{align}
  \vev{z^a \overline{z^b}} = -T^a_{ji}\,T^b_{kl}\,\vev{Z_{ij} \overline{Z_{kl}}}
\end{align}
evaluates to  
\begin{align}
  \vev{Z_{ij} \overline{Z_{kl}}}
  &= 
  \Bigvev{
    \Bigl(\xi_{ij} - \frac{1}{n}\,\xi_{pp}\,\delta_{ij}\Bigr) 
    \Bigl(\xi_{kl} - \frac{1}{n}\,\xi_{qq}\,\delta_{kl}\Bigr) }
\nonumber
\\
  &~~~ 
  + \Bigvev{
    \Bigl(\eta_{ij} - \frac{1}{n}\,\eta_{pp}\,\delta_{ij}\Bigr) 
    \Bigl(\eta_{kl} - \frac{1}{n}\,\eta_{qq}\,\delta_{kl}\Bigr) }
\nonumber
\\
  &= 2\sigma^2\,
  \Bigl( \delta_{ik}\,\delta_{jl} - \frac{1}{n}\,\delta_{ij}\,\delta_{kl} \Bigr),
\end{align}
so that we have 
\begin{align}
  \vev{z^a \overline{z^b}} = -2\sigma^2\,\tr (T^a T^b) = 2 \sigma^2 \delta^{ab}.
\end{align}

\subsection{$G = U(n)$}
\label{sec:Gaussian_U(n)}

In this case, 
$N = \dimG = n^2$.
A random variable $X = T_a x^a \in \mathfrak{u}(n)$ 
satisfying Eq.~\eqref{Gaussian_LieG} 
can be generated by the following two steps:

\noindent
\underline{(1)}: Generate two sets of independent Gaussian random variables 
$\xi = (\xi_{ij})$ and $\eta = (\eta_{ij})$,
\begin{align}
  \xi_{ij} \sim e^{-\xi_{ij}^2/(2\sigma^2)},
  \quad
  \eta_{ij} \sim e^{-\eta_{ij}^2/(2\sigma^2)}
  \quad
  (i,j = 1,\ldots,n).
\end{align}

\noindent
\underline{(2)}: Set $X = (X_{ij})$ as 
\begin{align}
  X = \frac{1}{2}\,(\xi - \xi^T) + \frac{i}{2}\,(\eta + \eta^T).
\end{align}

\subsection{$G^\bbC = U(n)^\bbC = GL(n,\bbC)$}
\label{sec:Gaussian_GL(n,C)}

In this case, 
$N = \dimG = n^2$.
A random variable $Z = T_a z^a \in \mathfrak{gl}(n,\bbC)$ 
satisfying Eq.~\eqref{Gaussian_LieGC} 
can be generated by the following two steps:

\noindent
\underline{(1)}: Generate two sets of independent Gaussian random variables 
$\xi = (\xi_{ij})$ and $\eta = (\eta_{ij})$,
\begin{align}
  \xi_{ij} \sim e^{-\xi_{ij}^2/(2\sigma^2)},
  \quad
  \eta_{ij} \sim e^{-\eta_{ij}^2/(2\sigma^2)}
  \quad
  (i,j = 1,\ldots,n).
\end{align}

\noindent
\underline{(2)}: Set $Z = (Z_{ij})$ as 
\begin{align}
  Z = \xi + i \eta.
\end{align}

\section{Proof of Eq.~\eqref{DaDb}}
\label{sec:DaDb}

To prove this equation, 
we first expand the function $f(e^\xi e^\eta U)$ 
with respect to $\xi,\,\eta \in \LieGC$ to second order:
\begin{align}
  f(e^\xi e^\eta U) 
  &= 
  f(e^\eta U) + \xi^a\,D_a f(e^\eta U)
  + \frac{1}{2}\,\xi^a \xi^b\,D_b D_a f(e^\eta U) + O(\xi^3)
\nonumber
\\
  &= f(U) + (\xi+\eta)^a\,D_a f(U) 
  + \Bigl[ \frac{1}{2}\,(\xi^a \xi^b + \eta^a \eta^b)
    + \xi^a \eta^b \Bigr]\,D_b D_a f(U) + \cdots,
\end{align}
from which we find the difference 
\begin{align}
  f(e^\xi e^\eta U) - f(e^\eta e^\xi U)
  = -\xi^a \eta^b\,\bigl[ D_a D_b f(U) - D_b D_a f(U) \bigr] + \cdots. 
\label{formula1_proof1}
\end{align}
On the other hand, 
the left-hand side can be rewritten using the BCH formula as 
\begin{align}
  f(e^{\xi + \eta + (1/2)[\xi,\eta] + \cdots} U)
  - f(e^{\xi + \eta - (1/2)[\xi,\eta] + \cdots} U)
  &= \tr \bigl( [\xi,\eta]\,Df(U) \bigr) + \cdots
\nonumber
\\
  &= \xi^a \eta^b\,C_{ab}{}^c D_c f(U) + \cdots.
\label{formula1_proof2}
\end{align}
By comparing the coefficients of the $\xi^a \eta^b$ term 
in Eqs.~\eqref{formula1_proof1} and \eqref{formula1_proof2}, 
we obtain the commutation relation \eqref{DaDb}.

\section{Proof of Eqs.~\eqref{T_V/2_U}--\eqref{T_K_pi}}
\label{sec:T_V/2_U-T_K_pi}

Equations \eqref{T_V/2_U}, \eqref{T_V/2_pi} and \eqref{T_K_pi} 
follow directly from the Poisson brackets \eqref{U_U}--\eqref{pi_piH}, 
as shown below:
\begin{align}
  &\bullet~~
  T_{V/2}\,U = e^{(\epsilon/2)\,\{\ast,V\}}\,U
  = U,
\\
  &\bullet~~
  T_{V/2}\,\pi = e^{(\epsilon/2)\,\{\ast,V\}}\,\pi
  = \pi + \frac{\epsilon}{2}\,\{\pi,V\} 
  = \pi - \epsilon\,(DV)^\dagger,
\\
&\bullet~~
  T_K\,\pi = e^{\epsilon\,\{\ast,K\}}\,\pi
  = e^{\epsilon\,[\pi-\pi^\dagger,\ast]}\,\pi
  = e^{\epsilon\,(\pi-\pi^\dagger)}\,\pi\,e^{-\epsilon\,(\pi-\pi^\dagger)},
\end{align}
where we have used the formula \eqref{f_K}. 
In order to show Eq.~\eqref{T_K_U}, 
we notice that $T_K\,U$ can be rewritten in the form
\begin{align}
  T_K\,U 
  = f(\pi,\pi^\dagger)\,U
  \equiv \Bigl[ \sum_{n=0}^\infty\,\frac{\epsilon^n}{n!}\,
  f_n(\pi,\pi^\dagger) \Bigr]\,U.
\end{align}
Here, the coefficients $f_n(\pi,\pi^\dagger)$ are determined recursively 
with $f_0=1$ as follows: 
\begin{align}
  f_{n+1} = [\pi-\pi^\dagger, f_n] + f_n\,\pi,
\label{rec1}
\end{align}
because
\begin{align}
  f_{n+1}\,U = \{f_n\,U, K\} 
  = \{ f_n, K \}\,U + f_n\,\{U, K\}
  = \bigl( [\pi-\pi^\dagger, f_n] + f_n\,\pi\bigr)\,U,
\end{align}
where we have again used the formula \eqref{f_K}.
The right-hand side of Eq.~\eqref{rec1} can be further rewritten as  
\begin{align}
  (\pi-\pi^\dagger) f_n + f_n \pi^\dagger
  = (L_{\pi-\pi^\dagger} + R_{\pi^\dagger})\,f_n,
\label{rec2}
\end{align}
where $L_X$ and $R_X$ denote left and right multiplication by the matrix $X$, 
respectively,
\begin{align}
  L_X f \equiv X f,\quad
  R_X f \equiv f X,
\end{align}
which commute with each other, $L_X R_Y = R_Y L_X$ 
(for all $X,\,Y$). 
We thus have 
\begin{align}
  f_n(\pi,\pi^\dagger) = (L_{\pi-\pi^\dagger} + R_{\pi^\dagger})^n\,1,
\end{align}
which leads to 
\begin{align}
  \sum_{n=0}^\infty
  \frac{\epsilon^n}{n!}\,f_n(\pi,\pi^\dagger)
  = e^{\epsilon L_{\pi-\pi^\dagger} + \epsilon R_{\pi^\dagger}}\,1
  = e^{\epsilon L_{\pi-\pi^\dagger}}\,e^{\epsilon R_{\pi^\dagger}}\,1
  = e^{\epsilon(\pi-\pi^\dagger)}\,e^{\epsilon \pi^\dagger}.
\end{align}
We thus have derived the desired equation [Eq.~\eqref{T_K_U}]:
\begin{align}
  \bullet~~
  T_K\,U = e^{\epsilon(\pi-\pi^\dagger)}\,e^{\epsilon\pi^\dagger}\,U.
\end{align}

\section{Proof of Eqs.~\eqref{rev1}--\eqref{rev5}}
\label{sec:rev}

To demonstrate property (a) in Sect.~\ref{sec:unconstrained_discrete} [Eqs.~\eqref{rev1}--\eqref{rev5}], 
we first note the relation 
\begin{align}
  \tilde\pi_{1/2} - \tilde\pi_{1/2}^\dagger
  = -(\pi_{1/2} - \pi_{1/2}^\dagger),
\end{align}
which can be proved as 
\begin{align}
  \text{(lhs)} = 
  - e^{\epsilon\,(\pi_{1/2} - \pi_{1/2}^\dagger)}\,
  (\pi_{1/2} - \pi_{1/2}^\dagger)\,
  e^{-\epsilon\,(\pi_{1/2} - \pi_{1/2}^\dagger)}
  = \text{(rhs)}.
\end{align}
This relation allows Eq.~\eqref{rev3} to be rewritten as 
\begin{align}
  \pi_{1/2}
  = -e^{\epsilon\,(\tilde\pi_{1/2} - \tilde\pi_{1/2}^\dagger)}\,\tilde\pi_{1/2}\,
  e^{-\epsilon\,(\tilde\pi_{1/2} - \tilde\pi_{1/2}^\dagger)}.
\end{align}
Then, by using Eq.~\eqref{md1} we have 
\begin{align}
  \tilde\pi' &= -\pi
\nonumber
\\
  &= -\pi_{1/2} - \epsilon\,[DV(U)]^\dagger
\nonumber
\\
  &= e^{\epsilon\,(\tilde\pi_{1/2} - \tilde\pi_{1/2}^\dagger)}\,
  \tilde\pi_{1/2}\,e^{-\epsilon\,(\tilde\pi_{1/2} - \tilde\pi_{1/2}^\dagger)}
  - \epsilon\,[DV(\tilde{U}')]^\dagger,
\end{align}
which is the time-reversed form of Eq.~\eqref{md3}. 
Similarly, using Eq.~\eqref{md3}, 
we obtain
\begin{align}
  \tilde\pi_{1/2} 
  &= -e^{\epsilon\,(\pi_{1/2} - \pi_{1/2}^\dagger)}\,\pi_{1/2}\,
  e^{-\epsilon\,(\pi_{1/2} - \pi_{1/2}^\dagger)}
\nonumber
\\
  &= -\pi' - \epsilon\,[DV(U')]^\dagger
\nonumber
\\
  &= \tilde\pi - \epsilon\,[DV(\tilde{U})]^\dagger,
\end{align}
which is the time-reversed form of Eq.~\eqref{md1}.
To show the remaining part,
we note the relation%
\footnote{ 
  This can be proved as
  \begin{align}
    e^{-\epsilon\,(\tilde\pi_{1/2} - \tilde\pi_{1/2}^\dagger)}\,
    e^{-\epsilon\,\pi_{1/2}^\dagger}\,
    e^{-\epsilon\,(\pi_{1/2} - \pi_{1/2}^\dagger)}
    &= 
    e^{\epsilon\,(\pi_{1/2} - \pi_{1/2}^\dagger)}\,
    e^{-\epsilon\,\pi_{1/2}^\dagger}\,
    e^{-\epsilon\,(\pi_{1/2} - \pi_{1/2}^\dagger)}
    \nonumber
    \\
    &= \exp\bigl[
      -\epsilon\,e^{\epsilon\,(\pi_{1/2} - \pi_{1/2}^\dagger)}\,
      \pi_{1/2}^\dagger\,
      e^{-\epsilon\,(\pi_{1/2} - \pi_{1/2}^\dagger)}\,
    \bigr]
    \nonumber
    \\
    &= e^{\epsilon\,\tilde\pi_{1/2}^\dagger}.
  \nonumber
  \end{align}
} 
\begin{align}
  e^{-\epsilon\,\pi_{1/2}^\dagger}\,
  e^{-\epsilon\,(\pi_{1/2} - \pi_{1/2}^\dagger)}
  =
  e^{\epsilon\,(\tilde\pi_{1/2} - \tilde\pi_{1/2}^\dagger)}
  e^{\epsilon\,\tilde\pi_{1/2}^\dagger}.
\end{align}
Then, we have 
\begin{align}
  \tilde{U}' &= U
\nonumber
\\
  &= e^{-\epsilon\,\pi_{1/2}^\dagger}\,
  e^{-\epsilon\,(\pi_{1/2} - \pi_{1/2}^\dagger)}\,U'
\nonumber
\\
  &= e^{\epsilon\,(\tilde\pi_{1/2} - \tilde\pi_{1/2}^\dagger)}
  e^{\epsilon\,\tilde\pi_{1/2}^\dagger}\,\tilde{U},
\end{align}
which is the time-reversed form of Eq.~\eqref{md2}.

\section{Proof of Eq.~\eqref{ene_consv2}}
\label{sec:ene_consv2}

To prove property (c) in Sect.~\ref{sec:constrained_discrete} 
[Eq.~\eqref{ene_consv2}], 
we introduce intermediate variables, 
\begin{align}
  \bar\pi 
  &\equiv \pi_{1/2} + \epsilon\,[DV(U)]^\dagger
  ~(\,= \pi - \epsilon\,\lambda),
\\
  \bar\pi' 
  &\equiv e^{\epsilon\,(\pi_{1/2} - \pi_{1/2}^\dagger)}\,\pi_{1/2}\,
  e^{-\epsilon\,(\pi_{1/2} - \pi_{1/2}^\dagger)}
  - \epsilon\,[DV(U')]^\dagger
  ~(\,= \pi' + \epsilon\,\lambda').
\end{align}
We then have 
\begin{align}
  \pi_{1/2} &= \bar\pi - \epsilon\,[DV(U)]^\dagger
\\
  U' &= e^{\epsilon\,(\pi_{1/2} - \pi_{1/2}^\dagger)}\,
  e^{\epsilon\,\pi_{1/2}^\dagger}\,U,
\\
  \bar\pi' &= e^{\epsilon\,(\pi_{1/2} - \pi_{1/2}^\dagger)}\,\pi_{1/2}\,
  e^{-\epsilon\,(\pi_{1/2} - \pi_{1/2}^\dagger)}
  - \epsilon\,[DV(U')]^\dagger.
\end{align}
This has the same form as Eqs.~\eqref{md1}--\eqref{md3}, 
and thus we have  
\begin{align}
  H(U',\bar\pi') = H(U,\bar\pi) + O(\epsilon^3). 
\end{align}
Furthermore, we have
\begin{align}
  H(U,\bar\pi) 
  &= H(U, \pi-\epsilon\,\lambda)
  = \frac{1}{2}\,\vev{\pi-\epsilon\,\lambda, \pi-\epsilon\,\lambda}
  + V(U)
\nonumber
\\
  &= H(U,\pi) + \frac{\epsilon^2}{2}\,\vev{\lambda,\lambda}
\nonumber
\\
  &= H(U,\pi) + \frac{\epsilon^2}{2}\,
  \vev{\lambda_\textrm{cont},\lambda_\textrm{cont}} + O(\epsilon^3),
\end{align}
and 
\begin{align}
  H(U',\bar\pi') 
  &= H(U', \pi'+\epsilon\,\lambda')
  = \frac{1}{2}\,\vev{\pi'+\epsilon\,\lambda', \pi'+\epsilon\,\lambda'}
  + V(U')
\nonumber
\\
  &= H(U',\pi') + \frac{\epsilon^2}{2}\,\vev{\lambda',\lambda'}
\nonumber
\\
  &= H(U',\pi') + \frac{\epsilon^2}{2}\,
  \vev{\lambda_\textrm{cont},\lambda_\textrm{cont}} + O(\epsilon^3).
\end{align}
Here, we have used the identities 
$\vev{\pi,\lambda} = \vev{\pi',\lambda'} = 0$  
and the expansions $\lambda = \lambda_\textrm{cont} + O(\epsilon)$,
$\lambda' = \lambda_\textrm{cont} + O(\epsilon)$,
where $\lambda_\textrm{cont}$ is the Lagrange multiplier in the continuum version 
[Eq.~\eqref{lambda_cont}].
We thus have 
\begin{align}
  H(U',\pi') - H(U,\pi)
  = H(U',\bar\pi') - H(U,\bar\pi) + O(\epsilon^3)
  = O(\epsilon^3).
\end{align}

\section{Proof of Eq.~\eqref{rattle_symplecticity}}
\label{sec:symplecticity}

Recall that the RATTLE update on $T\calS$ 
is given by Eqs.~\eqref{rattle1}--\eqref{rattle3}, 
which we reproduce here: 
\begin{align}
  \pi_{1/2} &= \pi - \epsilon\,[DV(U)]^\dagger
  - \epsilon\,\lambda,
\label{rattle1b}
\\
  U' &= e^{\epsilon\,(\pi_{1/2} - \pi_{1/2}^\dagger)}\,
  e^{\epsilon\,\pi_{1/2}^\dagger}\,U,
\label{rattle2b}
\\
  \pi' &= e^{\epsilon\,(\pi_{1/2} - \pi_{1/2}^\dagger)}\,\pi_{1/2}\,
  e^{-\epsilon\,(\pi_{1/2} - \pi_{1/2}^\dagger)}
  - \epsilon\,[DV(U')]^\dagger
  - \epsilon\,\lambda',
\label{rattle3b}
\end{align}
where the Lagrange multipliers 
$\lambda \in N_U\calS$ and $\lambda' \in N_{U'}\calS$ are determined 
such that $U' \in \calS$ and $\pi' \in T_{U'}\calS$, respectively. 
In this appendix, 
we prove that the symplectic potential 
$a_\calS \equiv \vev{\pi,\theta_\calS}$ transforms as 
\begin{align}
  a'_\calS = a_\calS + \frac{\epsilon}{2}\,d\bigl[
  \tr \pi_{1/2}^\dagger \pi_{1/2}
  - V(U) - V(U')
  \bigr],
\label{claim}
\end{align}
from which the symplecticity of the RATTLE update follows:
\begin{align}
  \omega'_\calS = da'_\calS = da_\calS = \omega_\calS.
\end{align}

We first note that $a'_\calS$ can be rewritten as follows:  
\begin{align}
  a'_\calS 
  &= \vev{\pi', \theta'_\calS}
  = \vev{
  e^{\epsilon\,(\pi_{1/2} - \pi_{1/2}^\dagger)}\,\pi_{1/2}\,
    e^{-\epsilon\,(\pi_{1/2} - \pi_{1/2}^\dagger)}
    - \epsilon\,[DV(U')]^\dagger
    - \epsilon\,\lambda'
    ,\,\theta'_\calS}
\nonumber
\\
  &= \vev{
   e^{\epsilon\,(\pi_{1/2} - \pi_{1/2}^\dagger)}\,\pi_{1/2}\,
   e^{-\epsilon\,(\pi_{1/2} - \pi_{1/2}^\dagger)}
  ,\,\theta'_\calS
  } 
  - \frac{\epsilon}{2}\,dV(U'),
\label{a'}
\end{align}
where we have used Eq.~\eqref{df} 
and $\vev{\lambda',\theta'_\calS} = 0$. 
The first term in Eq.~\eqref{a'} can be expressed as follows, 
using the unitarity of 
$e^{\epsilon(\pi_{1/2}-\pi_{1/2}^\dagger)}$ 
and the cyclicity of the trace: 
\begin{align}
  &\frac{1}{2}\,\tr\bigl[
    \bigl(
    e^{\epsilon\,(\pi_{1/2}-\pi_{1/2}^\dagger)}\,\pi_{1/2}\,
    e^{-\epsilon\,(\pi_{1/2}-\pi_{1/2}^\dagger)}\bigr)^\dagger\cdot
    d(e^{\epsilon\,(\pi_{1/2}-\pi_{1/2}^\dagger)}\,
      e^{\epsilon\, \pi_{1/2}^\dagger} U)\cdot
    U^{-1}\,e^{-\epsilon\,\pi_{1/2}^\dagger}\,
      e^{-\epsilon\,(\pi_{1/2}-\pi_{1/2}^\dagger)}
\nonumber
\\
  &~~~~~~~~ + \text{(h.c.)}
    \bigr]
\nonumber
\\
  &= \frac{1}{2}\,\tr\bigl[
    \pi_{1/2}^\dagger\,
    e^{-\epsilon\,(\pi_{1/2}-\pi_{1/2}^\dagger)}\cdot
    d(e^{\epsilon\,(\pi_{1/2}-\pi_{1/2}^\dagger)}\,
    e^{\epsilon\, \pi_{1/2}^\dagger} U)\cdot
    U^{-1}\,e^{-\epsilon\,\pi_{1/2}^\dagger}
    + \text{(h.c.)}
  \bigr]
\nonumber
\\
  &= \frac{1}{2}\,\tr\bigl[
  \pi_{1/2}^\dagger\,dU U^{-1}
  + \pi_{1/2}^\dagger\,
  e^{-\epsilon\,(\pi_{1/2}-\pi_{1/2}^\dagger)}\cdot
  d(e^{\epsilon\,(\pi_{1/2}-\pi_{1/2}^\dagger)}\,
  e^{\epsilon\, \pi_{1/2}^\dagger})\cdot
  e^{-\epsilon\,\pi_{1/2}^\dagger}
  + \text{(h.c.)}
  \bigr]
\nonumber
\\
  &= \vev{\pi_{1/2}, \theta_\calS}
  + \frac{1}{2}\,\tr\bigl[
    \pi_{1/2}^\dagger\,e^{-\epsilon\,(\pi_{1/2}-\pi_{1/2}^\dagger)}\,
    \bigl( d e^{\epsilon\,(\pi_{1/2}-\pi_{1/2}^\dagger)} \bigr)
     + \text{(h.c.)} \bigr]
\nonumber
\\
  &~~~~~~~~ + \frac{1}{2}\,\tr\bigl[
    \pi_{1/2}^\dagger\,
    (d e^{\epsilon\,\pi_{1/2}^\dagger})\,e^{-\epsilon\,\pi_{1/2}^\dagger}
     + \text{(h.c.)} \bigr]
\nonumber
\\
  &\equiv \vev{\pi_{1/2}, \theta_\calS} + \frac{1}{2}\,[\text{(A)} + \text{(B)}].
\label{first_term_eval}
\end{align}
Here, we have
\begin{align}
  \vev{\pi_{1/2}, \theta_\calS}
  &= \vev{\pi - \epsilon\,[DV(U)]^\dagger 
      - \epsilon \lambda,\, \theta_\calS}
\nonumber
\\
  &= a_\calS -\frac{\epsilon}{2}\,dV(U),
\label{a_half}
\end{align}
where we have again used Eq.~\eqref{df} and $\vev{\lambda,\theta_\calS} = 0$. 
We thus have the expression 
\begin{align}
  a'_\calS 
  &= \vev{\pi_{1/2}, \theta_\calS}
     + \frac{1}{2}\,[\text{(A)} + \text{(B)}] 
     - \frac{\epsilon}{2}\,dV(U') 
\nonumber
\\
  &= a_\calS + \frac{1}{2}\,[\text{(A)} + \text{(B)}] 
     - \frac{\epsilon}{2}\,d \bigl[ V(U) + V(U') \bigr]. 
\label{a'_2}
\end{align}
The terms (A) and (B) can be rewritten as 
\begin{align}
  \text{(A)}
  &= \tr\bigl[
    \pi_{1/2}^\dagger\,e^{-\epsilon\,(\pi_{1/2}-\pi_{1/2}^\dagger)}\,
    \bigl( d e^{\epsilon\,(\pi_{1/2}-\pi_{1/2}^\dagger)} \bigr)
    + 
    \bigl( d e^{-\epsilon\,(\pi_{1/2}-\pi_{1/2}^\dagger)} \bigr)\,
    e^{\epsilon\,(\pi_{1/2}-\pi_{1/2}^\dagger)}\,\pi_{1/2}
  \bigr]
\nonumber
\\
  &= -\tr\bigl[
    (\pi_{1/2}-\pi_{1/2}^\dagger)\,
      e^{-\epsilon\,(\pi_{1/2}-\pi_{1/2}^\dagger)}\,
    \bigl(d e^{\epsilon\,(\pi_{1/2}-\pi_{1/2}^\dagger)}\bigr)
  \bigr],
\\
  \text{(B)}
  &= \tr\bigl[
    \pi_{1/2}^\dagger\,
    (d e^{\epsilon\,\pi_{1/2}^\dagger})\,e^{-\epsilon\,\pi_{1/2}^\dagger}
    + 
    e^{-\epsilon\,\pi_{1/2}}\,\bigl(d e^{\epsilon\,\pi_{1/2}}\bigr)\,\pi_{1/2}
    \bigr]
\nonumber
\\
  &= \tr\bigl[
    \pi_{1/2}^\dagger\,e^{-\epsilon\,\pi_{1/2}^\dagger}\,
    \bigl(d e^{\epsilon\,\pi_{1/2}^\dagger}\bigr)
    + 
    \pi_{1/2}\,e^{-\epsilon\,\pi_{1/2}}\,\bigl(d e^{\epsilon\,\pi_{1/2}}\bigr)
    \bigr].
\end{align}
Note that both (A) and (B) are composed of 
terms of the form $\tr[ X e^{-X} d e^X]$, 
for which the following identity holds:%
\footnote{ 
  This can be proved as follows:
  \begin{align}
    \text{(lhs)}
    &= \int_0^1 dt\,\tr[X e^{-X} \,e^{(1-t)X}\,dX\,e^{tX} ]
    = \int_0^1 dt\,\tr[X dX]
    = \frac{1}{2}\,d\,\tr X^2.
  \nonumber
  \end{align}
} 
\begin{align}
  \tr[ X e^{-X} d e^X] = \frac{1}{2}\,d\, \tr X^2.
\end{align}
We thus have 
\begin{align}
  \text{(A)} + \text{(B)}
  = \frac{\epsilon}{2}\,d\,\tr \bigl[
    -(\pi_{1/2}-\pi_{1/2}^\dagger)^2
    + (\pi_{1/2}^\dagger)^2 + \pi_{1/2}^2
  \bigr]
  = \epsilon\,d\,\tr \pi_{1/2}^\dagger \pi_{1/2}.
\end{align}
Substituting this into Eq.~\eqref{a'_2}, 
we obtain Eq.~\eqref{claim}.

\section{Proof of Eq.~\eqref{wv_force}}
\label{sec:wv_force}

We prove that $[DV(U)]^\dagger$ can be set to the form \eqref{wv_force}, 
following the argument in Ref.~\cite{Fukuma:2023eru} 
(see also Ref.~\cite{Fukuma:2020fez}). 

We start from the expression
\begin{align}
  (D V)^\dagger 
  = \frac{1}{2}\,(D S)^\dagger
  + W'(t)\,(D t)^\dagger
  = \frac{1}{2}\,\xi + W'(t)\,(D t)^\dagger.
\end{align}
We decompose $(D t)^\dagger \in T_U G^\bbC$ 
at $U\in \Sigma_t \,(\subset \calR)$ 
into the form%
\footnote{ 
  The flow-time coordinate $t$ is originally defined on $\calR$. 
  Here, we extend it smoothly to a neighborhood of $\calR$ 
  in $G^\bbC$ and regard it as a real-valued function 
  $t(U,U^\dagger)$. 
  Different extensions change $(Dt)^\dagger$ 
  only by a component in $N_U\calR$, 
  which does not affect the constrained dynamics, as explained below. 
} 
\begin{align}
  (D t)^\dagger 
  = (D t)^\dagger_\parallel + (D t)^\dagger_\perp
  = v + c\,\xi_n + (D t)^\dagger_\perp
\end{align}
with $v = E_a v^a \in T_U\Sigma_t\,(\subset T_U\calR)$. 
Note that $\xi_n\in N_U\Sigma_t\cap T_U\calR$ and 
$(D t)^\dagger_\perp\in N_U\calR$. 
In the following, 
we show that (1)~$v=0$ and (2)~$c=1/(2\,\vev{\xi_n,\xi_n})$. 
These results establish Eq.~\eqref{wv_force}, 
because $(D t)^\dagger_\perp\in N_U\calR$ 
(appearing only in $Z$ [Eq.~\eqref{wv_rattle1}]) can be absorbed 
into $\lambda$ in Eq.~\eqref{wv_rattle2}, 
and $(D t')^\dagger_\perp\in N_{U'}\calR$ 
drops out under the projection Eq.~\eqref{wv_rattle4}.

\underline{(1)} 
For any $U\in\Sigma_t$, any $u = E_a u^a \in T_U\Sigma_t$,
and an infinitesimally small $\epsilon$, 
we have 
\begin{align}
  t(U,U^\dagger) 
  = t(e^{\epsilon u}U, (e^{\epsilon u}U)^\dagger ) + O(\epsilon^2)
  = t(U,U^\dagger)
    + \epsilon\,\tr [u\, D t + u^\dagger \,(D t)^\dagger]
    + O(\epsilon^2),
\end{align}
and thus,
\begin{align}
  0 = \tr [u\, D t + u^\dagger \,(D t)^\dagger]
    = 2\,\vev{u,(D t)^\dagger}
    = 2\,\vev{u,v}
    = 2\,u^a \gamma_{ab} v^b.
\end{align}
This means that $v=0$ due to the nondegeneracy of $\gamma_{ab}$. 

\underline{(2)} 
Using orthogonality, 
$c$ is given by 
\begin{align}
  c = \frac{\vev{\xi_n, (D t)^\dagger}}
      {\vev{\xi_n,\xi_n}}.
\end{align}
Here, noting that
\begin{align}
  1 &= [t(U,U^\dagger)]^\centerdot
    = \tr [ (\dot{U} U^{-1}) Dt 
      + (\dot{U} U^{-1})^\dagger\,(D t)^\dagger ]
\nonumber\\
  &= 2\,\vev{\xi,(D t)^\dagger}
  = 2\,\vev{\xi_n,(D t)^\dagger}
  \quad(\because v=0),
\end{align}
we have $\vev{\xi_n,(D t)^\dagger} = 1/2$.
We thus obtain $c = 1/(2\,\vev{\xi_n,\xi_n} )$.

\baselineskip=0.9\normalbaselineskip



\end{document}